\documentclass{aa}  
\usepackage{graphicx}
\usepackage{txfonts}
\PassOptionsToPackage{hyphens}{url}\usepackage[breaklinks]{hyperref}
\usepackage{xurl}
\usepackage{changes}
\usepackage{natbib}
\usepackage{pdflscape}
\usepackage{longtable}
\usepackage{supertabular,booktabs}
\usepackage{multicol}
\usepackage{longtable}

\usepackage{float}
\usepackage{subfig}

\newcommand{\mygi}{MyGIsFOS}
\newcommand{\logg}{\ensuremath{\log\,g}}
\def\teff{$T\rm_{eff}$}
\newcommand{\kms}{$\rm km s ^{-1}$}

\usepackage{hyperref}
\hypersetup{colorlinks=true,linkcolor=blue,citecolor=blue,filecolor=blue,urlcolor=blue}
\begin{document} 

\title{MINCE III. Detailed chemical analysis of the UVES sample\thanks{This paper is based on data collected with the Very Large Telescope (VLT) at the European Southern Observatory (ESO) on Paranal, Chile (ESO Program ID: 105.20ML.001, 106.21PG.001)}}
\titlerunning{MINCE III}

\author{
F.~Lucertini \inst{1} \and
L.~Sbordone \inst{1} \and
E.~Caffau    \inst{2,3} \and
P.~Bonifacio \inst{2,3} \and
L.~Monaco    \inst{4,3} \and
G.~Cescutti \inst{5,3,6} \and
R.~Lallement \inst{2} \and 
P.~Fran\c{c}ois \inst{2,8} \and
E.~Spitoni \inst{3} \and
C.~J.~Hansen \inst{7} \and
A.~J.~Korn \inst{9} \and
A.~Ku\v{c}inskas \inst{10}\and
A.~Mucciarelli \inst{11,12} \and
L. Magrini \inst{13} \and
L.~Lombardo \inst{7} \and
M.~Franchini \inst{3} \and
R.~F.~de Melo \inst{14}
}

\institute{
ESO - European Southern Observatory, Alonso de Cordova 3107, Vitacura, Santiago, Chile
\and
GEPI, Observatoire de Paris, Universit\'{e} PSL, CNRS,  5 Place Jules Janssen, 92190 Meudon, France
\and INAF-Osservatorio  Astronomico  di  Trieste,  Via  G.B.  Tiepolo  11,34143 Trieste, Italy
\and
Universidad Andres Bello, Facultad de Ciencias Exactas, Departamento de F{\'i}sicas y Astronom{\'i}a - Instituto de Astrof{\'i}sica, Autopista
Concepci\'on-Talcahuano, 7100, Talcahuano, Chile
\and Dipartimento di Fisica, Sezione di Astronomia, Università di Trieste, Via G. B. Tiepolo 11, 34143 Trieste, Italy
\and INFN, Sezione di Trieste, Via A. Valerio 2, 34127 Trieste, Italy 
\and
Institute for Applied Physics, Goethe University Frankfurt,  
Max-von-Laue-Strasse 1, 60438 Frankfurt am Main, Germany.
\and
UPJV, Universit\'e de Picardie Jules Verne, P\^ole Scientifique, 33 rue St Leu, 80039, Amiens, France
\and
Division of Astronomy and Space Physics, Department of Physics and Astronomy, Uppsala University, Box 516, 75120 Uppsala, Sweden
\and
Institute of Theoretical Physics and Astronomy, Vilnius University, Saul\.{e}tekio al. 3, 10257 Vilnius, Lithuania
\and
 Dipartimento di Fisica e Astronomia “Augusto Righi”, Alma Mater Studiorum, Universitá di Bologna, Via Gobetti 93/2,
40129 Bologna, Italy
\and
INAF – Osservatorio di Astrofisica e Scienza dello Spazio di Bologna, Via Gobetti 93/3, 40129 Bologna, Italy
\and
INAF – Osservatorio Astrofisico di Arcetri, Largo E. Fermi, 5, 50125, Firenze, Italy
\and
Goethe University Frankfurt, Institute for Applied Physics, Max-von-Laue-Str. 12, 60438 Frankfurt am Main, Germany
}

   \date{Received November 11, 2024; accepted January 17, 2025}

  \abstract
{The MINCE (Measuring at Intermediate Metallicity Neutron-Capture Elements) project aims to provide high quality neutron-capture abundances measurements in several hundred stars at intermediate metallicity, $-2.5 <$ [Fe/H] $<-1.5$. This project will shed light on the origin of the neutron-capture elements 
and the chemical enrichment of the Milky Way.}
{The goal of this work is to chemically characterize the second sample of the MINCE project and compare the abundances with the galactic chemical evolution model at our disposal.}
{We performed a standard abundance analysis based on 1D LTE model atmospheres on high-resolution and high-signal-to-noise-ratio UVES spectra.}
{We provide the kinematic classification (i.e., thin disk, thick disk, thin-to-thick disk, halo, Gaia Sausage Enceladus, Sequoia) of 99 stars and the atmospheric parameters for almost all stars.
We derive the abundances for light elements (from Na to Zn) and neutron-capture elements (Rb, Sr, Y, Zr, Ba, La, Ce, Pr, Nd, Sm, Eu) in a subsample of 32 stars in the metallicity range $-2.5 <$ [Fe/H] $< -1.00$. 
In the subsample of 32 stars, we identify 8 active stars exhibiting (inverse) P-Cygni profile and one Li-rich star, CD\,28-11039. We find a general agreement between the chemical abundances and the stochastic model computed for the chemical evolution of the Milky Way halo for the elements Mg, Ca, Si, Ti, Sc, Mn, Co, Ni, Zn, Rb, Sr, Y, Zr, Ba, La, and Eu .}
{The MINCE project has already significantly increased the number of neutron-capture elements measurements in the intermediate metallicity range. 
The results from this sample are in perfect agreement with the previous MINCE sample. 
The good agreement between the chemical abundances and the chemical evolution model of the Galaxy supports the nucleosynthetic processes adopted to describe the origin of the n-capture elements.
}

   \maketitle
\section{Introduction}
One of the astronomical ways to constrain the Galaxy's formation and evolution is a detailed chemical analysis of its stellar populations.
The main ingredients for this purpose are a complete census of the stellar populations hosted by the Galaxy, and a deep investigation of the different nucleosynthesis processes.
The goal of the MINCE project (Measuring at Intermediate Metallicity Neutron-Capture Elements, see \citealt{mince1}) is to fill in with chemical abundances of mildly metal-poor stars ($\rm -2.5 < [Fe/H] < -1.5$).

The understanding of early Galactic neutrosynthesis processes and chemical evolution led to hunt the ever more metal-poor stars. This effort increased with the discovery of Carbon Enhanced Metal-Poor (CEMP) stars \citep{sneden1994, barbuy97,norris97,bonifacio98,hill2000} characterized by different neutron-capture (n-capture) elements content \citep{cohen2003, beers2005, hansen2012, roederer2014, yong2014}.
At the same time, the n-capture elements abundances in halo stars 
have been analyzed in several Galactic archaeology studies \citep[see e.g.][and references therein]{francois03,francois2007}.
As a result, the old giant stars with intermediate metallicity have been less examined by the scientific community.

The bulk of elements heavier than iron are produced through slow (s), rapid (r) or intermediate (i) neutron capture processes. The difference among these processes lie in the neutron density, which determines the probability of capturing a neutron before the radioactive beta decay \citep{burbidge1957}.
The s-process occurs in
the inter-shell region of asymptotic giant branch (AGB) stars with low- and intermediate-mass \citep{Gallino1998,Busso1999, Karakas2014}.
Both Supernovae and neutron star mergers have been proposed as sites hosting the r-process \citep{argast04, cote2018, watson2019, cavallo2021, cowan2021}.
Despite several suggested sites, the site of the i-process remains an open question \citep{hampel2016,hampel2019, denissenkov2017}.
For a recent review on the origin of the elements, see \citet{arcones_origin_2023}.\\
Another puzzle concerning n-capture elements abundances is their large star-to-star scatter at low metallicities. Indeed, it has been shown that the dispersion in [n-capture/Fe] increases as metallicity decreases.
This finding provides critical insights into the origin of n-capture elements \citep{hansen2014} and the processes of formation of the MW.

In summary, the  measurement of n-capture elements in intermediate metallicity stars will allow us to increase our knowledge on the contribution from the different n-capture processes. 
Consequently, this will provide new insights into the origin of the heavy elements and their progenitor stars may be achieved.
This will also shed light on the early interstellar medium (ISM) pollution process and its timescales \citep{cote2019}, which are fundamental details to improve the Galactic chemical evolution models. 

In this work we present the chemical abundances of the second MINCE sample (hereafter referred to as MINCE\,III), thus the results obtained for light elements (from Na to Zn) and n-capture elements (Rb, Sr, Y, Zr, Ba, La, Ce, Pr, Nd, Sm, Eu), and their comparison with the chemical evolution model of the Galaxy.

\section{Observations and data reduction}
The MINCE project exploits several facilities around the world to collect high-quality data, and provide precise atmospheric parameters and chemical abundances of stars at intermediate metallicity.

In this paper, we present a sample of 99 stars observed with UVES$@$ESO-VLT/UT2 \citep{dekker2000}, Paranal Observatory (Chile). The data were collected between September 2020 and August 2021.
The instrument was set with two configurations: Dichroic$\#1$, blue and red arms centered at 346 nm and 580 nm, respectively, and Dichroic$\#2$, blue and red arms centered at 437 nm and 760 nm, respectively. The final dataset for each star covers the spectral range 304-945 nm.
A slit width of 0.5 arcsec was used, providing a high
resolving power of R$\sim$65000 and R$\sim$75000 in blue and red spectra, respectively. 
The data were reduced using the UVES pipeline (version 5.10.13\footnote{https://www.eso.org/sci/software/pipelines/}). 
The final spectra are characterized by high signal-to-noise ratio (SNR), with mean values of 148 at 580 nm and 105 at 760 nm.

Table \ref{data} lists the coordinates and the $Gaia$ eDR3 \citep{gaiaedr3} magnitudes of the sample.

\subsection{Radial velocities}\label{RV}
The $Gaia$ DR3 \citep{gaiadr3} 
radial velocities (RV) of the targets used in this work are listed in Table \ref{data}, when available.
When the radial velocity was not available from $Gaia$ DR3 it was estimated directly from our UVES spectrum.

These values were used to shift the observed spectra to the rest frame.
However, in the case of six stars we did not find a good match between the synthetic and the observational spectra. For CD-28\,10387, we found an RV difference of 7.49 \kms. For this star, $Gaia$ DR3 provides a ruwe value of 3.63, suggesting a possible binary nature.
In the case of CD-29\,15930 and CD-31\,16658, we applied an additional shift of -4.99 \kms and -7.49 \kms, respectively. The final RV of CD-31\,16922 differs from the $Gaia$ one by 17.47 \kms. 
It is interesting to note that $Gaia$ estimates an error on the RV of about 9.08 \kms for this star. 
The largest shift corrections found are those corresponding to CD-44\,12644 (-29.95 \kms) and TYC\,5422-1192-1 (19.97 \kms).\\
Except for TYC\,5422-1192-1, we do not detect the contribution of a companion in the spectra of the above mentioned stars.
Thus, we can reasonably exclude that these stars are double-lined spectroscopic binary (SB2), and that the companion contributes to the spectrum to explain the disagreements with $Gaia$ RV.
It is however quite likely that these five stars are single lined spectroscopic binaries; the future release of $Gaia$ epoch radial velocities will help to elucidate this issue.

Taking into account the shift estimated, we corrected the $Gaia$ RV and reported the final results in Table \ref{data}, where the results for these stars are highlighted with the superscript $(a)$.

\section{Analysis}
\subsection{Stellar parameters}
Each stellar parameters derivation method leads to systematic errors in the estimation of chemical abundances.
Since the MINCE project has at its disposal several spectra, a homogeneous way to derive the stellar parameters is required to present consistent analysis.
For this reason, we obtained the atmospheric parameters as described in \cite{mince1}.

Starting from the $Gaia$ eDR3 \citep{gaiaedr3} photometry, we first obtained the dereddened $G$ magnitude and $G_{BP}-G_{RP}$ color of the targets. 
Extinction estimates were made for each target star according to the following scheme. The first step uses the extinction density maps presented by \cite{vergely_three_2022}. For each star the extinction density (in mag per pc) is integrated along the paths between the Sun and the star to provide the total extinction. Details about the maps can be found in \cite{vergely_three_2022}. In brief, they were computed by inversion of about 40 million individual extinction measurements for stars with precise distances, mainly $Gaia$ parallaxes. The inversion is regularised on the basis of a covariance kernel (or minimum cloud size) adapted to the volume density of the targets through a hierarchical scheme. In addition to the three maps presented in that work, for close targets we also used an additional unpublished map with a minimum kernel size of 5 pc (courtesy J.L. Vergely). The largest map has a volume of 10 kpc x 10 kpc x 800 pc centred on the Sun (largest dimensions along the Plane, 400 pc maximum distance from the Plane). Given the limits of the volumes covered by the mapping, distant stars and halo stars may be located outside the computational volume. In this case, we also searched for the extinction estimate from \cite{green2019}, if available. For all targets, we also computed an estimate of the total extinction up to large distance outside the dust disk by using Planck dust emission measurements \citep{planck2016}. Specifically, we converted the optical thickness at 353 GHz into an extinction in the visible using $\rm{A_V}= 3.1* 1.5 10^{4} \tau$ (353). We slightly refined this estimate, when possible,  based on Figure\,11 of \cite{remy2018}, using in addition the Planck temperature and the $\beta$ coefficient. For most of the targets located outside the maps, the Planck-based value and the extinction reached at the boundary of the map (or for a few cases the \citealt{green2019} value) are very close to each other (average difference of 0.05 mag, maximum difference of 0.2 mag), meaning that the target is located beyond the majority of the dust. In this case we preferred the maximal value, given the large distance of the targets. The extinction values adopted in this work are reported in Table \ref{atm_param}.

The first guess stellar parameters were obtained comparing synthetic colors with the dereddened $G_{BP}-G_{RP}$ color, the absolute $G$ magnitude and a first guess metallicity.
Running our code MyGIsFOS \citep{mygi} with the first guess of the stellar parameters, we extracted the metallicity. 

\begin{figure}
\centering
\includegraphics[trim= 0cm 0cm 0cm 0cm, clip,width=0.5\textwidth]{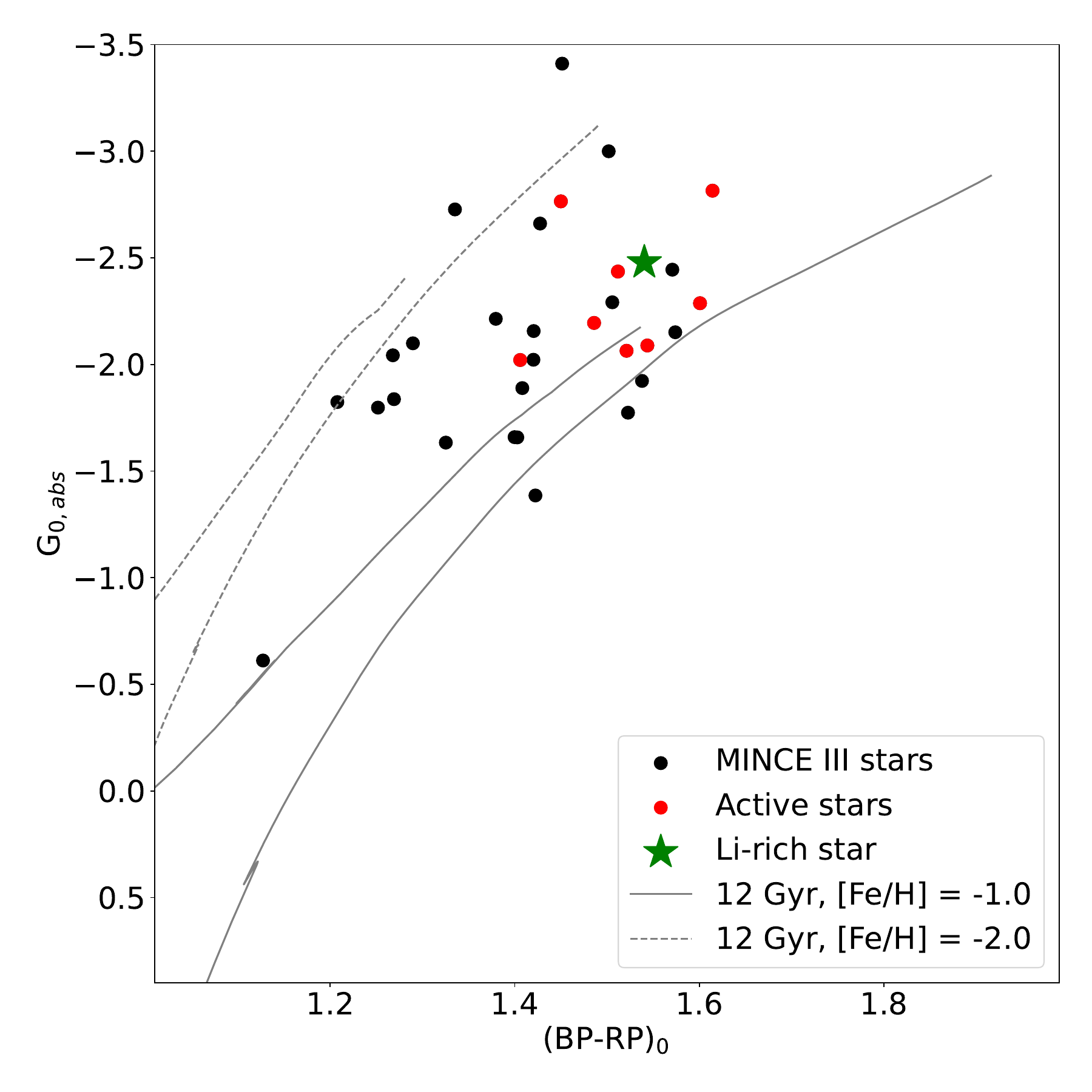}
\caption{$G_0$ versus $(G_{BP}-G_{RP})_0$ diagram of the sample. The stars exhibiting H$\alpha$ emission are reported in red. The green star symbol is the Li-rich star. The CMD is superimposed to the 12 Gyr BASTI isochrones \citep{BASTI21} with metallicities [Fe/H] = $-1.0$ (solid grey line) and [Fe/H] = $-2.0$ (dashed grey line).}
\label{fig:cmd}
\end{figure}

As discussed in \cite{mince1}, the initial selection of the sample allowed the presence both of stars that are more metal-rich than --1.0
and of stars with \teff\ $\le 4000$\,K, for both these kinds the spectral analysis is more complex due to line blending especially in the blue.
A first scrutiny of the spectra resulted in defining a 
sample of 32 stars for which we were confident that a complete chemical inventory could be derived.
Figure \ref{fig:cmd} shows the $G_{0}$ versus $(G_{BP}-G_{RP})_0$ color magnitude diagram (CMD) of the targets. We also highlight the stars exhibiting H$\alpha$ emission (red dots, see \ref{sec_Halpha}) and the Li-rich star (green symbol, see Sec.\,\ref{sec_Li_rich}).\\

For the remaining stars with effective temperature above 4000\,K we run \mygi\ limiting ourselves to the
437+760 spectral range, using the same grids and line selections detailed in \citet{rvs1,rvs2}.
For the stars with \teff $< 4000$\,K we limited the analysis to the interval 840\,nm -- 874\,nm, that corresponds
roughly to the range covered by $Gaia$ RVS and it is not contaminated by TiO. We computed a grid of synthetic spectra
based on the grid of ATLAS 9 models of Mucciarelli et al. (in preparation). The 
grid covers effective temperatures from 3750\,K to 5250\,K at steps of 125\,K surface
gravities from 0.0 to  2.5 at steps of 0.5\,dex, metallicities from --2.00 to +0.25 at steps of 0.25\,dex,
microturbulent velocities 1.2 and 3.0 \kms and [$\alpha$/Fe]= 0.0 and 0.4.
For the atomic lines we adopted the list used by $Gaia$ GP-Spec \citep{contursi21}.

For all stars,
we refined the atmospheric parameters
following the steps described above with the metallicities derived by MyGIsFOS, and then we run MyGIsFOS with the new stellar parameters. This process was iterated until it reached negligible changes in \teff \,(10 K) and \logg \,(0.05 dex). At any iteration, the microturbulence was derived from the calibration by \cite{mashonkina2017}.

The atmospheric parameters and the metallicity derived for 94 stars are listed in Table \ref{atm_param}. The 32  stars with a complete chemical
inventory are highlighted in bold. 

Five stars remain excluded from our analysis: two stars classified as RS CVn by  $Gaia$, the SB2 binary CD --52 2441, the fast rotator
CD --31 16922, and the fast rotator TY 5422-1192-1, that is a member of the Open Cluster NGC\,2423, of metallicity +0.12.

\subsection{Kinematics}\label{sec_kinematics}

We evaluated the kinematics of the target stars using the astrometric data (coordinates, proper motions, parallaxes) and radial velocities (RVs) from $Gaia$ DR3 \citep[][]{gaiadr3} together with the {\tt galpy} code\footnote{\url{http://github.com/jobovy/galpy}} \citep[][]{bovy15}. 

We adopted the default {\tt MWPotential2014} Milky Way potential \citep[][]{bovy15}, solar distance and circular velocity at the solar distance of r0=8\,kpc, v0=220\,\kms \citep{bovy12} and the solar motion from \citet[][]{sch10}. In order to evaluate the errors on the calculated quantities, similarly to \citet[][]{bonifacio21}, we employed the {\tt pyia} code \citep[][]{pyia18} to extract, for each star, one thousand realizations of the six input parameters, using multivariate Gaussians, using the errors in the input parameters and taking into account the covariance matrix between the astrometric parameters. 

Parallaxes were zero-point corrected following the prescription of \citet[][]{lindegren2021}\footnote{\url{https://gitlab.com/icc-ub/public/gaiadr3_zeropoint}}. The $Gaia$ DR3 catalog does not report RVs for stars TYC\,5422-1192-1 and CD-52\,2441. For these stars we adopted the RVs measured from the UVES spectra. A negative parallax is assigned to star CD-35\,14807 in $Gaia$ DR3. For this star we adopted the photogeometric distance from \citet[][]{bailer21}.

Following \citet[][]{bensby14}, our targets can be divided into 42 with thin disk kinematics, 22 belonging to the thick disk and one of transition type between them. 34 stars are classified as belonging to the halo. Among the halo stars, 6 and 3 are kinematically similar to the Gaia-Sousage-Enceladus \citep[GSE,][]{belokurov18,haywood18,helmi18} and the Sequoia \citep[Seq,][]{barba19,myeong19,villanova19} structures, respectively, according to the criteria introduced by \citet[][]{feuillet21}. We report in Tables\,\ref{kinematics} and \ref{dynamics} some of the property evaluated for the target stars and their classification.

Putting the focus on the 32 metal-poor stars, none of these belong to the thin disk or are in transition between the thin and the thick disk. The whole sample can be divided into halo (23) and thick disk (9) stars. Moreover, among the halo stars, 4 and 2 are kinematically similar to GSE and Seq, respectively.

\begin{figure*}
\centering
\includegraphics[trim= 0.5cm 0.5cm 0.5cm 0.5cm, clip,width=1\textwidth]{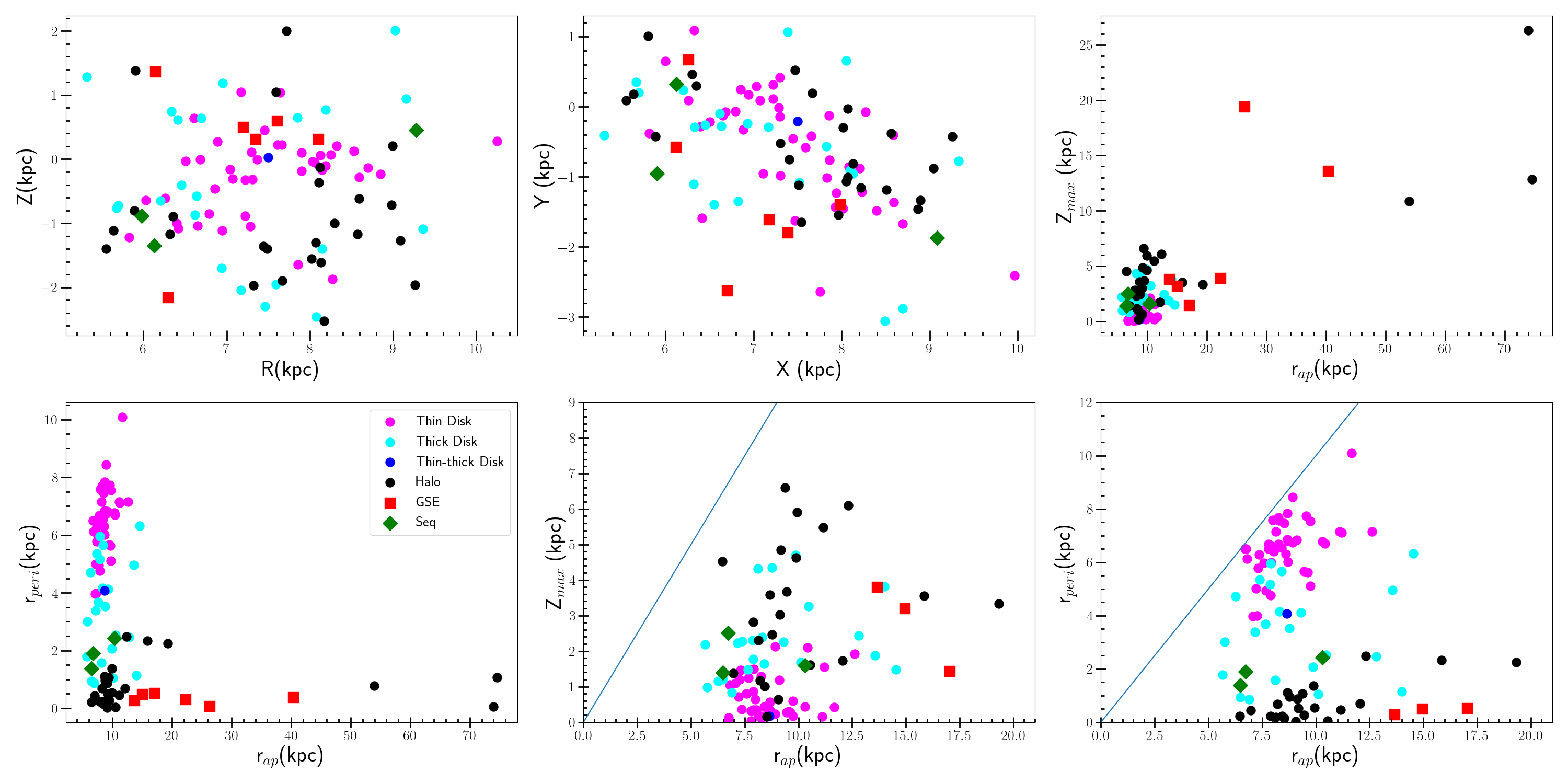}
\caption{Top-panels: left: Galactocentric Cartesian height over the galactic plane Z {\it} vs cylindrical galactocentric radius R. Middle: targets' Galactocentric Cartesian X {\it vs} Y coordinates. Right: maximum height over the galactic plane Z$_{max}$ {\it vs} apocentric distance r$_{ap}$. 
Bottom-panels: left: pericentric distance r$_{peri}$ {\it vs} r$_{ap}$. Middle: zoom-in of the top-right panel. The region above the line is forbidden. Right: zoom-in of the bottom-left panel. The stars with zero eccentricity lie on the line r$_{peri}$=r$_{ap}$. 
Targets classified as belonging to the thin (magenta) or thick disk (cyan), thin-thick transition (blue) and halo (black) are presented as filled circles. Candidate GSE and Seq stars are denoted with red squares and green diamonds, respectively.}
\label{fig:char}
\end{figure*}

The top panels of Fig.\,\ref{fig:char} present the target stars in the Galactocentric cartesian Z  {\it vs} cylindrical radius R plane (top-left), in the Galactocentric cartesian Y {\it vs} X plane (top-middle) and in the maximum height over the Galactic plane Z$_{max}$ {\it vs} the apocentric distance r$_{ap}$ plane (top-right).
The bottom-left panel presents stars in the pericentric distance r$_{peri}$ {\it vs} r$_{ap}$ plane.
The bottom-middle and bottom-right panels are a zoom-in of the top-right and bottom-left panels, respectively. The identity lines are also plotted in these panels to indicate the forbidden region above the line in the bottom-middle panel, and the locus of stars with zero eccentricity (r$_{peri}$=r$_{ap}$) in the bottom-right panel.  

Stars classified as belonging to the thin disk (magenta), thick disk (cyan), thin-thick transition (blue) and the halo (black) are marked as filled circles. Candidate GSE and Seq members are marked with red squares and green diamonds, respectively.

\begin{figure*}
\centering
\includegraphics[trim= 6cm 1cm 0.5cm 3cm, clip,width=1.1\textwidth]{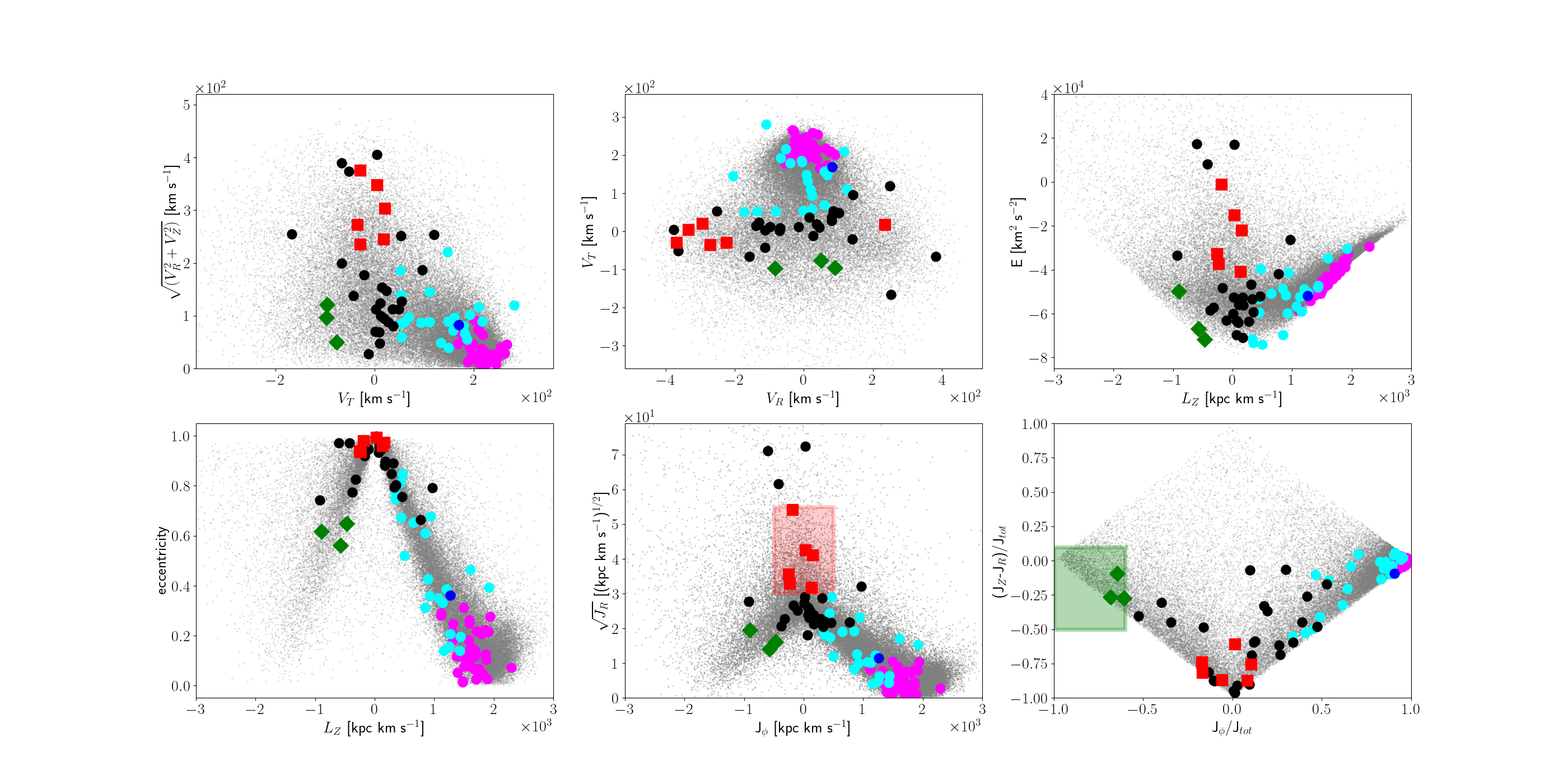}
\caption{Upper panels: left: Toomre diagram ($\sqrt{V_R^2 + V_Z^2}$ \,{\it vs}\, V$_T$). V$_R$, V$_Z$, V$_T$ are the stars' velocity components in galactocentric cylindrical coordinates. Middle: V$_R$ {\it vs} V$_T$. Right: Orbital energy E {\it vs} angular momentum L$_Z$. Bottom panels: left: orbital eccentricity {\it vs} L$_Z$. Middle: $\sqrt{J_R}$ {\it vs} J$_\phi$, where J${_R}$ and J$_\phi$=L$_Z$ indicate the radial and azimuthal actions. Right: action diamond, namely the difference between the vertical and radial actions (J$_Z$-J$_R$) {\it vs} J$_\phi$, both of them normalized to the total action J$_{tot}$=J$_R$+$|$J$_\phi|$+J$_Z$. The red (middle panel) and green (right panel) shaded areas indicate the criteria adopted here from \citet[][]{feuillet21} to select candidate GSE and Seq stars, respectively. Gray dots are stars from the ``good parallax sample" of \citet[][]{bonifacio21}. The colored symbols are the same as in Fig.\ref{fig:char}.}
\label{fig:char2}
\end{figure*}

Target stars are confined to distances lower than 3.7\,kpc from the Sun, they reach out to about 10\,kpc in the disk from the Galactic center and are found within 2.5\,kpc from the plane (top panels). The middle panels reveal that a few stars classified as halo and GSE reach large Z$_{max}$ and r$_{ap}$. Halo and GSE stars get to smaller r$_{peri}$ distances than Seq, thick, transition and thin disk stars. This is due to the progressively lower eccentricity of the orbits of these objects, as can be better appreciated looking at the distance from the identity line (zero eccentricity: r$_{peri}$=r$_{ap}$) of the different categories of objects in the bottom-right panel. Notice that Seq candidates have extremely retrograde orbits which are, however, less eccentric than halo and GSE stars (see bottom-left panel in Figure \ref{fig:char2}).

Excluding 3 halo and 2 GSE stars, which reach Z$_{max}>$10\,kpc, the rest of the halo stars are confined to Z$_{max}<$7\,kpc (bottom-left panel) and, therefore, belong to the inner halo population. Most of the thin disk stars have Z$_{max}<$1\,kpc, but a few stars extend to Z$_{max}\simeq$2\,kpc. Star classified as thick disk extend to Z$_{max}$ up to 4\,kpc and Seq stars to Z$_{max}<$3\,kpc. Notice here, that the classification between thin, thick, transition and halo stars is based on their current U$_{lsr}$,V$_{lsr}$ and W$_{lsr}$ velocity components. 

Figure\,\ref{fig:char2}, presents the target stars in several commonly used planes to characterize the stellar orbits and kinematics \citep[][]{lane22}, using the same symbols as in Fig.\,\ref{fig:char}. The background gray dot population is the ``good parallax sub-sample'' studied in \citet[][]{bonifacio21} which we use here for reference only. 

The top-left and top-middle panels present a version of the Toomre diagram (the square root of the sum of the squares of the radial V$_R$ and vertical V$_Z$ velocity components in galactocentric cylindrical coordinates {\it vs} the transverse velocity component, V$_T$) and the V$_T$  {\it vs} V$_R$ plane, first used to identify the ``Sausage'' structure by \citet[][]{belokurov18}. The top-right panel is the usual total orbital energy E {\it vs} angular momentum L$_Z$ plane. Bottom-panels present the eccentricity {\it vs} L$_Z$ (left), the square root of the radial action J$_R$ {\it vs} the azimuthal action (J$_\phi$=L$_Z$, middle) and the so called action diamond (right), namely the difference between the vertical and radial actions (J$_Z$-J$_R$) {\it vs} J$_\phi$, both normalized to the total action J$_{tot}$=J$_R$+$|$J$_\phi|$+J$_Z$. In the bottom-middle and bottom-right panels, the red and green shaded areas indicate the criteria adopted here from \citet[][]{feuillet21} to select candidate GSE and Seq stars, namely -500$<$L$_Z<$+500\,kpc\,\kms and 30$<\sqrt{J_R}<$55\,(kpc\,\kms)$^{1/2}$ for GSE and (J$_Z$-J$_R$)/J$_{tot}<0.1$ and J$_\phi$/J$_{tot}<-0.6$, for Seq, respectively.

The different classes of objects are perhaps better separated in these planes. 
Thin, transition and thick disk stars have prograde orbits (V$_T$ or L$_Z>$0). Moving from  thin to thick disk stars, the eccentricity of the orbits increases, together with J$_R$,  V$_R$ and V$_Z$, while L$_Z$=J$_\phi$, E and V$_T$ all decrease. Sequoia stars have the most retrograde orbits in the sample.

Halos stars have, as expected, together with GSE stars, the most eccentric orbits. Including the GSE and Seq candidates, 21 out of 34 halos stars have prograde orbits. These number change to 18 out of 25 on prograde orbits, excluding the 9 stars belonging to GSE (6) and Seq (3). GSE stars have prograde and retrograde orbits in equal number (3). Considering only stars having Z$_{max}<$7\,kpc (and therefore r$_{ap}<$50\,kpc,  $\sqrt{J_R}<$60\,kpc\,\kms, and E$<$0\footnote{Notice that 3 stars have E$>$0 in Fig.\ref{fig:char2}, upper right panel. These stars are, however, on bound orbits. In fact, their space velocities are lower than the escape velocities at their distances. It is a known and documented feature that ``galpy potentials do not necessarily approach zero at infinity". This is indeed the case of {\tt MWPotential2014}.}, see Figs.\,\ref{fig:char} and \ref{fig:char2}), then we are left with 17 out of 22 (77\%) stars on prograde orbits. This is consistent with these stars participating in a collapse process which gave rise to the inner halo and with the mostly prograde orbits (70\%) found in very metal-poor stars ([Fe/H]$<-$2) at $1<|Z|<$3\,kpc by \citet[][]{carter21} using data of the H3 survey.

\subsection{Abundances}

We estimated the chemical abundances of the elements up to zinc using MyGIsFOS.

Regarding n-capture elements measurements, we used the Wrapper for Turbospectrum and Fitprofile (henceforth WTF), a python code that computes line profile fits from a synthetic spectra grid computed on the fly. From an input file defining the region(s) for line fitting and continuum pseudonormalization, the input atmosphere model, the observed spectrum to fit and the ion and abundance range for the calculation, WTF calls turbospectrum \citep{alvarez98,plez2012} to compute a grid of small synthetic spectra around the region of interest, then launches Fitprofile \citep{thygesen16} to compute the best-fitting synthetic. In this case Fitprofile is run with the abundance of the element  of choice as the only free parameter. Finally WTF displays the fit result graphically. 

In the present case, ATLAS9 \citep{atlas05} models were computed for the parameters and the MyGIsFOS-derived abundance pattern for every star (elements for which we derived abundances in MyGIsFOS were put at the derived values, the others were set to the solar-scaled value for the star's metallicity). WTF retains the Fitprofile capability to fit different features for the same ion either as separate, or as a single global fit. In this work, if multiple lines were present, each one was fitted separately.

In Table \ref{linelist}, we provide the line list adopted for the star BD-11\,3235 as an example, while the line list for the other targets can be found in the online material.
The average abundances obtained are listed in Tables \ref{abu1_mince1} - \ref{abu5_mince1}. 
The error reported is the line-to-line scatter ($\sigma$). However, to get the final error, the uncertainties due to atmospheric parameters should be added in quadrature. \\
We estimated these last uncertainties for the star BD-11 3235 by varying $\Delta$\teff \, $= \pm 100$ K, $\Delta$ \logg \,$= \pm 0.2$ dex, and $\Delta \xi= \pm 0.2 $ \kms.
We considered this star because its atmospheric parameters are representative of the sample and all the elements have been measured for it. 
The typical errors due to atmospheric parameters are reported in Table \ref{err_atm_par} for all chemical species analyzed here. Similar results have been found for the first sample of metal-poor giant stars analyzed by the MINCE collaboration \citep[][hereafter the MINCE\,I sample]{mince1, mince2}. 
The solar abundances adopted in this work are listed in Table \ref{tab:solarabbo}.

Finally, we want to highlight that we used $\rm [X/Fe] = [X/\ion{Fe}{i}]$ or $\rm [X/Fe] = [X/\ion{Fe}{ii}]$ when X is a neutral or ionised species, respectively. 

The sulfur abundances A(S) were measured following a different procedure. 
In the wavelength ranges covered by the spectral frames centered at 580 and 760 nm lie the Sulfur (S) lines of Mult. 6, 8, and 1. 
These lines are of high excitation, thus, they become weak at low temperatures and in the metal-poor regime.
We rejected the S lines at bluer wavelengths (Mult.\,6, 8) because too weak, and we considered only the strongest features at 920 nm (Mult.\,1). \\
The contamination due to telluric lines in this wavelength region makes difficult the standard workflow of MyGIsFOS. 
In particular, the pseudonormalization of the observed spectrum is hampered when a telluric line falls in a continuum region. 
Moreover, since MyGIsFOS does not take into account the presence of telluric lines, blended or shape distorted lines are not recognized by the code. Therefore, depending on the degree of telluric contamination, MyGIsFOS rejects S lines or overestimates the abundances.
For these reasons, we decided to derive A(S) from equivalent width (EW). \\
We first compared the observed spectra of our stars with that of a B-type star to evaluate the suitability of Mult. 1 lines. 
S lines contaminated by telluric lines were rejected. 
This reduced our sample to a total of 23 stars. \\
Using the deblending option of the IRAF\footnote{\url{https://iraf-community.github.io/}} task {\tt splot}, we took into account the contribution in EW from the telluric and S lines. \\
The conversion from EW to abundances was computed through the code GALA \citep{mucciarelli2013}, that is a user-friendly wrapper
for WIDTH9 \citep{atlas05}. The results obtained for sulfur are reported in Table \ref{tab_S_mince}.

\subsection{Non-local thermodynamic equilibrium effects}
 Among the elements analyzed in this work, some are known for being sensitive to deviation from LTE, or NLTE effects.
In this work, the chemical investigation was performed under the assumption of LTE, similar to what done in the first two
papers of the series: \citet{mince1} and \cite{mince2}.
The intervals in metallicity, effective temperature and surface gravity spanned by the present sample
is similar to that of the sample analysed in \citet{mince1} and \cite{mince2}.
We defer a more complete analysis of the NLTE effects to a future investigation on the complete sample.
To provide an indication of the magnitude of NLTE effects we are dealing with, in \cite{mince1}
we took advantage of the thorough investigation by \citet{hansen2020}. 
In Table 9 of \citet{mince1}, we provided the NLTE corrections from \citet{hansen2020}
for BD\,$-10$\,3742 
(\teff,\,\logg,\,[Fe/H],\,$\xi$: $4678$\,K, $1.38$, $-1.96$, $1.9$\,\kms) and 
BD\,$-12$\,106 ($4889$\,K, $2.03$, $-2.11$, $1.5$\,\kms).
These two stars have parameters close to those of BD\,$+07$\,4625 and BD\,$+39$\,3309 in the MINCE\,I sample
and to those of 
CD\,$-52$\,976 (4655\,K, 1.29, $-1.90$, 1.97\,\kms) 
and BD\,$-19$\,3663 (4767\,K, 1.82, $-2.51$, 1.82\,\kms) in the present sample.
We therefore refer the reader to the discussion in \citet{mince1} and table\,9 to have an indication
of the NLTE corrections we should be dealing with.
Another relevant paper is 
\citet{mataspinto2021} who 
provide NLTE corrections for two stars with stellar parameters similar to the stars here investigated.
From this comparison, we conclude that for all elements, except \ion{Na}{i} and \ion{Co}{i},
the NLTE corrections are smaller than the typical uncertainty on the abundances and they are positive, as NLTE corrections for Fe, becoming even smaller when using [X/Fe]. 
Therefore, the comparison of LTE abundances with chemical evolution models is meaningful. 
There are not many NLTE computations available for neutron capture elements, but
a discussion of NLTE corrections for Sr, Eu and Ba can be found in \cite{mince2}.

\begin{table}
\caption{Sensitivity of abundances on atmospheric parameters.}
\label{err_atm_par}
    \begin{tabular}{llll}
         \hline \hline
         Element&$\Delta$\teff &$\Delta$\logg &$\Delta \xi$\\
         &100 K&0.2 dex&0.2 \kms\\
         \hline
         Na I&0.08&0.02&0.01\\
         Mg I&0.04&0.06&0.03\\
         Al I&0.06&0.01&0.00\\
         Si I&0.04&0.03&0.00\\
         S I&0.18&0.09&0.03\\
         Ca I&0.12&0.02&0.03\\
         Sc II&0.04&0.14&0.02\\
         Ti I&0.18&0.01&0.01\\
         Ti II&0.02&0.14&0.11\\
         V I&0.20&0.02&0.01\\
         Cr I&0.14&0.02&0.01\\
         Cr II&0.04&0.06&0.02\\
         Mn I&0.17&0.02&0.01\\
         Fe I&0.09&0.01&0.04\\
         Fe II&0.07&0.06&0.02\\
         Co I&0.11&0.02&0.00\\
         Ni I&0.08&0.03&0.03\\
         Cu I&0.17&0.01&0.00\\
         Zn I&0.07&0.06&0.04\\
         Sr I&0.19&0.02&0.02\\
         Y II&0.03&0.12&0.06\\
         Zr II &0.01&0.08&0.02\\
         Ba II &0.03&0.08&0.18\\
         La II&0.08&0.11&0.07\\
         Ce II&0.13&0.11&0.02\\
         Pr II&0.08&0.13&0.05\\
         Nd II&0.07&0.11&0.02\\
         Sm II&0.05&0.14&0.07\\
         Eu II&0.03&0.14&0.04\\
         \hline \hline
         \multicolumn{4}{l}{Note. Sulfur results were obtained with GALA.}
    \end{tabular}
\end{table}

\begin{table}
\caption{Solar abundances used throughout this paper.}
\label{tab:solarabbo}
\begin{tabular}{lll}
\hline
Element & A(X) & Reference \\
\hline
Na & 6.30 & \citet{Lodders09} \\
Mg & 7.54 & \citet{Lodders09} \\
Al & 6.47 & \citet{Lodders09} \\
Si & 7.52 & \citet{Lodders09} \\
S  & 7.16 & \citet{caffau2011a} \\
Ca & 6.33 & \citet{Lodders09} \\
Sc & 3.10 & \citet{Lodders09} \\
Ti & 4.90 & \citet{Lodders09} \\
V  & 4.00 & \citet{Lodders09} \\
Cr & 5.64 & \citet{Lodders09} \\
Mn & 5.37 & \citet{Lodders09} \\
Fe & 7.52 & \citet{caffau2011a} \\
Co & 4.92 & \citet{Lodders09} \\
Ni & 6.23 & \citet{Lodders09} \\
Cu & 4.21 & \citet{Lodders09} \\
Zn & 4.62 & \citet{Lodders09} \\
Rb & 2.60 & \citet{Lodders09} \\
Sr & 2.92 & \citet{Lodders09} \\
Y  & 2.21 & \citet{Lodders09} \\
Zr & 2.58 & \textbf{\citet{Lodders09} }  \\
La & 1.14 & \citet{Lodders09} \\
Ce & 1.61 & \citet{Lodders09} \\
Pr & 0.76 & \citet{Lodders09} \\
Nd & 1.45 & \citet{Lodders09} \\
Sm & 1.00 & \citet{Lodders09} \\
Eu & 0.52 & \citet{Lodders09} \\
Ba & 2.17 & \citet{Lodders09} \\
\hline
\end{tabular}
\end{table}

\section{Discussion}
In the following section are shown the results obtained for the MINCE III stars. 

\subsection{The stars exhibiting H$\alpha$ emission}\label{sec_Halpha}
We realised that some stars in the sample show P-Cygni and/or inverse P-Cygni profile, a signature of stellar activity. \\
In Figure\,\ref{fig:cmd}, the stars with activity (red dots) are among the brightest and reddest in the sample, as expected by \citet[][see their Figure\,1]{smith88}. 

\begin{figure*}
\centering
\includegraphics[trim= 6cm 0.5cm 2cm 1cm, clip,width=1.1\textwidth]{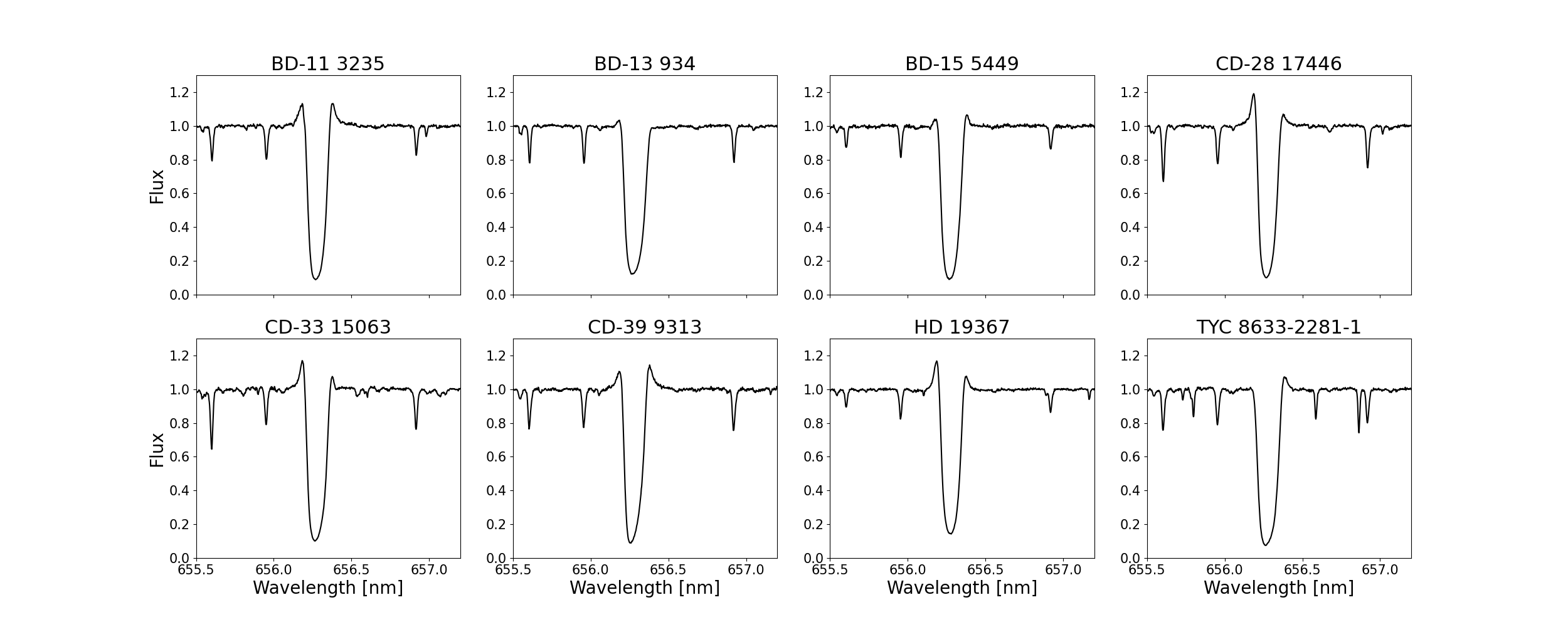}
\caption{Spectra of the H$\alpha$ line of the stars exhibiting P-Cygni and/or inverse P-Cygni profile.} 
\label{fig:Halpha}
\end{figure*}

The stars in question are: BD-11 3235, BD-13 934, BD-15 5449, CD-28 17446, CD-33 15063, CD-39 9313, HD 19367 and TYC 8633-2281-1.
The spectra of the H$\alpha$ line of these stars are shown in Figure \ref{fig:Halpha}. 
BD-11 3235 is the only star with symmetric H$\alpha$ emission wings. 
In BD-13 934, CD-28 17446, CD-33 15063 and HD 19367, the blue wing is stronger than the red one, while the remaining stars show the opposite profile.

Finally, for these active stars, one must be cautious not to use lines whose contribution also comes from the chromosphere, such as the \ion{Ca}{ii} triplet lines, the \ion{Na}{i}-D lines or the \ion{K}{i} line at 769.9\,nm, lines that have not been used in this investigation.

\subsection{The Li-rich star: CD-28 10039}\label{sec_Li_rich}

The star CD--28\,10039 has $\rm A(Li)=1.1$, which is not expected for a star of this \logg.
According to our kinematic results, this star belongs to the thick disk.

To check the evolutionary state of this star, we considered two BASTI isochrones \citep{BASTI21} characterized by [Fe/H] = $- 1.0$ and $-2.0$, [$\alpha$/Fe] = +0.4, and Age = 12 Gyr.
In Figure \ref{fig:cmd}, CD--28\,10038 (green star symbol) lies close to the asymptotic-giants branch of the isochrone with [Fe/H]=$-1.0$.

\begin{figure}
\centering
\includegraphics[trim= 0.5cm 0cm 0cm 1cm, clip,width=0.5\textwidth]{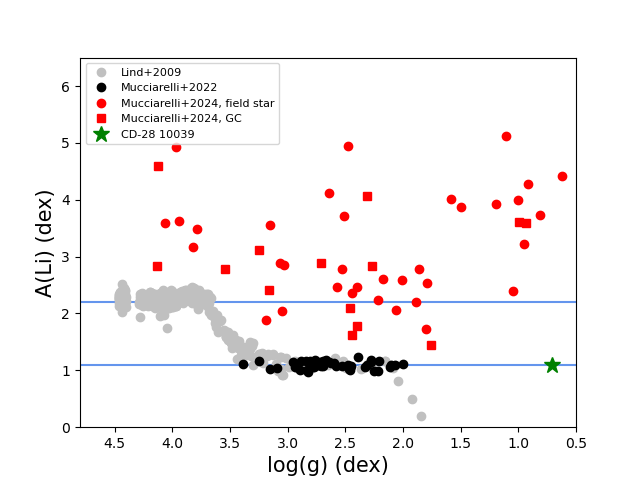}
\caption{A(Li) versus \logg\, diagram for CD--28\,10039 in comparison with literature data: \citet[][grey]{lind2009}, \citet[][black]{mucciarelli2022}, \citet[][red]{mucciarelli2024}. The blue lines are the Spite (A(Li)$\sim2.2$, \citealt{spite1982}) and RGB (A(Li)$\sim 1.1$, \citealt{mucciarelli2012}) Plateaus.}
\label{fig:ali_vs_logg}  
\end{figure}

We reported CD--28\,10039 in the A(Li) versus \logg\, diagram (green star symbol in Figure \ref{fig:ali_vs_logg}).
The evolution of Li with \logg\, is characterized by the Spite Plateau (A(Li)$\sim 2.2$\,dex, \citealt{spite1982}), followed by the First Dredge-Up (FDU) drop until the RGB Plateau (A(Li)$ \sim 1$\,dex, \citealt{mucciarelli2012, mucciarelli2022}), which is followed by another drop due to the RGB bump (RGBb) mixing episode.
In this diagram, the unexpected position of CD--28\,10039, slightly above the RGB Plateau at low \logg, reveals that the Li enhancement of this star is not due to standard stellar evolution \citep{iben1967} or non-canonical mixing processes \citep{charbonnel2020, magrini2021}.

\begin{figure}
\centering
\includegraphics[trim= 0.5cm 0cm 0cm 1cm, clip,width=0.5\textwidth]{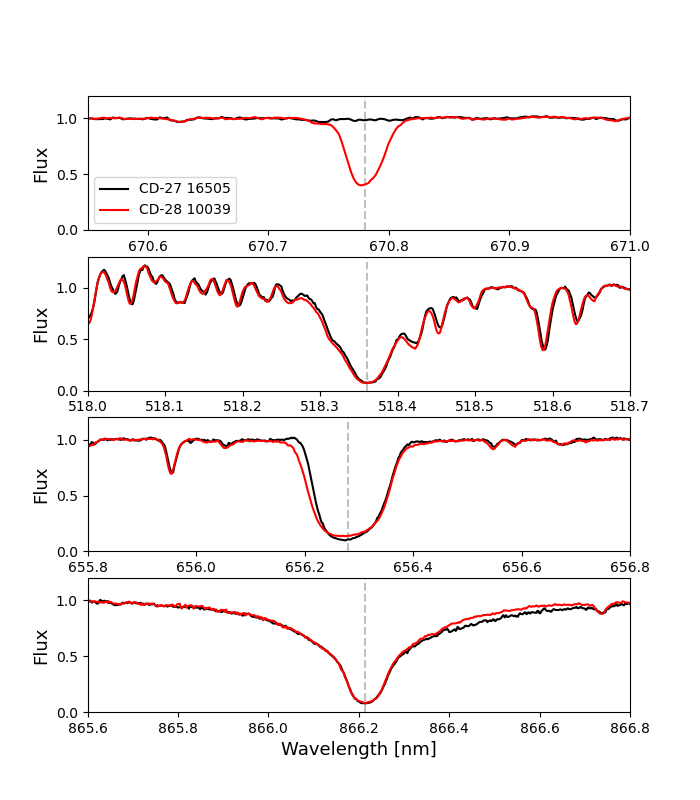}
\caption{Comparison of the spectra of the Li-rich star CD--28\,10039 (red) with that of CD--27\,1605 (black) in different wavelength regions. From the top to bottom: Li doublet at 670.78 nm, reddest line of the Mg II triplet at 518.36 nm, H$\alpha$ at 656.28 nm, and the reddest line of the Ca II triplet at 866.21 nm.}
\label{fig:li-rich_spectra}  
\end{figure}

The spectra of CD--28\,10039 (red) are compared with those of the star with similar parameters CD--27\,16505 (black) in Fig.\,\ref{fig:li-rich_spectra}.
The top panel shows the evident difference in strength of the Li line at 670.78 nm between the two stars. The reddest line of the Mg II triplet at 518.36 nm seems similar in both stars (second panel from the top), suggesting that the Li enhancement of CD--28\,10032 is not due to rotation effects.
The H$\alpha$ and the reddest line of the Ca II triplet at 866.21 nm are reported in the last two panels. From these two features it is possible to notice slight differences between the Li-rich and the reference stars.
The H$\alpha$ of the Li-rich star seems shallower, with asymmetrical core and blue wing. 
On the other hand, the red wing of the reddest line of the Ca II triplet appears shallower in the Li-rich stars than in the reference one.
These anomalous profiles could be compatible with chromospheric activity and mass loss \citep{meszaros2008}.
However, the H$\alpha$ blue wing of CD--27\,16505 recall a slight emission profile, so it could be possible that we are comparing two active stars.

From the chemical point of view, CD--28\,10039 has A(Y) = 1.17 $\pm 0.00$ and A(Ba) = 1.34 $\pm$ 0.09. 
These values are not discrepant with those of the sample.
For instance, we obtained A(Y) = 1.01 $\pm$ 0.10 and A(Ba) = 1.14 $\pm$ 0.00 in CD--27\,16505.
Comparing the abundances of the $s$-process elements, in the Li-rich star, the ratio [Rb/Y] = $-0.06$ dex suggests a low neutron density at the sites of these elements \citep{smith1984}. 
Moreover, we find a similar $r$-process elements enrichment for the Li-rich star, A(Eu)=0.04 dex, and CD--27\,16505, A(Eu)=0.05 dex.

\begin{figure}
\centering
\includegraphics[trim= 0.5cm 0cm 0cm 1cm, clip,width=0.5\textwidth]{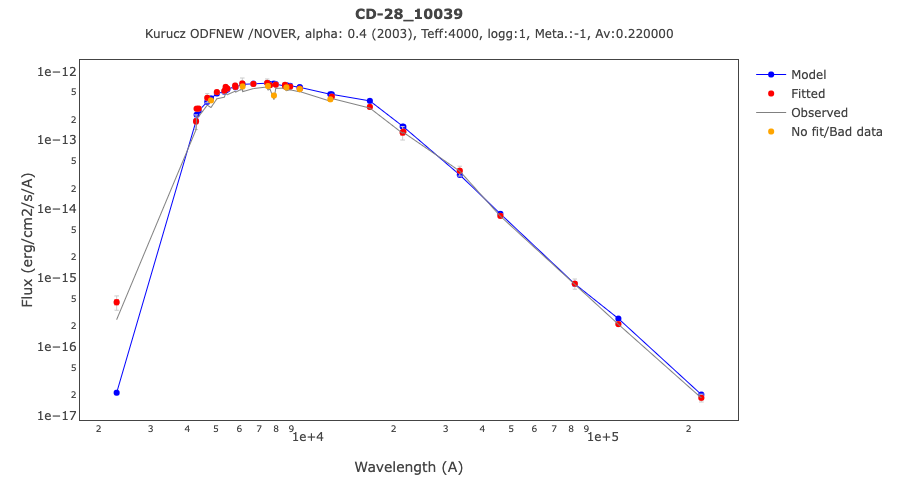}
\caption{Spectral energy distribution of CD--28\,10039 obtained from the VOSA service. The model obtained with the parameters indicated in the toplabel is shown in blue. An apparent NUV excess is detected from the GALAX photometry.}
\label{fig:sed}  
\end{figure}

We present in Figure \ref{fig:sed} the Spectral Energy Distribution (SED, red and yellow points) of CD--28\,10039, which we constructed using the Virtual Observatory SED Analyzer \citep[VOSA\footnote{\url{http://svo2.cab.inta-csic.es/theory/vosa/}},][]{vosa} service. The SED was generated adopting the stellar distance and extinction, namely d=2.14\,kpc and $A_V=0.22$\,mag. We fitted the SED using Kurucz ATLAS9 model atmospheres calculated with new opacity distribution functions and no overshooting \citep[][]{kurucz03}, allowing for a narrow range of parameters, around those we adopted here for the star. The best fitting model (blue points and line) obtained with the parameters indicated in the top-label of the Figure, seems a fair representation of the stellar SED all over the wavelength range, but for the Near Ultra Violet (NUV) GALEX photometry at 2303.37$\AA$. In fact, CD--28\,10039 appears to present a significant NUV excess, which may be indicating the presence of a hot companion, like a white dwarf. 

On the other hand, the radial velocity measured from the UVES spectra and that reported in $Gaia$ DR3, are very similar (RV difference of less than 0.5\,\kms) to each other. Furthermore, the $Gaia$ DR3 astrometry quality parameters are not suggestive of binarity: RUWE=1.13, astrometric\_gof\_al=2.79, astrometric\_excess\_noise = 0.09\,mas, although astrometric\_excess\_noise\_sig = 10.1.  

Finally, $Gaia$ DR3 does not report variability information for CD--28\,10039 (phot\_variable\_flag="NOT\_AVAILABLE"). However, the star is reported as variable in G, G$_{BP}$ and G$_{RP}$ (VarF=VVV), according to the analysis of \citet[][]{maiz23}.

\subsection{$\alpha$-elements}
\begin{figure*}
\centering
\includegraphics[width=0.9\textwidth]{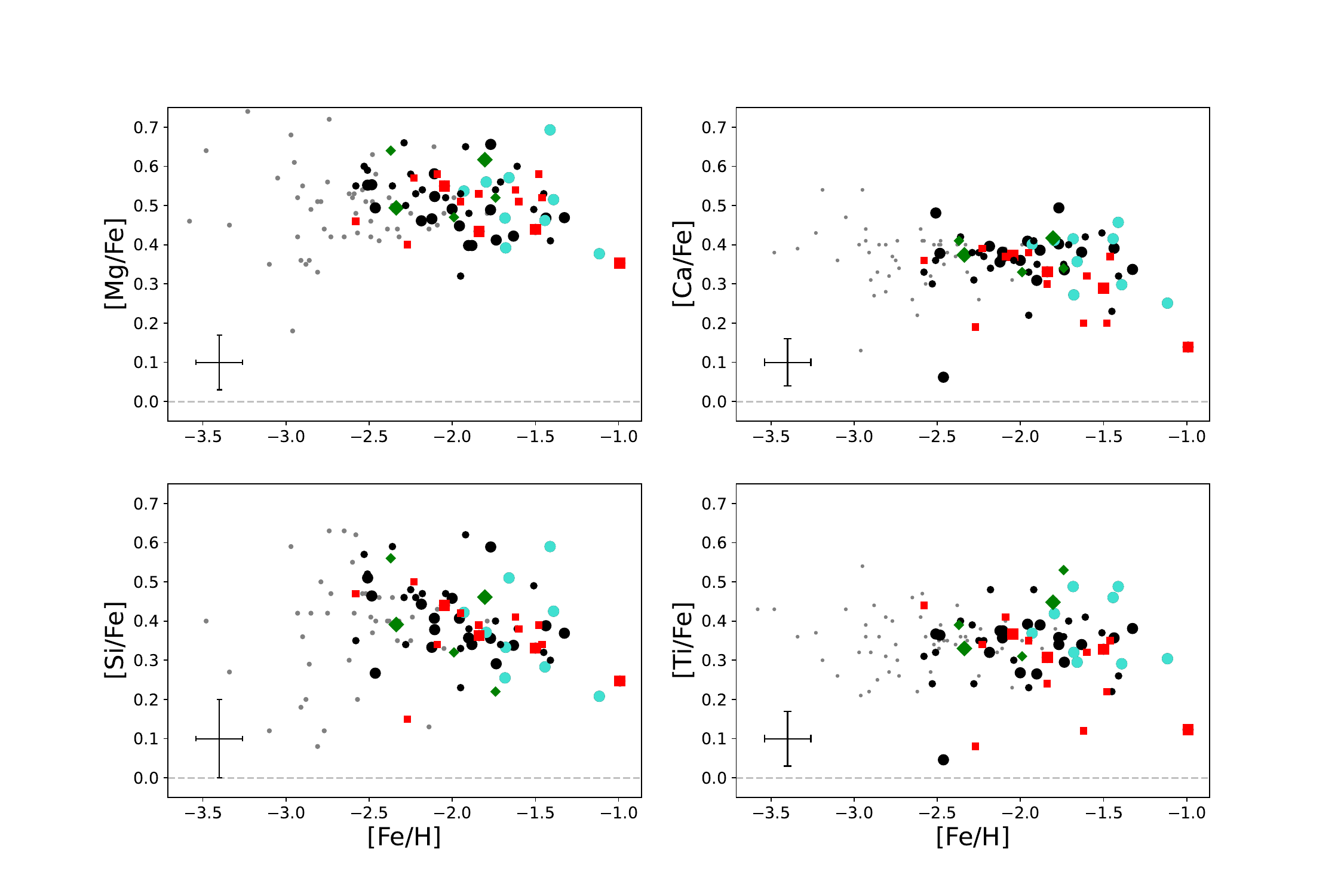}
\caption{$\alpha$-elements abundances measured in the MINCE III stars. Black and cyan points are Galactic stars belonging to the halo and thick-disk, respectively. Candidate Gaia-Sausage-Enceladus (red squares) and Sequoia (green diamonds) stars are also shown. The results of \cite{mince1} for the same substructures in the MINCE I sample are reported with same colors and smaller symbols. The gray points are the results of \cite{lombardo2022}} 
\label{fig:alpha}
\end{figure*}

Figure \ref{fig:alpha} shows our results for the $\alpha$-elements Mg, Si, Ca, and Ti. 
The stars belonging to the Galactic halo (black dots), Galactic thick-disk (cyan dots), GSE (red squares), and Sequoia (green diamonds) substructures are distinguished by colors and symbols. The good agreement with the first results of the MINCE series \citep{mince1} is shown adopting the same colors and smaller symbols. 
As comparison, we reported the results of \cite{lombardo2022} obtained from the CERES survey in gray.
This legend will be maintained for the rest of the section.\\
The stars of the different substructures seem to share the same plateau trend in all panels in the metallicity range $-2.5 <$ [Fe/H] $<-1.5$. Moreover, due to the low number of GSE, and Sequoia stars, it is not possible to draw further conclusions about the differences between MW, GSE and Sequoia members.

All the $\alpha$-elements are enhanced with respect to the [Fe/H] abundances with sample averages and standard deviations of $\rm \langle[Mg/Fe]\rangle= 0.49 \pm 0.08$, $\rm\langle[Si/Fe]\rangle= 0.39 \pm 0.09$, $\rm\langle[Ca/Fe]\rangle = 0.36 \pm 0.09$ and $\rm\langle[Ti/Fe]\rangle= 0.34 \pm 0.09$.

We also estimated the [\ion{Ti}{ii}/\ion{Fe}{ii}] ratio. In particular, the mean abundances ratio is 0.42 $\pm$ 0.14, and the mean difference between [\ion{Ti}{ii}/\ion{Fe}{ii}] and [\ion{Ti}{i}/\ion{Fe}{i}] is 0.08 $\pm$ 0.14. The NLTE effects for \ion{Ti}{i} lines \citep{mashonkina2016a} may explain the fact that the ionization equilibrium of Ti is not established, although the offset is not very significant.
These results are in agreement, within errors, with those in the literature \citep[see e.g.][]{mince1, lombardo2022}.

\begin{figure*}
\centering
\includegraphics[width=0.8\textwidth]{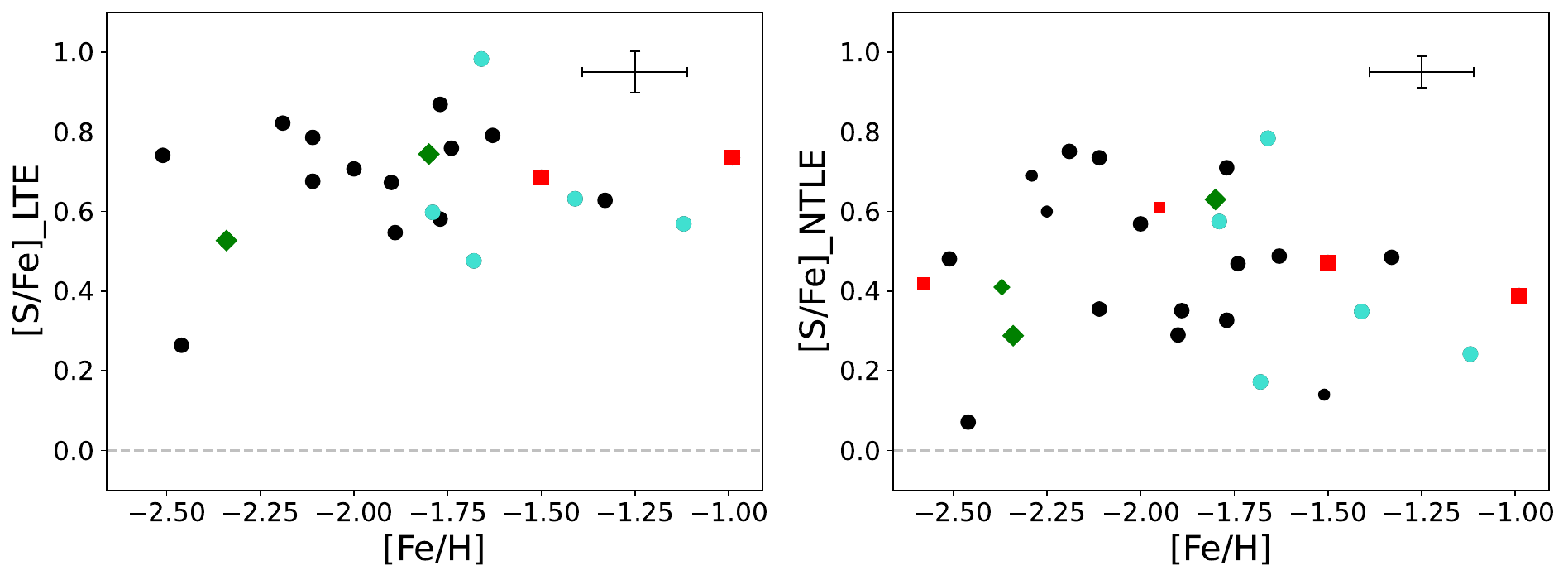}
\caption{[S/Fe] versus [Fe/H] LTE (left panel) and NLTE (right panel) diagrams of the MINCE sample,  the MINCE I sample \citep{mince1} is identified by smaller symbols. Colors and symbols are the same in Figure \ref{fig:alpha}.} 
\label{fig:s_lte_vs_nlte}
\end{figure*}

Figure \ref{fig:s_lte_vs_nlte} compares the sulfur results before (left panel) and after (right panel) NLTE corrections.
In the left panel, we can guess the presence of a plateau at the value [S/Fe]$_{\rm LTE} \sim$ 0.64. This quantity is considerably higher than both the usual [$\alpha$/Fe] $\sim$ 0.40 and [S/Fe] $\sim$ 0.30 values in the literature \citep{caffau2010, nissen2007}. 
After NLTE corrections, the [S/Fe] ratios present a larger scatter but the plateau value, $\rm \langle[S/Fe]\rangle= 0.45 \pm 0.19$, is in better agreement with the typical [$\alpha$/Fe] abundance ratio.
The large scatter in [S/Fe] is introduced by the difficulties in the A(S) estimation, such as the telluric contamination and the NLTE corrections.  

In Figures \ref{fig:alpha} and \ref{fig:s_lte_vs_nlte}, we notice a peculiar star with low $\alpha$-elements content.
The MW halo star CD--38\,13823 is Mg-rich ([Mg/Fe]=0.49), as expected for its metallicity ($\rm [Fe/H]=-2.46$), slightly poor in [Si/Fe]=0.27 and [S/Fe]=0.26, and with solar ratios of [Ca/Fe]=0.06 and [Ti/Fe]=0.06.

\subsection{Light-odd elements}

\begin{figure*}
\centering
\includegraphics[width=\textwidth]{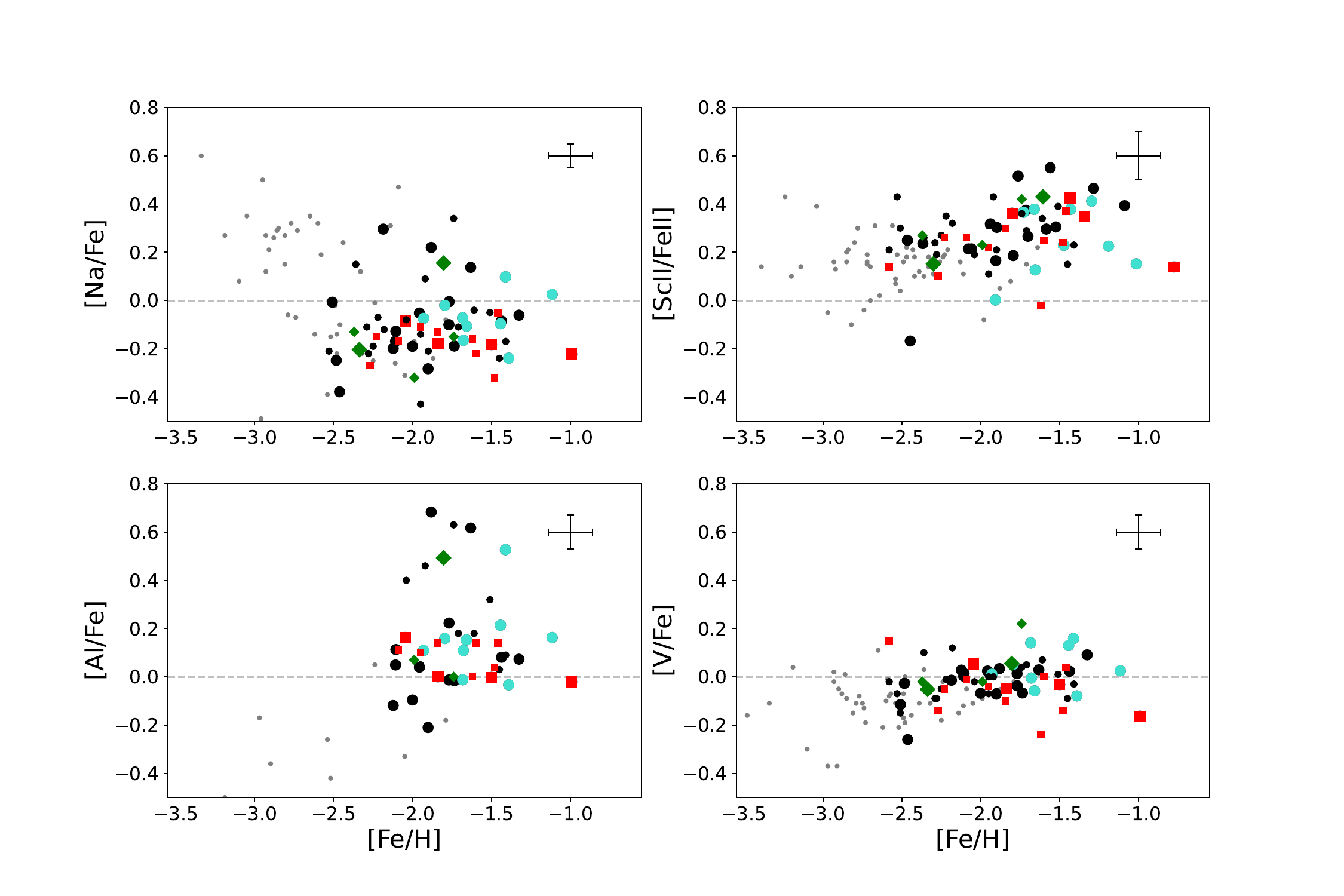}
\caption{Light-odd elements abundances measured in the MINCE stars, the MINCE\,I sample \citep{mince1} is identified by smaller symbols. 
Colors and symbols are the same as in Figure \ref{fig:alpha}.} 
\label{fig:light-odd}
\end{figure*}

Figure \ref{fig:light-odd} shows our results for the light-odd elements Na, Al, Sc, and V. 
The  stars of the different kinematic components  are identified with the same colors and symbols as in Fig.\,\ref{fig:alpha}.
In the following we shall comment on which lines were used to determine the abundances in the MINCE\,III sample.
    
The Na abundances were derived from the Na I lines at 498.3\,nm, 568.2\,nm 568.8\,nm, the Na D 
resonance lines at 588.9\,nm (D1) and 589.5\,nm (D2), and 616.1\,nm. 
The large dispersion $\rm <[Na/Fe]>= -0.08 \pm 0.15$ may be due to the LTE deviations of these lines.
In our sample in three halo stars, two thick-disk stars and one Sequoia candidate, the Na is enhanced over iron. 
However, the limited size of the sample does not allow us to draw any robust conclusion about the differences between MW and GSE stars.

The Al abundances were derived from the Al I resonance lines at 394.4\,nm and 396.1\,nm as well as from 
the Al I lines at 555.6\,nm, and the doublet at 669.6\,nm and 669.8\,nm. 
However, the Al I resonance lines were rejected for most of the stars to estimate the final Al abundances.
Inspection of Fig.\,\ref{fig:light-odd} shows that the [Al/Fe] ratio is quite scattered spanning a range of almost 0.8\,dex.
The mean value is
<[Al/Fe]>= 0.16\,dex  with a standard deviation  of 0.26\,dex. 
The  dispersion is driven by four Al-enhanced stars, two of which belong to  the MW halo, one to the MW thick-disk, and 
one to GSE. The MINCE\,I sample shows similar scatter and dispersion.
In Fig.\,\ref{fig:light-odd}, the [Na/Fe] and [Al/Fe] ratios do not show clear trends with the metallicity.

We determined the mean abundance ratios of Sc and V ions to the corresponding Fe ions.
The ratios and their corresponding standard deviations are: 
$\rm \langle[\ion{Sc}{i}/\ion{Fe}{i}]\rangle = -0.01 \pm 0.13$, 
$\rm \langle[\ion{Sc}{ii}/\ion{Fe}{ii}] \rangle = 0.28 \pm 0.15$, 
$\rm\langle [\ion{V}{i}/\ion{Fe}{i}] \rangle = -0.01 \pm 0.08$, and 
$\rm\langle [\ion{V}{ii}/\ion{Fe}{ii}] \rangle = 0.02 \pm 0.15$. 
Thus, while the mean difference between [\ion{Sc}{ii}/\ion{Fe}{ii}] and [\ion{Sc}{i}/\ion{Fe}{i}] is 
about $0.29 \pm 0.17$, the one between [\ion{V}{ii}/\ion{Fe}{ii}] and [\ion{V}{i}/\ion{Fe}{i}] is only of $0.03 \pm 0.17$. 
These results agree within errors with those obtained by \cite{lombardo2022} in a sample of lower metallicity giant stars. 
We notice a slight increase in the [\ion{Sc}{II}/Fe] and [\ion{V}{I}/Fe] ratios with metallicity.

In Fig.\,\ref{fig:light-odd}, CD--38\,13823 shows a low [Sc/Fe] ratio of about $-0.17$, and it is slightly poor in $\rm [Na/Fe]=-0.38$ and $\rm [V/Fe]=-0.26$.
We could not measure the Al abundance for this star.

\subsection{Iron-peak elements}
 
\begin{figure*}
\centering
\includegraphics[width=\textwidth]{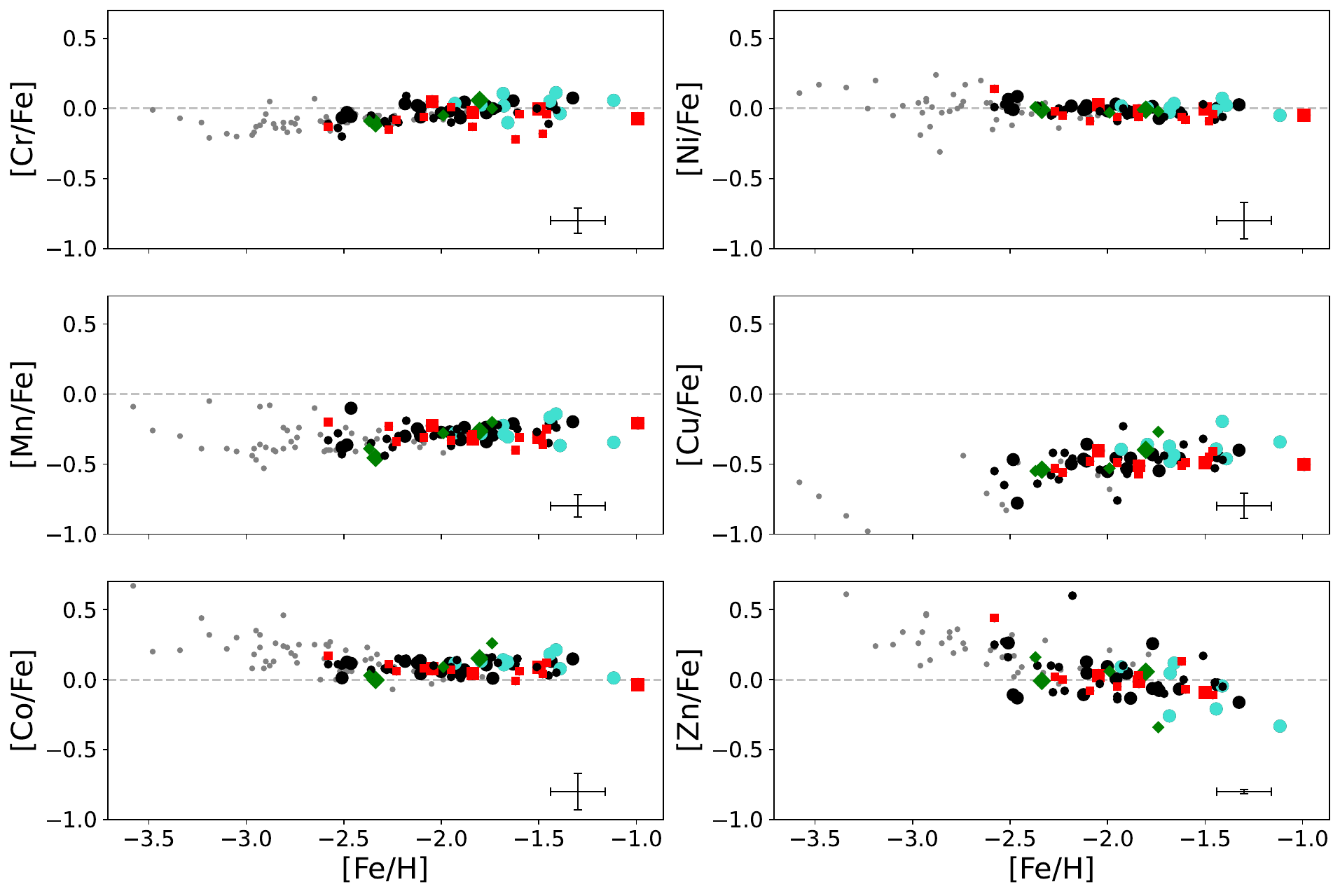}
\caption{Iron-peak elements abundances measured in the MINCE stars, the MINCE\,I sample \citep{mince1} is identified by smaller symbols. 
Colors and symbols are the same as in Figure \ref{fig:alpha}.} 
\label{fig:iron-peak}
\end{figure*}

Figure \ref{fig:iron-peak} shows our results for the iron-peak elements Cr, Mn, Co, Ni, Cu, and Zn. 
The first five of these elements 
are produced in nuclear statistical equilibrium in SNe of all types, like iron, 
so 
their ratio to iron are expected to be close to  the solar value.
The situation is less clear for Zn, that may have contribution from $\alpha-$rich freeze-out and 
from neutron captures \citep[see][for a discussion on the nucleosynthesis of Zn]{duffau2017}.
In the MINCE\,III sample Cr, Ni, Co, and Zn have mean abundance ratios close to solar:  
$\rm\langle[Cr/Fe]\rangle= 0.00 \pm 0.06$, $\rm \langle[Co/Fe]\rangle= 0.09 \pm 0.06$, $\rm\langle[Ni/Fe]\rangle= 0.00 \pm 0.03$, and $\rm\langle [Zn/Fe]\rangle= -0.02 \pm 0.13$. Co and Ni display flat trends, while Cr and Zn exhibit increasing and decreasing trends with metallicity, respectively. Previous studies have found similar behavior for these elements \citep{cayrel2004b,bonifacio2009, ishigaki2013, mince1,lombardo2022}.

The iron-peak elements showing exceptional behavior are Mn and Cu. 
Indeed, we observe sub-solar trends for these elements, which slightly increasing with metallicity. 
Similar behaviors have been noted by \cite{ishigaki2013} and \cite{mince1}. 

Manganese is produced by both SNe Ia and SNe II, like the other iron peak elements; however, the contribution by SNe Ia at solar metallicity is higher compared to iron, leading to a speculate trend to that of the $\alpha$-elements. Moreover, Mn yields from SNe Ia are the most dependent to the explosion mechanism among the iron peak elements and its evolution can be used to evaluate the presence and contribution of different SNe Ia channels \citep{Cescutti17,Seitenzahl13}.
The nucleosynthesis of Cu is also peculiar. According to \cite{timmes1995}, the explosion of massive stars at solar metallicity produces yields 5 times larger than the same explosion at low metallicity. Therefore it is the metal dependencies of SNe II that produce the rise rather than the SNe type-Ia. 
The behavior of Mn may also be due to the strong NLTE effects on the lines of these elements \citep{bergemann2008}.
However, 
the use of NLTE corrections decreases the trend with metallicity, but does not cancel it, 
suggesting that there is a role of SNe Ia in Mn nucleosynthesis \citep{Eitner20}.
Among the stars of the MINCE\,III sample, CD--38\,13823 is the most Mn-rich, $\rm [Mn/Fe]=-0.10$, and the most Cu-poor, $\rm [Cu/Fe]=-0.78$. This star does not show peculiar values for the other iron-peak elements in comparison to the other targets.

The mean abundance ratios and their standard deviations for Mn and Cu ions to the corresponding Fe ions
are: $\rm\langle[\ion{Cr}{i}/\ion{Fe}{i}]\rangle= 0.00 \pm 0.06$, $\rm \langle[\ion{Cr}{ii}/\ion{Fe}{ii}]\rangle= 0.11 \pm 0.12$, $\rm\langle[\ion{Mn}{i}/\ion{Fe}{i}]\rangle= -0.27 \pm 0.07$, $\rm\langle[\ion{Mn}{ii}/\ion{Fe}{ii}]\rangle= -0.23 \pm 0.06$. Thus, the ionisation equilibrium is reached in LTE
for both Cr and Mn. 
Our results are in perfect agreement with those found by \cite{lombardo2022}.

\subsection{Neutron capture elements}

\begin{figure*}
\centering
\includegraphics[width=\textwidth]{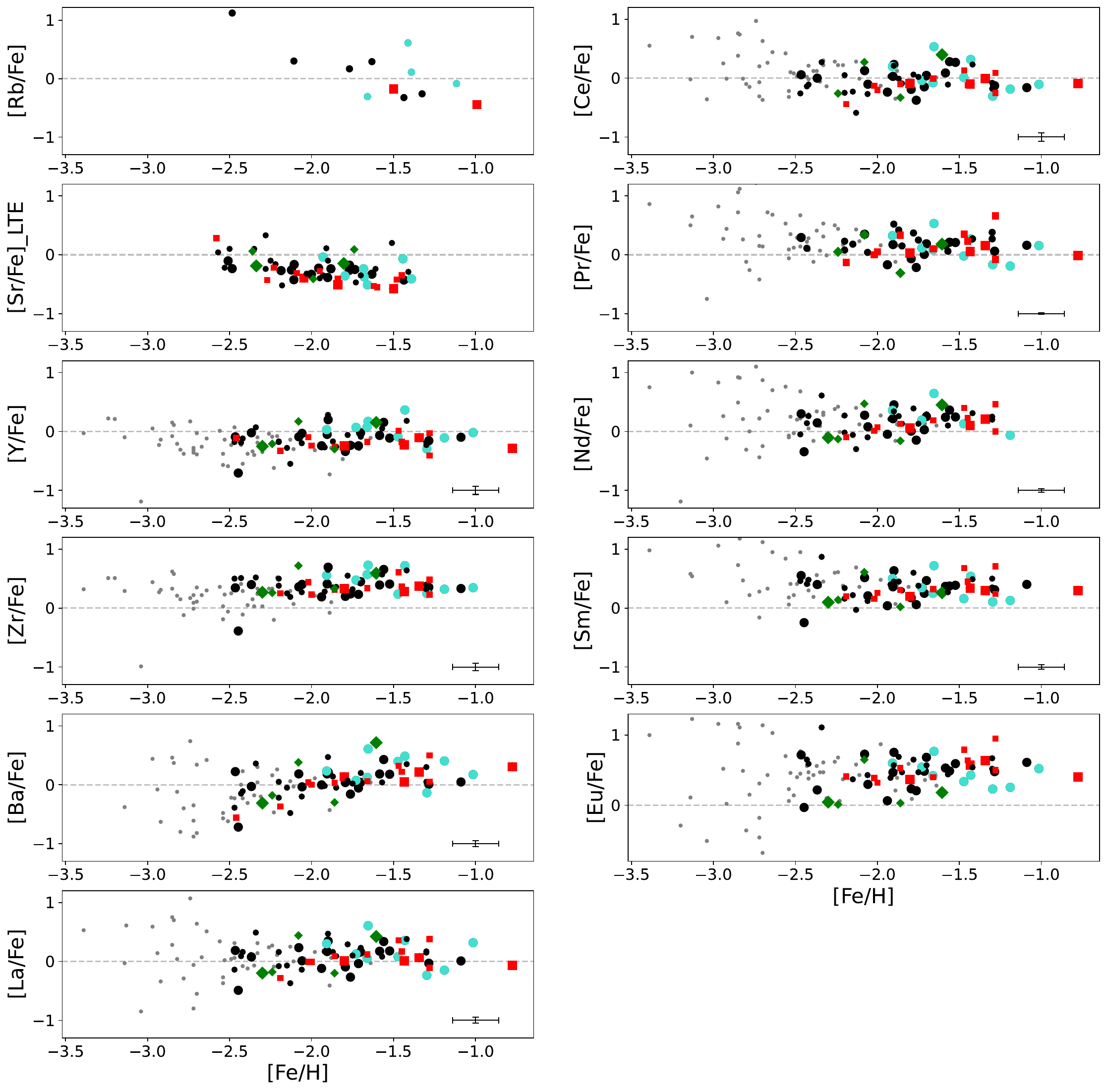}
\caption{Neutron capture elements abundances measured in the MINCE stars, the MINCE\,I sample \citep{mince2} is identified by smaller symbols. Colors and symbols are the same as in Figure \ref{fig:alpha}. The stars of the CERES survey analyzed by \cite{lombardo2021, lombardo2023} are reported in gray.}
\label{fig:n-capture}
\end{figure*}

Figure \ref{fig:n-capture} displays our results for the n-capture elements Rb, Sr, Y, Zr, Ba, La, Ce, Pr, Nd, Sm and Eu. The different MW substructures are shown with the same colors and symbols as in
Figure \ref{fig:alpha}. The gray points are from \cite{lombardo2022, lombardo2023}.
Our results are generally in agreement with the analysis of the  MINCE\,I sample \citep[smaller symbols][]{mince2}.

We observed flat trends for all the n-capture elements with the exception of Sr and Ba. 
Sr decreases with increasing  metallicity, while Ba
increases. 
Similar behaviors have also been noted in other studies \citep{francois2007,ishigaki2013,lombardo2022}.
In particular, we found sub-solar mean abundances ratios and standard deviation for $\rm\langle[Sr/Fe]\rangle=-0.41 \pm 0.28$.
The mean abundances of $\rm\langle[Y/Fe]\rangle=-0.09 \pm 0.19$, $\rm\langle[La/Fe]\rangle =0.08 \pm 0.22$ and $\rm\langle[Ce/Fe]\rangle = 0.01 \pm 0.21$ are close to solar.
Finally, we found mean abundances above solar 
for $\rm\langle[Zr/Fe]\rangle=0.38 \pm 0.21$, $\rm\langle[Ba/Fe]\rangle=0.13 \pm 0.26$, 
$\rm\langle[Pr/Fe]\rangle=0.11 \pm 0.17$, $\rm\langle[Nd/Fe]\rangle=0.16 \pm 0.21$, 
$\rm\langle[Sm/Fe]\rangle= 0.31 \pm 0.19$, and $\rm\langle[Eu/Fe]\rangle=0.43 \pm 0.21$.

Considering both the MINCE\,I \citep{mince2} and the present samples, we 
find an indication of a slight increase in the n-capture elements scatter with decreasing metallicity.
This is more visible in the trends of Ba, La and Eu.
The comparison with the results of \cite{lombardo2022, lombardo2023} allows to better visualize this effect.
The MW, GSE and Sequoia stars seem to behave similarly for all elements (see Fig.\,\ref{fig:n-capture}). 
However, a larger sample for each substructure is required to draw firmer conclusions.

The $s$-process elements content of CD--38\,13823 is low: $\rm [Y/Fe]=-0.70$, $\rm[Zr/Fe]=-0.39$, $\rm [Ba/Fe]=-0.72$, $\rm[La/Fe]=-0.49$.
On the other hand, in comparison to the other stars of the sample, it is also the poorest one in $\rm [Sm/Fe]=-0.24$ and $\rm[Eu/Fe]=-0.03$.

\begin{figure}
\centering
\includegraphics[trim= 3.2cm 0.8cm 4.5cm 3cm, clip,width=0.5\textwidth]{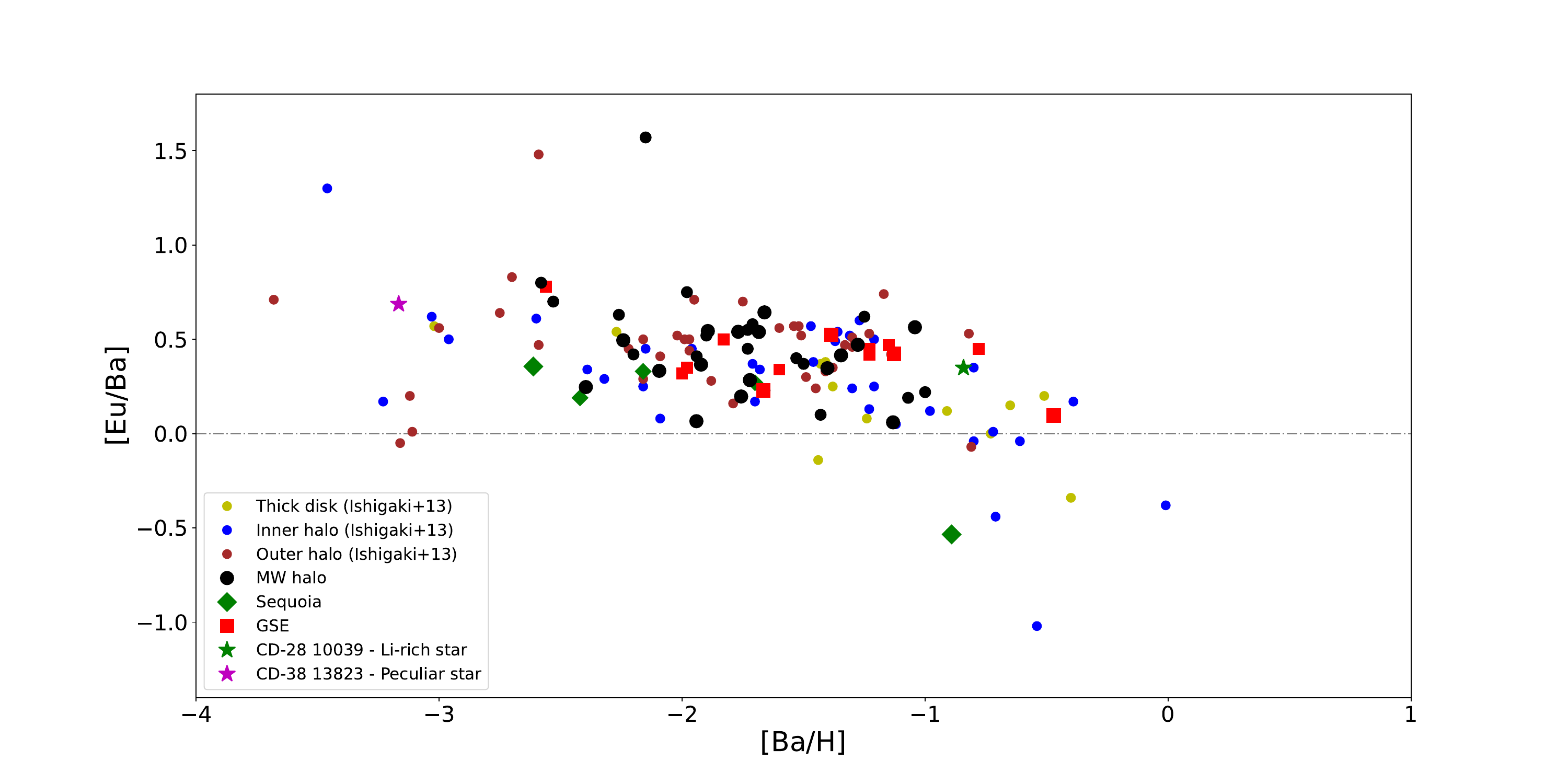}
\caption{Abundance ratio of [Eu/Ba] as a function of [Ba/H] for MINCE stars. MW halo, Seq and GSE stars are reported as black dots, green diamond symbols, and red square symbols, respectively. The results for MINCE I stars \citep{mince2} are represented with same color and smaller symbols. The Li-rich star CD-28 10039 (green) and CD-38 13823 (magenta) are highlighted with star symbols. We added the samples from \cite{ishigaki2013} for comparison.}
\label{fig:euba_vs_ba}
\end{figure}

The [Eu/Ba] versus [Ba/H] diagram in Fig.\,\ref{fig:euba_vs_ba} 
allowed us to discriminate the origin of the n-capture elements enrichment. 
In this plot, the (almost) Eu-free stars lie at $\rm [Ba/H] \leq - 3$ \citep{cescutti2015, cavallo2021}, 
the enrichment in the range $\rm -2.5 < [Ba/H] < -1.5$ may be
attributed to rotating massive stars, while AGB stars are likely responsible for the enrichment at $\rm [Ba/H] > -1$.
The $r$-process pollution occurs at $\rm [Eu/Ba] > 1$.
Our sample of stars appears to define a flat trend up to [Ba/H] $\sim -1.0$.
The Li-rich star, CD-28\,10039, lies in the area polluted by AGB stars, while the position of the peculiar star, CD--38\,13823, supports its low Eu-content.

\section{Chemical evolution model}

We compare our results 
with a stochastic chemical evolution model \citep{cescutti14,cescutti2015}
that traces Rb, Sr, Y, Zr, Ba, La, and Eu as well
as the $\alpha$ 
elements Mg, Ca, Si, Ti, the light-odd element Sc, and the iron peak elements Mn, Co, Ni and Zn. 

The chemical evolution model presented in Figures \ref{fig:models1} -\ref{fig:models2} is the same as
described in \citet{cescutti14}. 
The model describes the evolution of the Galactic halo and assumes a stochastic formation of stars, it 
consists of 100 realizations, each of them with the same parameters of chemical evolution
(infall, outflow, star formation efficiency, initial mass function, etc). 
In each region, the following gas infall law with a primordial chemical composition was applied:  

\begin{equation}
\frac{dGas_{in}(t)}{dt} \propto e^{-(t-t_{o})^{2}/\sigma_{o}^{2}}, 
\end{equation}
where $t_{o}$ is set to 100 Myr and $\sigma_{o}$ is 50 Myr.
Similarly, the star formation rate  is defined as 

\begin{equation}
SFR(t) \propto ({\rho_{gas}(t)})^{1.5},
\end{equation}

where $\rho_{gas}(t)$ represents  the gas mass density within the  volume under consideration.
Additionally, the model includes an outflow
from the system: 
 \begin{equation}
\frac{dGas_{wind}(t)}{dt}  \propto SFR(t).
\end{equation}

In each volume, at each time step, the masses of the stars are assigned with a random function, weighted according to the same initial mass function. In this way in each region, at each time step, the mass of gas turned into stars is the same, but the total number and mass distribution of the stars are different. This produces different chemical abundances, stronger if there are large differences among the yields of the stellar masses \citep{cesc08}. The other stochastic events are the enrichment by magneto-rotational driven supernovae \citep{nishimura15}, which are the events producing n-capture elements through the $r$-process in our model. Similar results can be obtained also assuming neutron star mergers with very short time delay \citep[see][]{cescutti2015}. Therefore, some volumes may be enriched early on by an r-process event, while others are only enriched later, producing a dispersion in the traced n-capture elements.  
In this model, besides the enrichment due to $r$-process events, the n-capture elements are considered produced by rotating massive stars, assuming the yields by \citet{Frischknecht16}. The use of different yields does not alter substantially the outcome \citep{Rizzuti21}. The other source of neutron capture elements, asymptotic giant branch stars, is also included by adopting yields from \citet{cristallo11}. All the details can be found in \citet{cescutti14}.
The nucleosynthesis for the remaining elements, namely $\alpha$- and iron-peak elements, is based on \citet{francois04}. They slightly altered the original yields presented in \citet{WW95} to match, with a standard chemical evolution model, the abundances in extremely metal-poor stars measured in \citet{cayrel2004b}. The only variation for these yields concerned nickel. The original results produced a [Ni/Fe]$\sim$ 0.3 at [Fe/H]=$-$2, visible in \citet{francois04} as well as \citet{mince1}. 
According to the present results, nickel should be decreased by a factor of two in a new implementation of this  model.

The comparison of our chemical evolution model for elements Mg to Zn of the MINCE sample (that includes
also the star in the MINCE\,I sample), 
shown in Fig.\,\ref{fig:models1}\, and Fig.\,\ref{fig:models2}, is by and large,
satisfactory. Some concern is on Zn, for which the observed dispersion in 
abundance ratios is larger than what predicted by the model. Another
cause of concern is for \ion{Co}{i}, if a NLTE correction of the order of +0.5 dex,
as suggested by \citet{hansen2020} is applied, the [Co/Fe] ratios are all well above
the model predictions. We do not provide a comparison of the model with the data for 
\ion{Na}{i}, because this element is not currently included in our model.
We refrain to perform any quantitative comparison of the model with the observations, 
and defer such an exercise to when the sample of analysed stars will be larger.

The primary goal of MINCE is to provide abundance ratios for neutron-capture elements in intermediate metal-poor stars 
$(\rm -2.5 < [Fe/H] < -1.5)$ , offering crucial constraints for model development to distinguish between different nucleosynthesis scenarios.
In Figure  \ref{fig:models2}, we compare model predictions with the MINCE III data for neutron-capture elements (X = Rb, Sr, Y, Zn, Ba, La, Eu) presented in this study. Notably, the spread in model predictions in the [X/Fe] versus [Fe/H] space shrinks at higher metallicities, and within the metallicity range of the MINCE data, the model’s predicted [X/Fe] dispersion aligns well with observations. This agreement supports the validity of the adopted nucleosynthesis prescriptions 
and the model assumptions. At the same time, iron peak elements and $\alpha-$elements dispersion are very reduced in the abundance measurements and compatible at this metallicity with the one proposed by the model results.  
Moreover, this is the first time that we can confirm the model predictions for rubidium shown in \citet{cescutti15}, at that time there were not stellar abundances measurements for this element in halo stars.
We also note that europium predictions are slightly too high compared to the measured data. This is most likely connected to the  simplified assumptions for the r-process yields that scale among the elements as the observed pattern in r-process-rich stars. This offset was visible also in the original results  \citep{cescutti14}. Barium shows some stars located above the model results, but lanthanum, which has a very similar nucleosynthesis, the comparison between abundance measurements and model results is excellent. Small offsets are visible for Zr and Y too.

\section{Summary and conclusions}
The MINCE project investigates the nucleosynthetic processes that lead to the chemical elements production in the intermediate metallicity range, $-2.5 <$ [Fe/H] $<-1.5$, with a particular focus on the bulk of elements heavier than iron (Z>30).
In this work, we presented the analysis carried out of high resolution and high signal-to-noise UVES data of 94 stars.
Of these we provide a detailed chemical inventory for 32 stars, while for the remaining stars we only provide
[Fe/H] and atmospheric parameters.
For five stars, no abundances or atmospheric parameters are provided. 
Remarks on individual stars can be found in Appendix\,\ref{remarks}.

All 99  stars observed were kinematically characterized and divided into thin disk (42), thick disk (22), thin-to-thick disk (1) and halo (34), where 6 and 3 halo stars belong to GSE and Sequoia, respectively.
Among the subsample of 32 stars with detailed chemical analysis,
 23 stars belong to the Galactic halo 9 to the thick-disk, 4 to GES 
 and 2 to Sequoia.

We derived 
high precision abundances for light elements (from Na to Zn) and n-capture elements (Rb, Sr, Y, Zr, Ba, La, Ce, Pr, Nd, Sm, Eu) in the 32 metal-poor stars. These results along with those in the first two papers of the MINCE serie (I -\citealt{mince1}, II-\citealt{mince2}) already represent a significant increase of n-capture elements measurements in the intermediate metallicity range.
However, the low number of GSE and Sequoia candidates does not allow us to draw firm conclusions on the differences between these sub-structures and the MW halo.

Among the brightest and reddest stars, 8 of them exhibit (inverse) P-Cygni profile. 
We also identified one low gravity Li-rich star, CD\,28-10039, with A(Li) = 1.1. This last belongs to the thick disk, and its Li enhancements is not due to the standard stellar evolution or non-canonical mixing processes. According to the SED of this star, the NUV excess may suggest the presence of a hot companion.

Finally, we compared our results with stochastic chemical evolution model of the MW halo.
The events considered by the model for the production and the enrichment of n-capture elements are the stochastic formation of stars in the Galactic halo, magneto-rotational driven supernovae yields, rotating massive stars and asymptotic giant branch stars yields.
The good agreement between the chemical abundances and the chemical evolution model proves that the nucleosynthesis processes adopted to describe the origin of the n-capture elements are reliable.

\begin{acknowledgements}
Support for the author F.L. is provided by CONICYT- 118 PFCHA/Doctorado Nacional año 2020-folio 21200677.
We gratefully acknowledge support from the French National Research Agency (ANR) funded project ``Pristine'' (ANR-18-CE31-0017).
PB acknowledges support   from the ERC advanced grant N. 835087 -- SPIAKID. 
This work has made use of data from the European Space Agency (ESA) mission
{\it Gaia} (\url{https://www.cosmos.esa.int/gaia}), processed by the {\it Gaia}
Data Processing and Analysis Consortium (DPAC,
\url{https://www.cosmos.esa.int/web/gaia/dpac/consortium}). Funding for the DPAC
has been provided by national institutions, in particular the institutions
participating in the {\it Gaia} Multilateral Agreement.
This work was also partially supported by the European Union (ChETEC-INFRA, project no. 101008324)
This research has used the SIMBAD database, operated at CDS, Strasbourg, France. 
This publication makes use of VOSA, developed under the Spanish Virtual Observatory (\url{https://svo.cab.inta-csic.es}) project funded by MCIN/AEI/10.13039/501100011033/ through grant PID2020-112949GB-I00.
VOSA has been partially updated by using funding from the European Union's Horizon 2020 Research and Innovation Programme, under Grant Agreement nº 776403 (EXOPLANETS-A). GC
acknowledges the grant PRIN project No. 2022X4TM3H ‘Cosmic POT’ from
Ministero dell’Università e della Ricerca (MUR). 
A.M. acknowledges support from the project “LEGO– Reconstructing the building blocks of the Galaxy by chemical tagging”
(PI: A. Mucciarelli). granted by the Italian MUR through contract PRIN 2022LLP8TK\_001.
\end{acknowledgements}

\bibliographystyle{aa}
\bibliography{biblio}
\begin{appendix} 
\onecolumn

\section{Additional figures}
\vspace{1cm}

\begin{figure*}[h]
\centering
\subfloat{\includegraphics[width=0.35\textwidth]{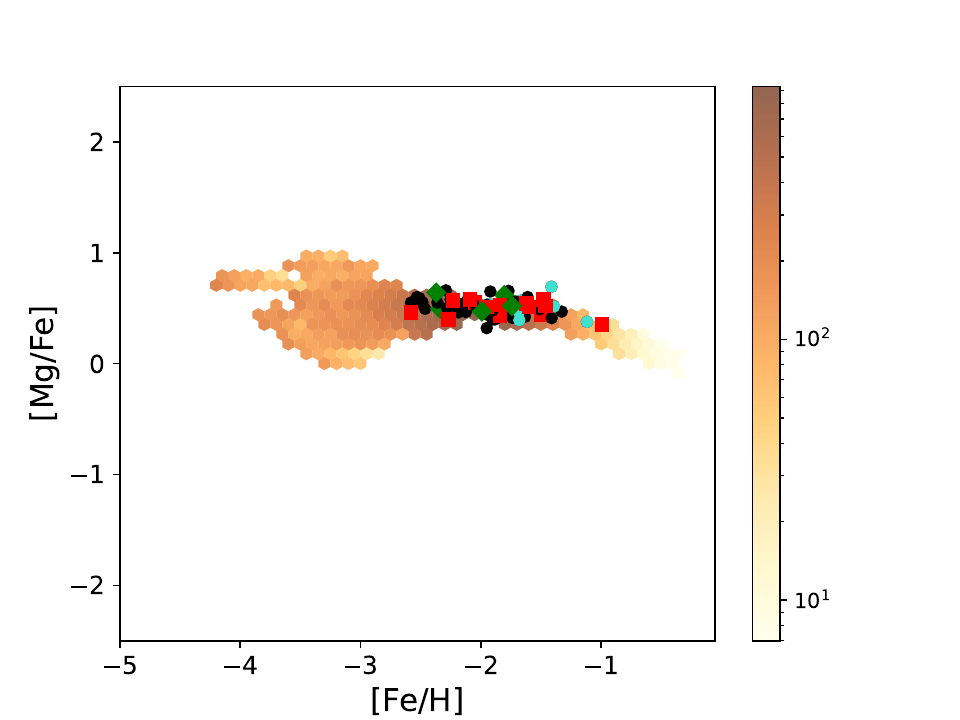}}
\subfloat{\includegraphics[width=0.35\textwidth]{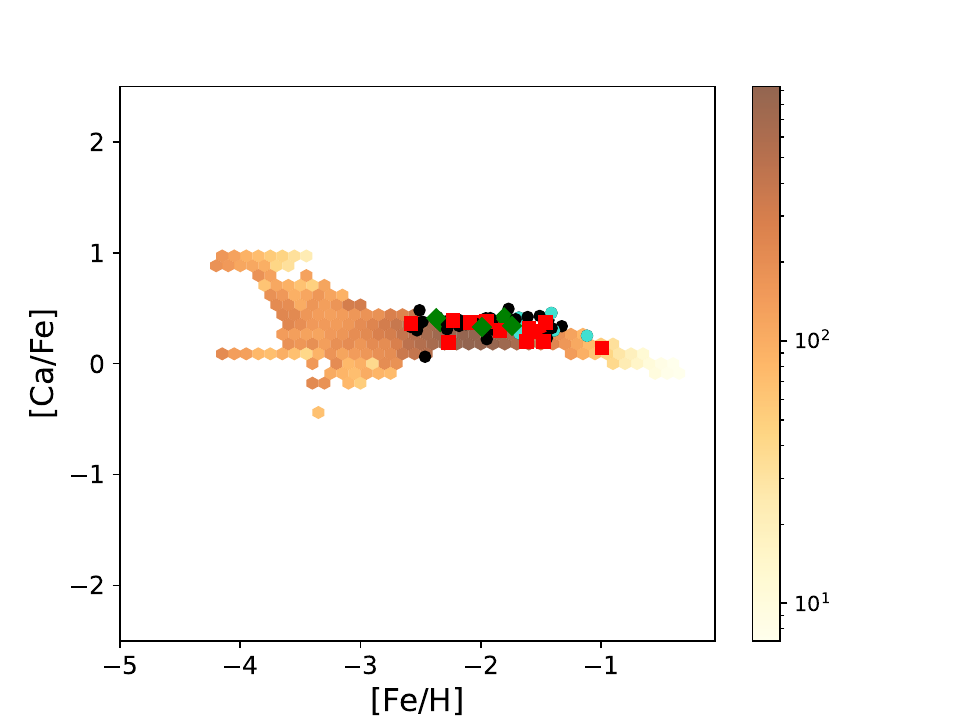}}\\ 
\subfloat{\includegraphics[width=0.35\textwidth]{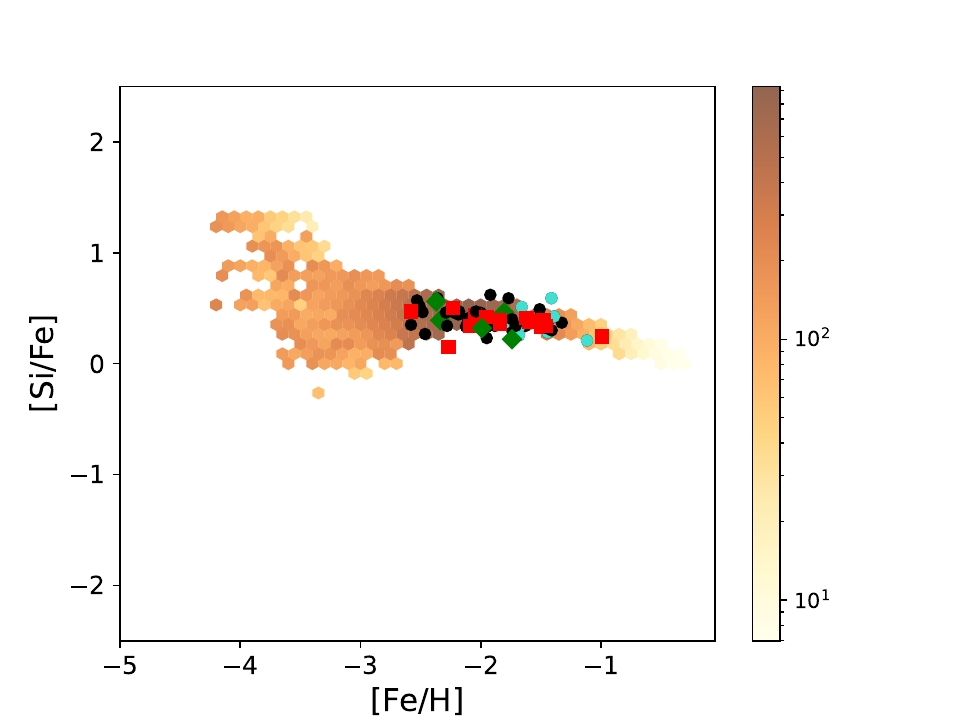}}
\subfloat{\includegraphics[width=0.35\textwidth]{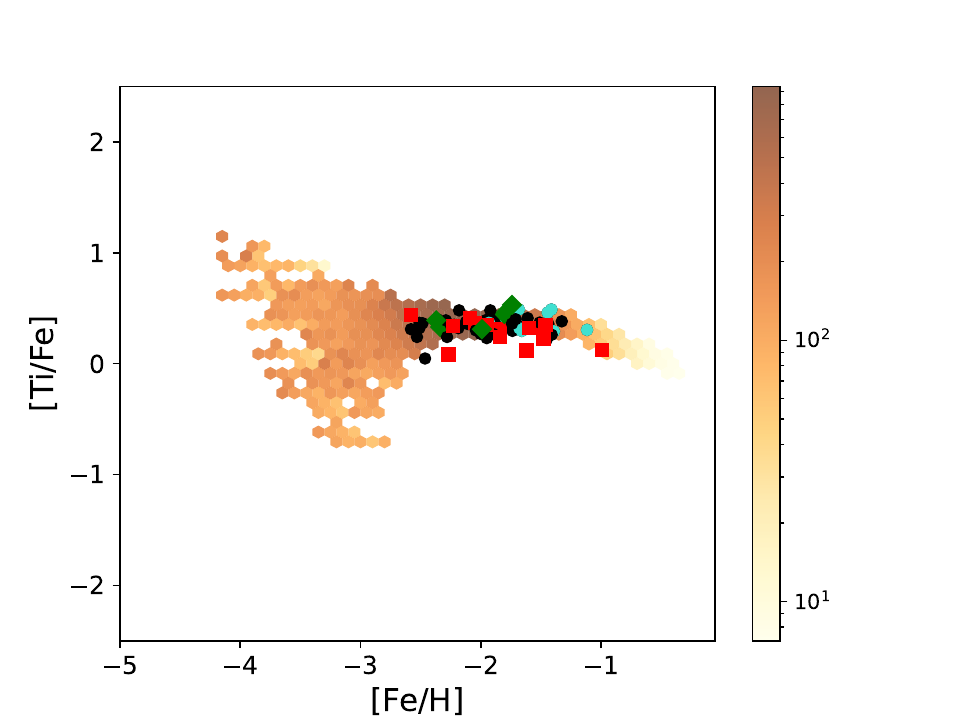}} \\
\subfloat{\includegraphics[width=0.35\textwidth]{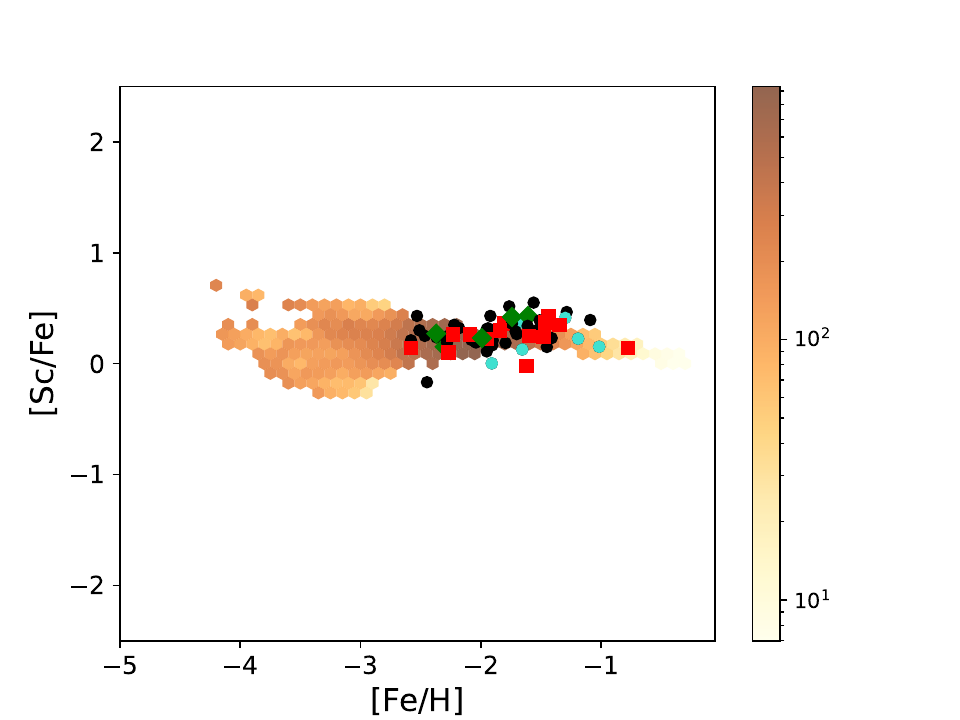}} 
\subfloat{\includegraphics[width=0.35\textwidth]{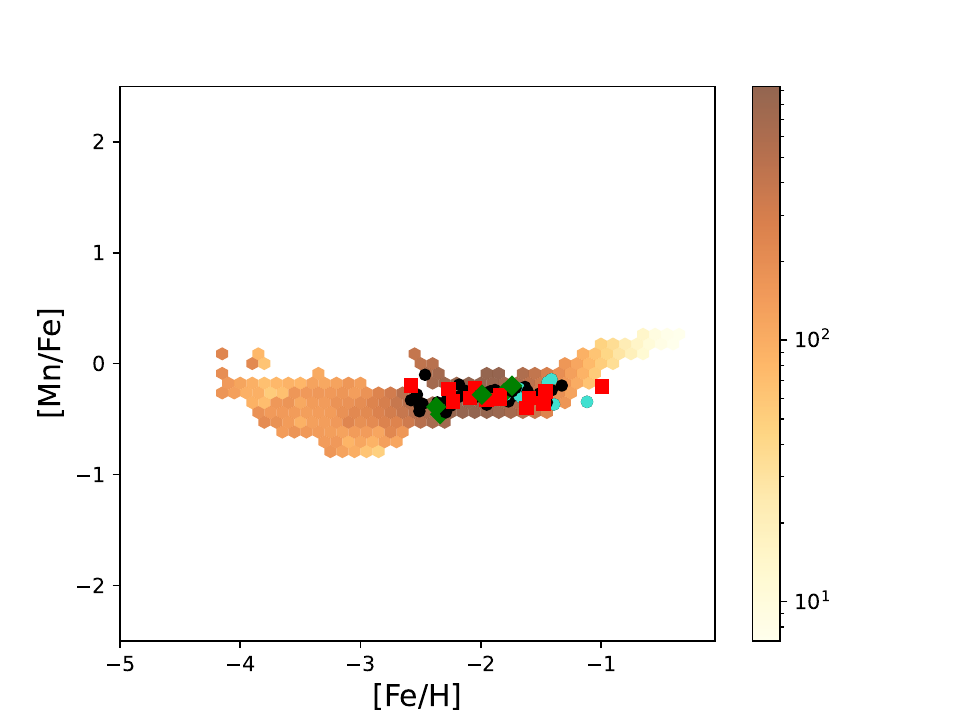}}\\
\subfloat{\includegraphics[width=0.35\textwidth]{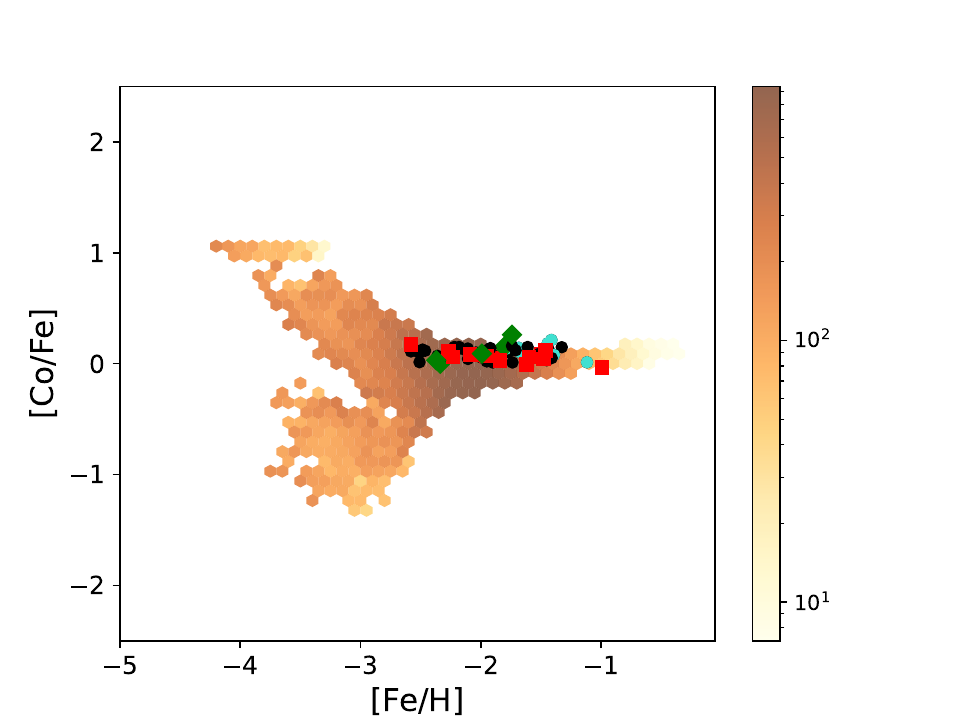}}
\subfloat{\includegraphics[width=0.35\textwidth]{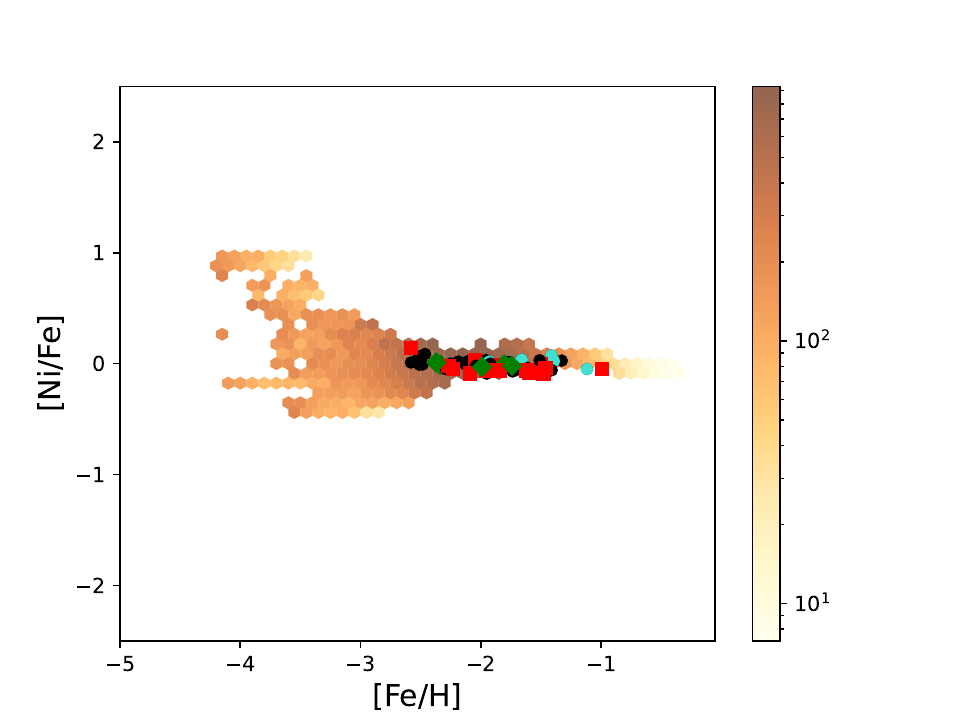}}  
\caption{Comparison between the chemical abundances in MINCE I and III stars and the chemical evolution model for Mg, Ca, Si, Ti, Sc, Mn, Co and Ni. Colors and symbols are the same as in Figure \ref{fig:alpha}}.
\label{fig:models1}
\end{figure*}

\begin{figure*}
\centering
\subfloat{\includegraphics[width=0.35\textwidth]{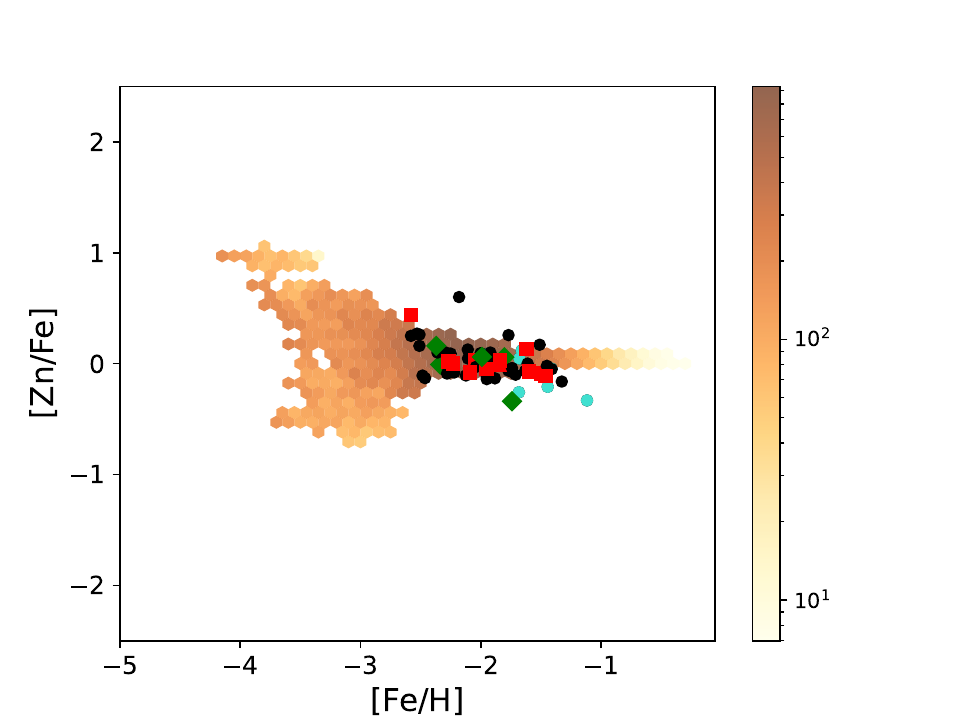}}
\subfloat{\includegraphics[width=0.35\textwidth]{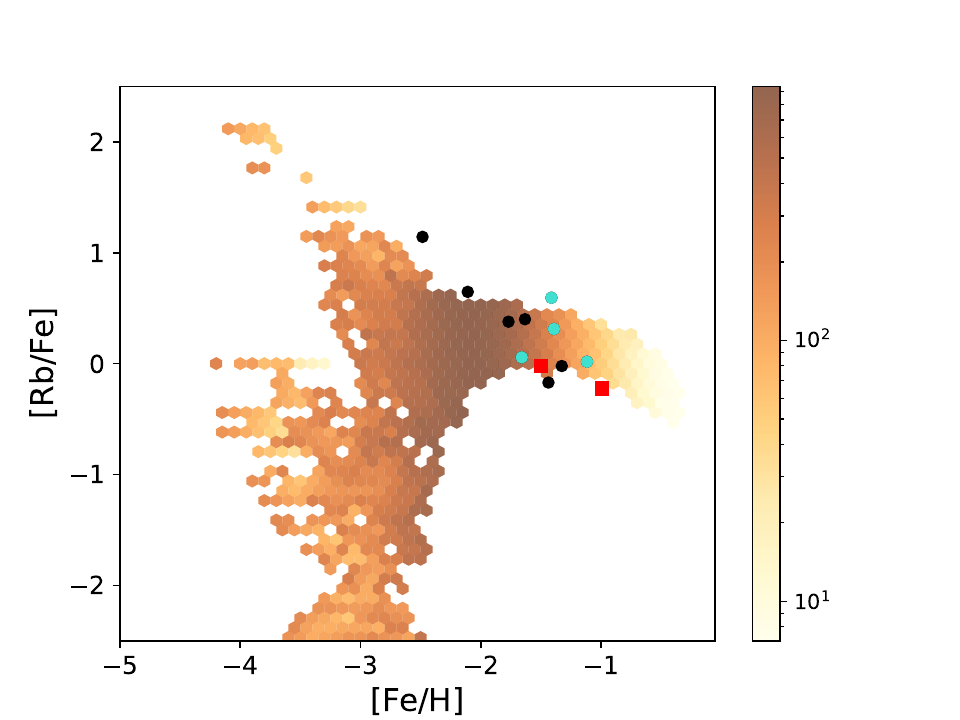}}\\
\subfloat{\includegraphics[width=0.35\textwidth]{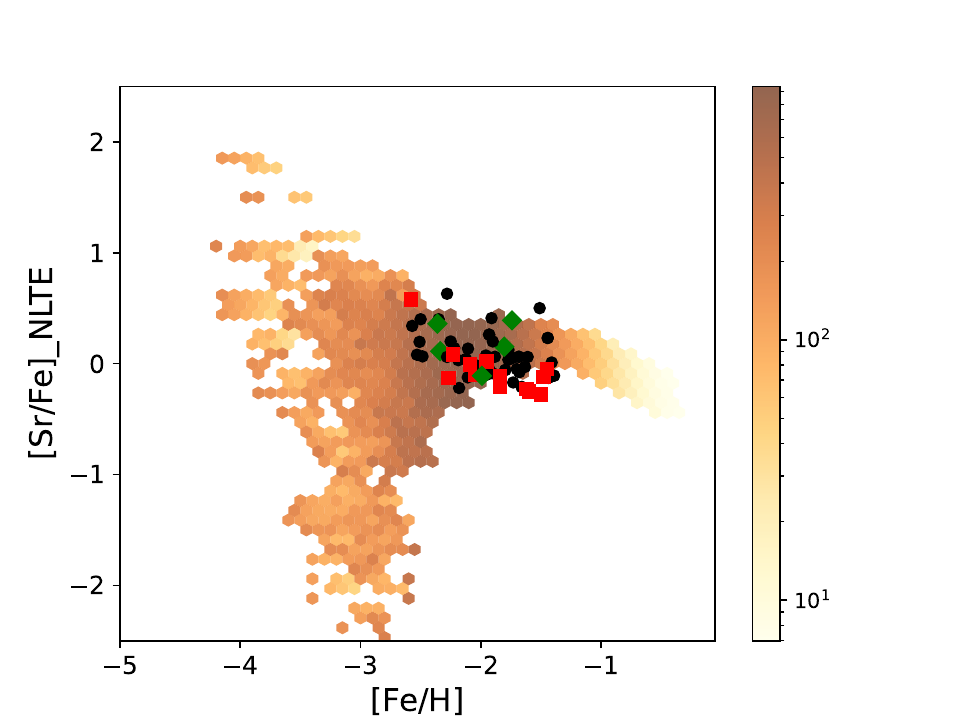}}
\subfloat{\includegraphics[width=0.35\textwidth]{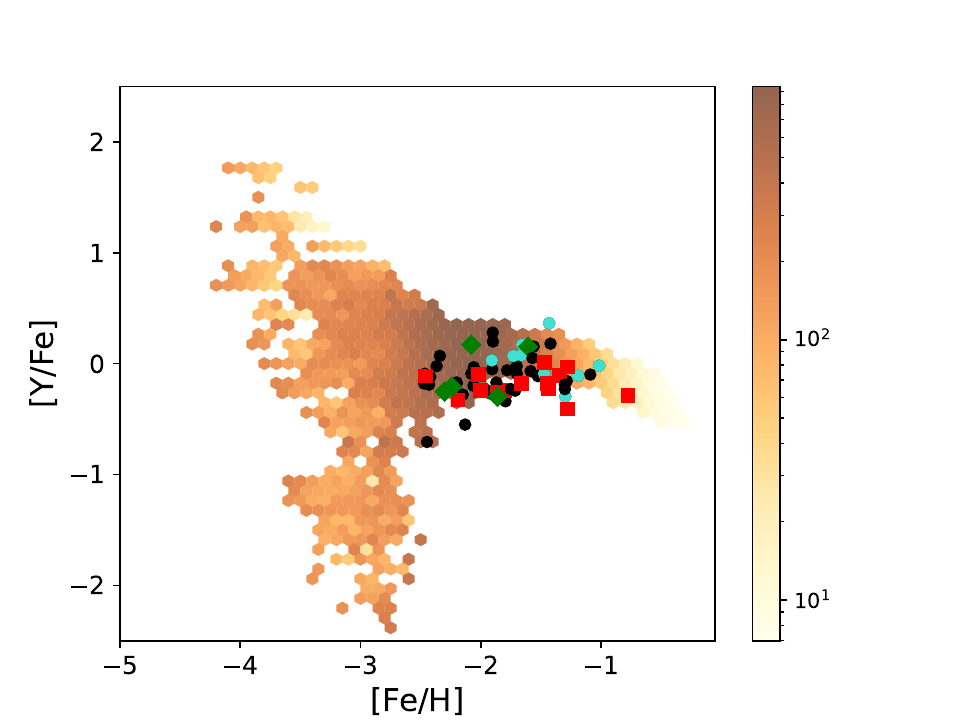}}\\
\subfloat{\includegraphics[width=0.35\textwidth]{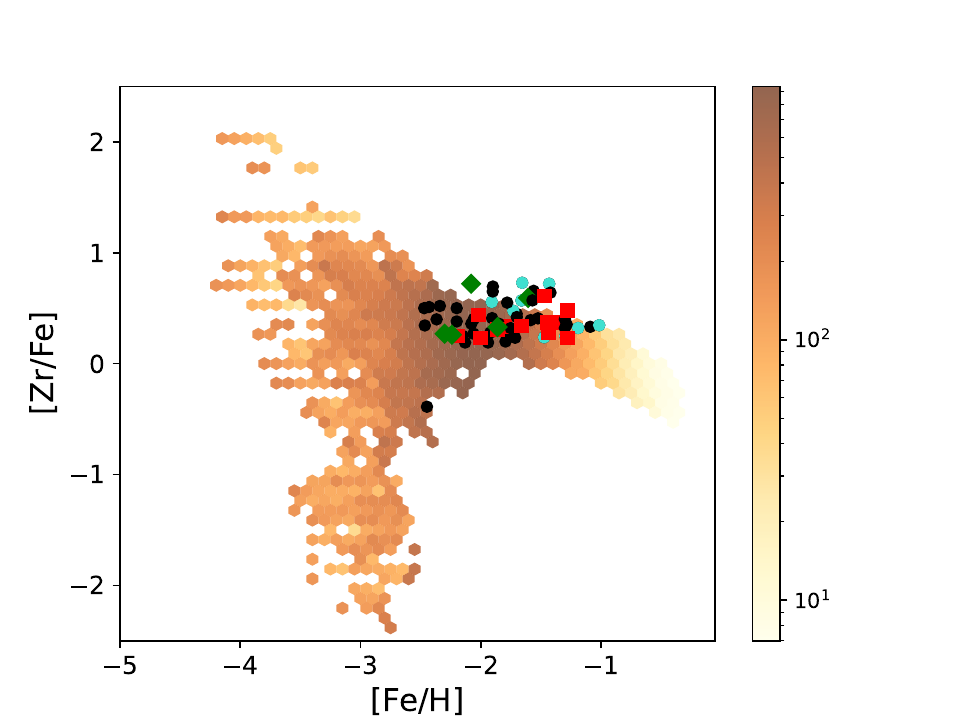}}
\subfloat{\includegraphics[width=0.35\textwidth]{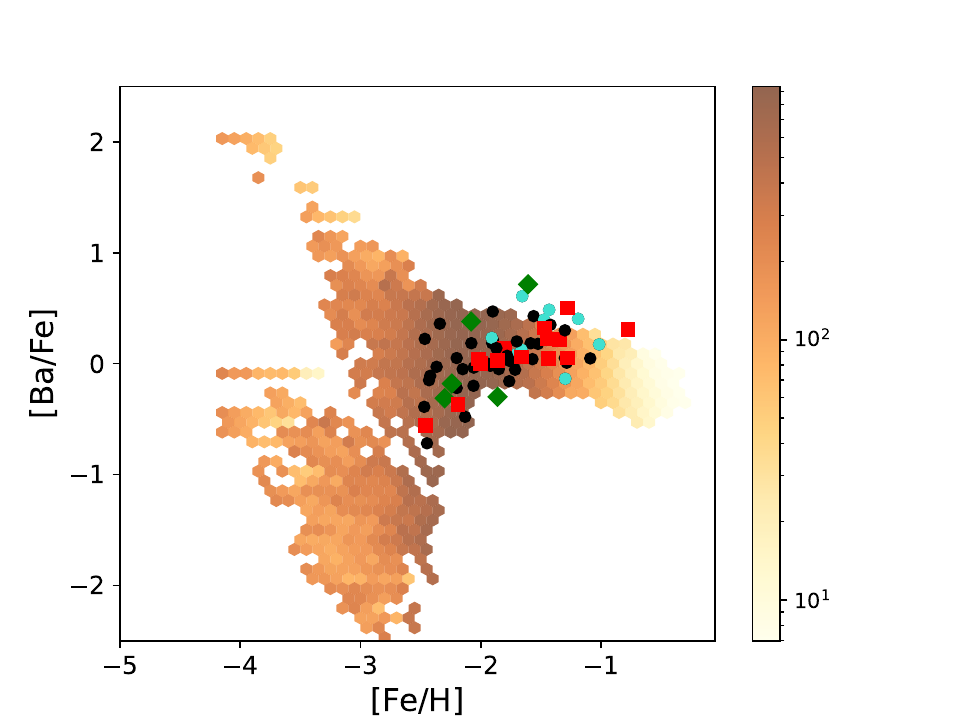}}\\
\subfloat{\includegraphics[width=0.35\textwidth]{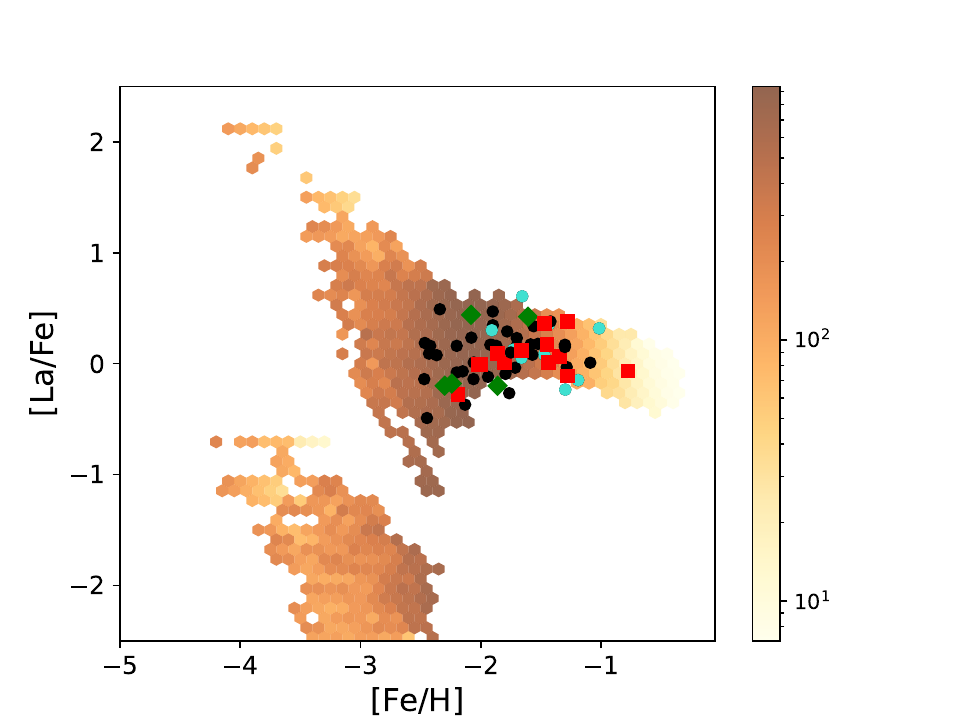}} 
\subfloat{\includegraphics[width=0.35\textwidth]{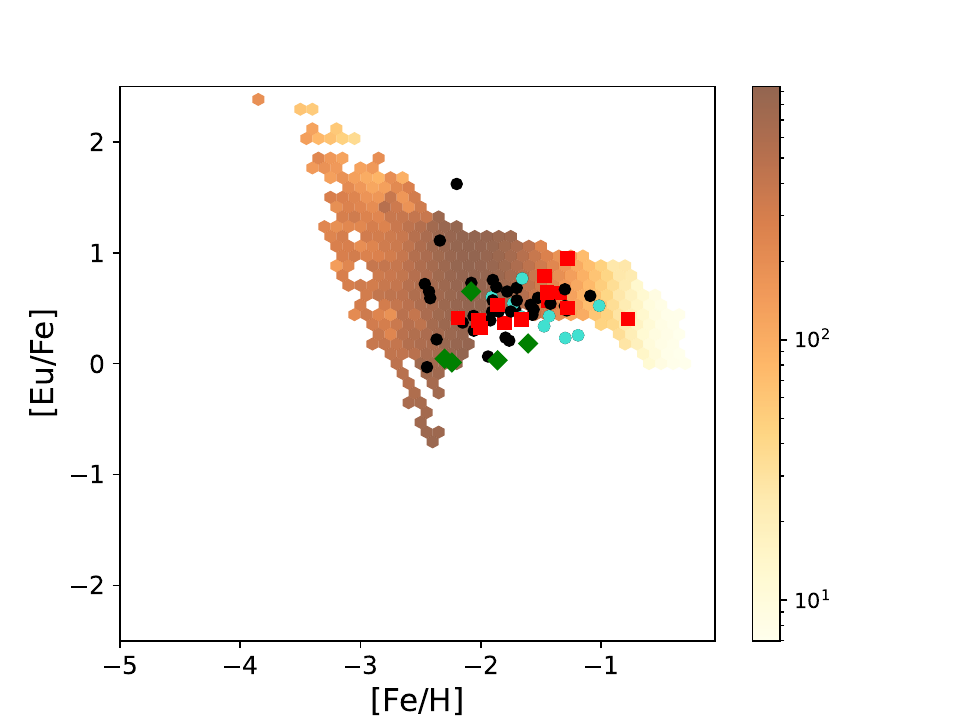}}
\caption{Comparison between the chemical abundances in MINCE I and III stars and the chemical evolution model for Zn, Rb, Sr, Y, Zr, Ba, La, Eu. Colors and symbols are the same as in Figure \ref{fig:alpha}}.
\label{fig:models2}
\end{figure*}

\newpage

\section{Additional tables}
\vspace{1cm}

\topcaption{\label{data} Coordinates, $Gaia$ magnitudes and radial velocities of the sample.}
\tablefirsthead{ \toprule \toprule 
\multicolumn{1}{c}{N}&\multicolumn{1}{c}{Star}&\multicolumn{1}{c}{RA}&\multicolumn{1}{c}{DEC}&\multicolumn{1}{c}{$G$}&\multicolumn{1}{c}{$G_{\rm BP}-G_{\rm RP}$}&\multicolumn{1}{c}{RV}&\multicolumn{1}{c}{$\sigma_{\rm RV}$}\\ 

\multicolumn{1}{c}{}&\multicolumn{1}{c}{}&\multicolumn{1}{c}{[hh:mm:ss]}&\multicolumn{1}{c}{[$^\circ:':''$]}&\multicolumn{1}{c}{[mag]}&\multicolumn{1}{c}{[mag]}&\multicolumn{1}{c}{km s$^{-1}$}&\multicolumn{1}{c}{km s$^{-1}$}\\
\midrule}
\tablehead{%
\multicolumn{8}{l}%
{\bfseries \tablename \thetable{--continued.}} \\
\toprule
\multicolumn{1}{c}{N}&\multicolumn{1}{c}{Star}&\multicolumn{1}{c}{RA}&\multicolumn{1}{c}{DEC}&\multicolumn{1}{c}{$G$}&\multicolumn{1}{c}{$G_{\rm BP}-G_{\rm RP}$}&\multicolumn{1}{c}{RV}&\multicolumn{1}{c}{$\sigma_{\rm RV}$}\\ 

\multicolumn{1}{c}{}&\multicolumn{1}{c}{}&\multicolumn{1}{c}{[hh:mm:ss]}&\multicolumn{1}{c}{[$^\circ:':''$]}&\multicolumn{1}{c}{[mag]}&\multicolumn{1}{c}{[mag]}&\multicolumn{1}{c}{[km s$^{-1}$]}&\multicolumn{1}{c}{[km s$^{-1}$]}\\  
\midrule}
\tabletail{%
\midrule \multicolumn{8}{r}{{Continue on the next page}} \\}
\tablelasttail{%
\\\midrule
\bottomrule
\multicolumn{8}{l}{\textbf{Note.} $^{(a)}$RV corrected by our findings (see subsection \ref{RV}).} \\
\multicolumn{8}{l}{\textbf{Note.} $^{(b)}$RV derived by the UVES spectrum (see subsection \ref{RV}).} \\}
\begin{supertabular}{llllrrrr}
 1&BD-11 3235 & 12:05:09.45 & $-$12:04:53.1 & 9.41 & 1.54 & 118.66                  & 0.24\\
 2&BD-12 5650 & 20:07:39.97 & $-$11:44:04.6 & 9.03 & 1.47 & $-$7.89                   & 0.15\\
 3&BD-13 934 & 04:37:15.68 & $-$13:38:48.1 & 9.05 & 1.48 & 80.89                    & 0.19\\
 4&BD-13 3195 & 10:41:59.98 & $-$14:41:14.4 & 9.03 & 1.4 & 129.82                   & 0.16\\
 5&BD-13 3488 & 12:06:16.06 & $-$14:33:42.6 & 9.02 & 1.68 & 72.06                   & 0.13\\
 6&BD-13 6352 & 23:09:38.58 & $-$12:17:27.4 & 9.63 & 1.78 & 6.15                    & 0.37\\
 7&BD-14 52 & 00:22:00.67 & $-$13:59:08.9 & 9.66 & 1.56 & 59.75                     & 0.24\\
 8&BD-15 4109 & 15:27:20.71 & $-$16:17:57.8 & 9.22 & 1.65 & $-$30.24                  & 0.22\\
 9&BD-15 5449 & 19:44:42.15 & $-$15:31:01.1 & 9.35 & 1.51 & 125.96                  & 0.37\\
 10&BD-16 2232 & 07:59:46.83 & $-$17:23:08.4 & 9.44 & 1.38 & 510.25                  & 0.29\\
 11&BD-17 3143 & 10:25:45.93 & $-$17:59:20.5 & 9.8 & 1.45 & 44.59                    & 0.33\\
 12&BD-17 4251 & 15:06:44.19 & $-$17:37:39.6 & 9.16 & 1.71 & 352.18                  & 0.41\\
 13&BD-18 5150 & 18:58:39.07 & $-$18:46:52.4 & 8.49 & 1.54 & 25.84                   & 0.13\\
 14&BD-19 6363 & 22:55:38.46 & $-$18:45:54.0 & 9.71 & 1.13 & 37.6                    & 3.29\\
 15&CD-23 1855 & 04:18:21.57 & $-$23:23:39.1 & 9.63 & 1.27 & 53.83                   & 0.17\\
 16&CD-23 11064 & 13:25:48.58 & $-$23:59:13.0 & 9.36 & 1.97 & $-$47.95                 & 0.21\\
 17&CD-24 613 & 01:26:26.97 & $-$24:21:03.3 & 9.62 & 1.96 & $-$4.02                    & 2.16\\
 18&CD-24 1384 & 03:02:13.68 & $-$23:48:37.4 & 9.23 & 1.67 & 179.45                  & 0.13\\
 19&CD-26 7083 & 09:27:48.16 & $-$26:48:30.9 & 8.82 & 1.29 & 105.36                  & 0.13\\
 20&CD-27 14182 & 19:40:22.01 & $-$27:37:48.4 & 9.21 & 1.71 & $-$4.85                  & 0.24\\
 21&CD-27 15535 & 21:37:36.08 & $-$26:56:48.0 & 9.28 & 1.79 & 40.97                  & 0.13\\
 22&CD-27 16505 & 23:59:31.55 & $-$27:16:52.6 & 9.18 & 1.57 & $-$98.43                 & 0.3\\
 23&CD-28 10039 & 13:22:48.61 & $-$29:13:12.2 & 9.09 & 1.54 & 60.7                   & 0.17\\
 24&CD-28 10387 & 14:01:50.71 & $-$29:28:30.1 & 9.71 & 1.05 & $-$43.3$^{(a)}$          & 1.77\\
 25&CD-28 16762 & 20:34:20.77 & $-$28:32:33.3 & 9.55 & 1.42 & $-$90.89                 & 0.16\\
 26&CD-28 17446 & 21:46:45.62 & $-$27:46:43.4 & 9.44 & 1.6 & 96.97                   & 0.34\\
 27&CD-29 9391 & 11:52:09.82 & $-$30:16:19.7 & 9.43 & 1.94 & $-$0.18                   & 0.97\\
 28&CD-29 15930 & 19:17:19.19 & $-$29:28:02.9 & 9.38 & 1.57 &$-$10.5$^{(a)}$           & 0.45\\
 29&CD-31 16658 & 19:24:40.24 & $-$31:03:08.6 & 9.28 & 1.87 & 18.51$^{(a)}$          & 0.18\\
 30&CD-31 16922 & 19:39:27.38 & $-$31:20:59.4 & 8.95 & 1.57 & 19.69$^{(a)}$          & 9.08\\
 31&CD-31 17277 & 20:04:18.55 & $-$31:17:17.4 & 9.18 & 1.72 & 111.21                    & 0.51\\
 32&CD-32 4154 & 07:34:02.70 & $-$32:37:53.2 & 9.23 & 1.73 & 56.74                      & 0.14\\
 33&CD-32 13158 & 17:41:14.77 & $-$32:08:03.2 & 6.91 & 1.03 & $-$15.71                    & 0.13\\
 34&CD-32 14894 & 19:08:54.80 & $-$32:25:49.6 & 9.31 & 1.82 & 93.59                     & 0.24\\
 35&CD-33 2721 & 06:05:11.04 & $-$33:08:47.2 & 8.89 & 1.43 & 219.28                     & 0.25\\
 36&CD-33 15063 & 20:37:09.52 & $-$33:30:29.8 & 9.51 & 1.52 & $-$83.15                    & 0.27\\
 37&CD-34 242 & 00:42:17.06 & $-$33:23:34.3 & 9.41 & 1.88 & 11.86                       & 1.41\\
 38&CD-34 3626 & 07:28:22.20 & $-$34:16:43.9 & 8.62 & 1.37 & 39.16                     & 0.13\\
 39&CD-34 5597 & 09:07:05.26 & $-$34:36:22.8 & 8.98 & 1.41 & 45.47                      & 0.14\\
 40&CD-35 4882 & 08:38:16.96 & $-$36:16:12.3 & 8.95 & 1.53 & 58.45                      & 0.14\\
 41&CD-35 13334 & 19:15:27.67 & $-$34:54:44.0 & 9.44 & 1.91 & $-$84.73                    & 0.29\\
 42&CD-35 13661 & 19:44:11.70 & $-$35:26:28.2 & 9.55 & 1.37 & 59.37                     & 0.26\\
 43&CD-35 14807 & 21:29:53.08 & $-$35:10:08.6 & 9.73 & 1.53 & 14.21                     & 0.18\\
 44&CD-36 518 & 01:21:26.75 & $-$35:41:02.3 & 9.45 & 1.33 & 154.19                      & 0.16\\
 45&CD-36 11584 & 17:31:01.16 & $-$36:45:38.0 & 7.57 & 1.06& $-$15.71$^{(b)}$             & \\
 46&CD-38 13823 & 20:03:11.21 & $-$38:27:01.0 & 9.85 & 1.33 &$-$218.19                    & 0.28\\
 47&CD-39 6037 & 10:00:24.45  & $-$39:59:00.1 & 9.08 & 1.54 & 203.46                     & 0.22\\
 48&CD-39 9313 & 14:53:41.90 & $-$40:29:56.2 & 9.29 & 1.41 & 136.67                     & 0.44\\
 49&CD-40 471 & 01:52:29.17 & $-$39:39:54.5 & 9.19 & 1.5 & 81.68                        & 0.36\\
 50&CD-41 4744 & 09:02:29.24 & $-$42:25:06.9 & 8.84 & 1.48 & 19.46                      & 0.18\\
 51&CD-41 7816 & 13:25:56.60 & $-$41:40:29.2 & 8.89 & 1.48 & 18.38                      & 0.14\\
 52&CD-41 14843 & 22:20:13.37 & $-$41:02:47.4 & 9.3 & 1.67 & $-$8.63                      & 0.15\\
 53&CD-43 7161 & 11:35:54.19 & $-$43:41:57.8 & 9.24 & 1.51 & 189.2                      & 0.26\\
 54&CD-43 8736 & 14:01:43.85 & $-$43:56:33.8 & 9.31 & 1.79 & $-$2.15                      & 0.33\\
 55&CD-44 12644 & 18:29:45.64 & $-$44:24:28.2 & 9.08 & 1.27 & 34.60$^{(a)}$          & 0.16\\
 56&CD-44 13783 & 20:11:49.99 & $-$44:07:42.6 & 9.33 & 1.87 & 92.29                     & 0.25\\
 57&CD-44 13981 & 20:29:41.18 & $-$43:47:02.7 & 9.23 & 1.75 & $-$22.36                    & 0.18\\
 58&CD-44 15269 & 23:31:56.96 & $-$44:12:43.4 & 9.9 & 1.42 & 25.83                      & 0.23\\
 59&CD-46 8357 & 13:03:11.95 & $-$46:52:51.5 & 9.07 & 1.41 & $-$32.39                     & 5.24\\
 60&CD-46 13658 & 20:41:33.09 & $-$45:45:25.0 & 9.53 & 1.78 &$-$39.74$^{(b)}$             & \\
 61&CD-47 14271 & 22:38:03.11 & $-$46:32:32.0 & 9.57 & 1.88 & $-$36.33                    & 0.28\\
 62&CD-48 12928 & 19:09:18.11 & $-$48:35:45.5 & 8.85 & 2.08 & 7.46                      & 0.35\\
 63&CD-50 823 & 02:47:57.63 & $-$49:41:36.7 & 8.86 & 1.54 & 44.46                       & 0.23\\
 64&CD-50 877 & 02:57:28.07 & $-$49:31:26.2 & 8.73 & 1.98 & 109.99                      & 0.61\\
 65&CD-52 976 & 04:41:20.20 & $-$51:54:47.5 & 9.35 & 1.21 & 193.02                      & 0.16\\
 66&CD-52 2441 & 08:34:39.47 & $-$52:36:37.8 & 8.67 & 1.33 &16.85$^{(b)}$               & \\
 67&CD-52 4849 & 12:08:03.89 & $-$52:58:26.7 & 9.21 & 1.56 & 150.76                     & 0.34\\
 68&CD-57 1959 & 08:01:58.97 & $-$57:52:03.2 & 9.42 & 1.41 & 140.32                     & 0.14\\
 69&CD-58 294 & 01:24:37.75 & $-$57:33:46.9 & 9.61 & 2.01 & 104.44                      & 0.25\\
 70&CD-59 6913 & 18:36:34.19 & $-$59:42:35.5 & 9.14 & 1.66 & 31.47                      & 0.33\\
 71&CD-62 15 & 00:24:26.77 & $-$61:41:39.8 & 9.39 & 1.25 & 95.52                        & 0.18\\
 72&CPD-62 1126 & 08:54:42.79 & $-$62:49:48.0 & 8.78 & 1.45 & $-$4.04                     & 0.15\\
 73&HD 6518 & 01:05:49.23 & $-$26:39:45.2 & 8.92 & 1.39 & 128.24                        & 0.17\\
 74&HD 19367 & 03:06:25.12 & $-$15:00:12.9 & 9.22 & 1.45 & 141.72                       & 0.29\\
 75&HD 41020 & 06:00:37.04 & $-$40:22:06.0 & 8.61 & 1.69 & 164.25                       & 0.49\\
 76&HD 114621 & 13:12:56.14 & $-$63:56:09.6 & 8.35 & 1.43 & 1.22                        & 0.14\\
 77&HD 298008 & 10:23:46.94 & $-$50:01:54.2 & 8.83 & 1.52 & 7.2                         & 0.51\\
 78&HD 298296 & 09:11:01.43 & $-$52:25:17.1 & 8.69 & 1.4 & 26.64                        & 0.13\\
 79&TYC 5340-1656-1 & 05:29:17.35 & $-$11:40:37.6 & 9.62 & 1.42 & 46.67$^{(b)}$         & \\
 80&TYC 5422-1192-1 & 07:41:52.22 & $-$13:36:53.5 & 9.43 & 0.48 & 60.24$^{(b)}$         & \\
 81&TYC 5763-1084-1 & 20:35:42.25 & $-$12:35:36.5 & 9.82 & 1.91 &$-$54.96                 & 0.24\\
 82&TYC 6002-1724-1 & 08:24:38.17 & $-$17:55:23.6 & 9.75 & 1.29 & 207.81               & 0.2\\
 83&TYC 6108-150-1 & 12:20:06.99 & $-$22:14:35.3 & 9.82 & 1.21 & 109.24                & 2.59\\
 84&TYC 6195-815-1 & 15:56:04.95 & $-$19:59:52.7 & 9.58 & 1.76 & 13.01                 & 0.28\\
 85&TYC 6199-351-1 & 15:58:13.62 & $-$21:56:52.4 & 9.19 & 1.52 & 78.89                 & 0.19\\
 86&TYC 6310-1651-2 & 19:28:39.93 & $-$20:48:30.6 & 9.53 & 1.91 & 27.56                & 0.21\\
 87&TYC 7144-1122-1 & 08:36:40.01 & $-$34:15:50.7 & 9.28 & 1.81 & 22.53                & 0.13\\
 88&TYC 7351-734-1 & 16:13:05.07 & $-$34:23:59.3 & 8.99 & 1.74 & $-$46.3                 & 0.13\\
 89&TYC 7354-1196-1 & 16:46:09.99 & $-$34:08:43.1 & 8.82 & 1.31 & 4.47                 & 0.21\\
 90&TYC 7427-35-1 & 19:17:11.88 & $-$32:32:42.5 & 8.91 & 1.63 & $-$37.15                 & 0.14\\
 91&TYC 8025-237-1 & 00:01:55.06 & $-$48:45:17.1 & 9.51 & 1.75 & 60.1                  & 0.23\\
 92&TYC 8161-753-1 & 08:23:43.70 & $-$50:58:23.4 & 8.84 & 1.42 & 209.7                 & 0.18\\
 93&TYC 8178-334-1 & 09:12:04.35 & $-$52:08:22.1 & 8.87 & 1.5 &  16.59$^{(b)}$         & \\
 94&TYC 8263-2160-1 & 13:58:20.15 & $-$45:46:43.0 & 9.01 & 1.43 & 10.51                & 0.19\\
 95&TYC 8325-6844-1 & 16:31:24.40 & $-$46:32:20.0 & 9.21 & 1.58 & $-$65.07               & 0.46\\
 96&TYC 8394-14-1 & 19:48:11.54 & $-$48:35:11.9 & 9.29 & 1.67 & 1.92                   & 0.45\\
 97&TYC 8400-1610-1 & 20:04:23.01 & $-$49:06:31.6 & 9.56 & 1.97 & 7.83                 & 0.25\\
 98&TYC 8584-127-1 & 09:23:52.76 & $-$53:58:53.1 & 8.66 & 1.51 & 21.96                 & 0.23\\
 99&TYC 8633-2281-1 & 12:13:36.87 & $-$53:13:12.9 & 8.67 & 1.61 & 302.75               & 0.25\\
\end{supertabular}

\newpage
\topcaption{\label{atm_param} Reddening, atmospheric parameters and metallicity of the sample.} 
\tablefirsthead{ \toprule \toprule 
\multicolumn{1}{l}{N}&\multicolumn{1}{c}{Star}&\multicolumn{1}{c}{${\rm A_V}$}&\multicolumn{1}{c}{\teff}&\multicolumn{1}{c}{\logg}&\multicolumn{1}{c}{$\xi$}&\multicolumn{1}{c}{[Fe/H]}&\multicolumn{1}{l}{Comments}\\ 

\multicolumn{1}{l}{}&\multicolumn{1}{c}{}&\multicolumn{1}{c}{}&\multicolumn{1}{c}{[K]}&\multicolumn{1}{c}{[gcs]}&\multicolumn{1}{c}{[\kms]}&\multicolumn{1}{c}{}&\multicolumn{1}{l}{}\\
\midrule}
\tablehead{%
\multicolumn{8}{l}%
{\bfseries \tablename \thetable{--continued.}} \\
\toprule
\multicolumn{1}{l}{N}&\multicolumn{1}{l}{Star}&\multicolumn{1}{c}{${\rm A_V}$}&\multicolumn{1}{c}{\teff}&\multicolumn{1}{c}{\logg}&\multicolumn{1}{c}{$\xi$}&\multicolumn{1}{c}{[Fe/H]}&\multicolumn{1}{l}{Comments}\\ 

\multicolumn{1}{l}{}&\multicolumn{1}{c}{}&\multicolumn{1}{c}{}&\multicolumn{1}{c}{[K]}&\multicolumn{1}{c}{[gcs]}&\multicolumn{1}{c}{[\kms]}&\multicolumn{1}{c}{}&\multicolumn{1}{l}{}\\
\midrule}
\tabletail{%
\midrule \multicolumn{8}{r}{{Continue on the next page}} \\}
\tablelasttail{%
\\\midrule
\bottomrule
\multicolumn{8}{l}{}\\}
\begin{supertabular}{llcrrrrl}
1  &  \textbf{BD-11        3235} &                                \textbf{0.18} &     \textbf{4158}  &    \textbf{0.90} &      \textbf{1.93} &   \textbf{$-$2.11$\pm$ 0.14} & \\ 
2  &  BD-12        5650          &                     0.44                    &    4200                &       1.49           &    1.50                &  $-0.13\pm 0.19$                           & \\
3  &  \textbf{BD-13         934 }&                                \textbf{0.66} &    \textbf{4245}   &    \textbf{0.91} &      \textbf{1.97} &   \textbf{$-$1.93 $\pm$ 0.12} & --\\
4  &  \textbf{BD-13        3195} &                                \textbf{0.17} &     \textbf{4280}  &    \textbf{1.15} &      \textbf{1.81} &   \textbf{$-$1.41 $\pm$ 0.15} & --\\ 
5  &  BD-13        3488          &                                0.21          &  4000              &      1.36        &  1.50              &      $-0.39\pm 0.20$                     &            \\
6  &  BD-13        6352          &                                0.09          &  3928              &      0.61        &  1.90              &  $-1.28\pm 0.07$            & \\
7  &  BD-14          52          &                                0.06          &   4108             &          0.75    &  1.94              &  $-1.70\pm 0.17$            &  $Gaia$ SB1 \\
8  &  BD-15        4109          &                      0.34                    &   4048             &          0.53    &   2.04             &      $-2.01\pm 0.15$       & \\
9  &  \textbf{BD-15        5449 }&                                \textbf{0.58} &     \textbf{4230}  &    \textbf{0.80} &      \textbf{2.01} &   \textbf{$-$2.19 $\pm$ 0.12} & \\ 
10 &  \textbf{BD-16        2232} &                                \textbf{0.09} &     \textbf{4367}  &    \textbf{0.98 }&      \textbf{1.96} &   \textbf{$-$1.77 $\pm$ 0.12} & \\ 
11 & \textbf{BD-17        3143}  &                                \textbf{0.24} &     \textbf{4344}  &    \textbf{0.48} &     \textbf{2.19}  &   \textbf{$-$1.68 $\pm$ 0.14} & \\
12 &  BD-17        4251          &                               0.30           &    4000            &          0.53    &       1.97         &      $-1.47\pm 0.20$      &  \\
13 &  BD-18        5150          &                                0.57          &      4080          &  1.06            &   1.66             &      $-0.27\pm 0.16$        & \\
14 & \textbf{BD-19        6363}  &                                \textbf{0.09} &     \textbf{4767}  &    \textbf{1.82} &     \textbf{ 1.82 }&   \textbf{$-$2.51 $\pm$ 0.12} & \\ 
15 &  \textbf{CD-23        1855} &                                \textbf{0.14} &     \textbf{4551}  &    \textbf{1.23 }&      \textbf{1.96} &   \textbf{$-$1.96 $\pm$ 0.11} & \\ 
16 &  CD-23       11064          &                      0.28                    &   3748             &  0.82            &       1.73         &  $-0.83\pm 0.19$            &  \\
17 &  CD-24         613          &                                0.04          &  3758             &       0.81       &        1.82        &  $-0.54 \pm 0.16$           & \\
18 &  CD-24        1384          &                                0.06          &      4000          &      1.05        &   1.67             &      $-0.70\pm 0.16$        & \\
19 &  CD-26        7083          &                                0.41          &  4579              &          0.53    &       2.23         &        $-0.43\pm 0.08$      & \\
20 &  CD-27       14182          &                                0.50          &    4000            &  1.11            &  1.63              &  $-0.28\pm 0.20$            & \\
21 &  CD-27       15535          &                                0.08          &     3935           &  1.13            &  1.59              & $-0.41\pm 0.12$             & \\
22 &  \textbf{CD-27       16505 }&                                \textbf{0.05} &     \textbf{4046}  &    \textbf{0.80} &      \textbf{1.85} &   \textbf{$-$1.33 $\pm$ 0.16} & \\ 
23 & \textbf{CD-28       10039 } &                                \textbf{0.22 }&     \textbf{4100}  &    \textbf{0.71} &     \textbf{ 1.91} &   \textbf{$-$1.12 $\pm$ 0.20} & \\ 
24 & CD-28 10387                 &                               0.16           &       5043         &      2.25        &  1.58              &          $-0.28\pm 0.17$    & \\
25 & \textbf{CD-28       16762}  &                                \textbf{0.17} &     \textbf{4291}  &    \textbf{0.96} &     \textbf{ 1.93} &   \textbf{$-$1.63 $\pm$ 0.13} & \\
26 & \textbf{CD-28       17446}  &                                \textbf{0.07} &     \textbf{4063}  &    \textbf{0.75} &      \textbf{1.93} &   \textbf{$-$1.84 $\pm$ 0.17} & \\ 
27 &  CD-29        9391          &                                0.17          &    3780            &      0.71        &       1.75         &      $-0.46\pm 0.13$        & \\
28 & \textbf{CD-29       15930}  &                                \textbf{0.40} &     \textbf{4105}  &    \textbf{0.72} &      \textbf{1.96} &   \textbf{$-$1.79 $\pm$ 0.11} & \\ 
29 &  CD-31       16658          &                                0.26          &   3844             &  0.70            &       1.81         &   $-0.84\pm 0.16$           & \\
30 &  CD-31       16922          &                                0.31          &                    &                  &                    &                             & fast rotator \\
31 &  CD-31       17277          &                      0.68                    &  4000              &      0.74        &  1.84              & $-0.97\pm0.11$              & LPV\\
32 &  CD-32        4154          &                      0.27                    &      3993          &      0.97        &  1.69              &  $-0.62\pm 0.10$            & \\
33 & CD-32       13158           &                      3.06                    &      5129          &      1.65        &   1.87             &      $0.10\pm 0.15$         &  \\
34 & CD-32       14894           &                      0.35                    &    3892            &      0.53        & 1.92               &  $-1.05\pm 0.10$            & \\
35 & \textbf{CD-33       2721}   &                                \textbf{0.16} &     \textbf{ 4343} &    \textbf{0.78} &     \textbf{ 2.06 }&   \textbf{$-$1.89 $\pm$ 0.18} & \\
36 &  \textbf{CD-33       15063} &                                \textbf{0.12} &     \textbf{4140}  &    \textbf{0.90} &      \textbf{1.89} &   \textbf{$-$1.80 $\pm$ 0.16} & \\
37 &  CD-34         242          &                                0.04          &      3843          &  0.88            &   1.67             &  $-0.33\pm 0.09$            & \\
38 &  CD-34        3626          &                                0.36          &       4396         &      0.81        &  1.90              &      $-0.01\pm 0.17$        & \\
39 &  CD-34        5597          &                                0.66          &     4277           &  1.21            &  1.71              &      $-0.56\pm 0.15$       & \\
40 &  CD-35        4882          &                                0.30          &     4101           &  1.37            &   1.54             &      $-0.41\pm 0.16$       & \\
41 &  CD-35       13334          &                                0.24          &      3800          &  0.78            &  1.76              &      $-0.74\pm 0.11$        & \\
42 &  CD-35       13661          &                                0.60          &      4442          &      0.89        &   2.08             &      $-2.29\pm 0.15$        & $Gaia$ SB1 \\
43 &  CD-35       14807          &                                0.12          &     4083           &      1.24        &       1.62         &      $-0.75\pm 0.16$        & \\
44 & \textbf{CD-36         518 } &                                \textbf{0.07} &     \textbf{4424}  &    \textbf{1.24} &      \textbf{1.86} &   \textbf{$-$1.74 $\pm$ 0.11} & \\ 
45 &  CD-36       11584          &                      2.64                    &     5059           &      0.88        &  2.20              &      $0.10\pm 0.15$         & fast rotator\\
46 & \textbf{CD-38       13823}  &                                \textbf{0.22} &     \textbf{4515}  &    \textbf{0.85} &     \textbf{ 2.15 }&   \textbf{$-$2.46 $\pm$ 0.14} & \\
47 & \textbf{CD-39        6037 } &                                \textbf{0.23} &     \textbf{4082 } &    \textbf{0.92} &     \textbf{ 1.79} &   \textbf{$-$0.99 $\pm$ 0.16} & \\ 
48 & \textbf{CD-39        9313}  &                               \textbf{ 0.43} &     \textbf{4323}  &    \textbf{1.03} &      \textbf{1.93} &   \textbf{$-$1.68 $\pm$ 0.15} & \\ 
49 &  \textbf{CD-40         471} &                                \textbf{0.05 }&     \textbf{4323}  &    \textbf{0.63} &     \textbf{ 2.17} &   \textbf{$-$2.48 $\pm$0.14}  & \\ 
50 &  CD-41        4744          &              
0.33                    &      4168          &      1.41        &   1.55             &      $-0.31\pm 0.16$        & \\
51 &  CD-41        7816          &                                0.16          &      4199          &      1.80        &   1.35             &       $-0.0 7\pm 0.17$      & \\
52 &  CD-41       14843          &                                0.03          &     4000           &      1.13        &       1.63         &      $-0.59\pm 0.19$        & \\
53 &  \textbf{CD-43        7161} &                                \textbf{0.30} &     \textbf{4167}  &    \textbf{0.82} &     \textbf{1.93}  &   \textbf{$-$1.50 $\pm$ 0.13} & \\
54 &  CD-43        8736          &                                0.28          &    3919            &      0.67        &  1.87              &          $-1.21\pm 0.09$    & \\
55 & \textbf{CD-44       12644 } &                               \textbf{0.28}  &     \textbf{4564 } &    \textbf{1.15} &     \textbf{ 2.00} &   \textbf{$-$1.44 $\pm$ 0.12} & \\ 
56 &  CD-44       13783          &                                0.14          &  3848              &      0.86        &   1.73             &      $-0.76\pm 0.15$        & \\
57 &  CD-44       13981          &                               0.12           &    3970            &      0.96        &    1.71            &  $-0.71 \pm 0.14$           & \\
58 & \textbf{CD-44       15269}  &                                \textbf{0.05} &     \textbf{4265}  &    \textbf{1.00} &     \textbf{ 1.87} &  \textbf{$-$1.39 $\pm$ 0.15} & \\ 
59 &  CD-46        8357          &                               0.18           &                    &                  &                    &                             & RS CVn\\
60 &  CD-46       13658          &                                0.09          &    3936            &  0.94            &   1.70             &  $-0.68 \pm 0.14$           & \\
61 &  CD-47       14271          &                                0.04          &    3827            & 0.76             &  1.81              &      $-0.93\pm 0.20$        & \\
62 &  CD-48       12928          &                               0.24           &    3642            &  0.64            &  1.80              &      $-0.90\pm 0.20$        & LPV\\
63 &  CD-50         823          &                                0.11          &      4141          &  0.70            &  1.97              &      $-1.64\pm  0.18$       & $Gaia$ SB1 \\ 
64 &  CD-50         877          &                                0.08          &      3733          &  0.55            & 1.86               &    $-0.91\pm 0.12$          & LPV\\
65 & \textbf{CD-52         976 } &                                \textbf{0.06} &     \textbf{4655}  &    \textbf{1.29} &     \textbf{ 1.97} &   \textbf{$-$1.90 $\pm$ 0.09} & \\ 
66 &  CD-52        2441          &                      1.24                    &                    &                  &                    &                             & SB2\\
67 &  CD-52        4849          &                                0.52          &      4233          &      0.53        &   2.05             &      $-2.36\pm 0.17$        & $Gaia$ SB1 \\
68 & \textbf{CD-57        1959}  &                                 \textbf{0.50}&     \textbf{4342}  &   \textbf{ 1.09} &     \textbf{ 1.93} &   \textbf{$-$2.00 $\pm$ 0.12} & \\
69 &  CD-58         294          &                                0.07          &   3707             &  0.62            &   1.81             &  $-0.81\pm 0.12$            & LPV\\
70 &  CD-59        6913          &                                0.22          &   4044             &      0.52        &   2.04             &      $-2.03\pm 0.17$        &  \\
71 &  \textbf{CD-62          15} &                                \textbf{ 0.05}&     \textbf{4584}  &    \textbf{1.26} &      \textbf{1.97} &   \textbf{$-$2.11 $\pm$ 0.12} & \\ 
72 &  CPD-62        1126         &                               0.34           &      4224          &      1.26        &       1.64         &      $-0.49\pm 0.15$        & \\
73 &  \textbf{HD        6518}    &                                 \textbf{0.06}&     \textbf{4285}  &    \textbf{1.15} &     \textbf{ 1.81} &   \textbf{$-$1.44 $\pm$ 0.14} & \\ 
74 &  \textbf{HD       19367}    &                                 \textbf{0.13}&     \textbf{4345}  &    \textbf{0.74} &      \textbf{2.10} &   \textbf{$-$2.12 $\pm$ 0.14} & \\
75 &     HD       41020          &                                0.23          &   4000             &  0.76            &   1.80           &        $-1.10\pm 0.19$      & LPV\\
76 &     HD      114621          &                                0.80          &      4260          &      0.98        &   1.80             &      $-0.33\pm 0.15$        & \\
77 &     HD      298008          &                      0.21                    &     4141           &  1.48            &       1.48         &      $-0.07\pm 0.18$        & \\
78 &     HD      298296          &                                0.61          &   4311             &      1.00        &   1.80             &          $-0.35\pm 0.15$  & \\
79 &    TYC 5340-1656-1          &                               0.28           &                    &                  &                    &                             & RS CVn (?), SB2\\
80 &    TYC 5422-1192-1          &                        0.54                  &                    &                  &                    &                             & member of OC NGC 2423, fast rotator\\
81 &    TYC 5763-1084-1          &                               0.18           &  3801              & 0.75             &  0.77              & $-0.87\pm 0.12$             & \\
82 &  \textbf{TYC 6002-1724-1}   &                               \textbf{ 0.27} &     \textbf{4551}  &    \textbf{1.12} &     \textbf{ 2.05} &   \textbf{$-$2.34 $\pm$ 0.12} & \\ 
83 &    TYC  6108-150-1          &                                0.20          &    4617            &      1.52        &   1.84             &      $-1.94\pm 0.12$        & $Gaia$ SB1 \\
84 &    TYC  6195-815-1          &                       0.60                   &    3947            &          0.48    &       1.99         &      $-1.49\pm 0.09$        & LPV\\
85 &  \textbf{TYC  6199-351-1 }  &                       \textbf{0.71}          &     \textbf{4111}  &   \textbf{0.99}  &     \textbf{ 1.82} &   \textbf{$-$1.66 $\pm$ 0.16} & \\
86 &    TYC 6310-1651-2          &                        0.30                  &    3809            &      0.67        &       1.80         &  $-0.67\pm 0.12$             & \\
87 &    TYC 7144-1122-1          &                        0.44                  &    3912            &          0.93    &       1.69         &  $-0.45\pm 0.12$            & \\
88 &    TYC  7351-734-1          &                       0.76                   &    3996            & 1.07             &      1.59          &       $ -0.09\pm 0.16$      & \\
89 &    TYC 7354-1196-1          &                        1.03                  &       4456         &      1.73        &   1.55             &      $-0.41\pm 0.15$ & \\
90 &    TYC   7427-35-1          &                               0.26           &      4000          &      1.21        &  1.50              &          $-0.23\pm 0.19$    & \\
91 &    TYC  8025-237-1          &                                0.05          &    3958            &          0.74    &       1.85         &      $ -1.09\pm 0.11 $      & \\
92 &  \textbf{TYC  8161-753-1}   &                                \textbf{0.88} &     \textbf{4266}  &    \textbf{1.25} &     \textbf{ 1.80} &   \textbf{$-$1.77 $\pm$ 0.13} & \\ 
93 &    TYC  8178-334-1          &                                0.68          &      4160          &      1.10        &   1.66             &      $-0.10\pm 0.18$        & \\
94 &    TYC 8263-2160-1          &                                0.31          &    4278            &      0.69        &   1.94             &      $-0.27\pm 0.14$        & \\
95 &    TYC 8325-6844-1          &                                0.52          &      4060          &  1.88            &   1.24             &      $0.03\pm 0.17$         & \\
96 &    TYC   8394-14-1          &                                0.16          &   4000             &      0.53        &  1.97              &      $-1.49\pm 0.19$        & LPV fast rotator\\
97 &    TYC 8400-1610-1          &                                0.13          &    3748            &     0.78          &      1.78         & $-0.88 \pm 0.14$            & \\
98 &    TYC  8584-127-1          &                       1.08                   &    4144           &       0.95       &        1.76         &  $-0.58\pm 0.09$            & 300 Myr, 3.2M\sun \\
99 &   \textbf{TYC 8633-2281-1}  &                        \textbf{0.28}         &     \textbf{4118}  &    \textbf{0.58} &     \textbf{ 2.06} &   \textbf{$-$2.05 $\pm$0.19}  & \\ 
\end{supertabular}

\newpage

\topcaption{\label{kinematics} Targets' velocity components in galactocentric cylindrical coordinates (V$_R$,V$_T$,V$_Z$), apocentric and pericentric distances (r$_{ap}$, r$_{peri}$), orbital eccentricities (ecc.) and maximum height over the galactic plane (Z$_{max}$).}
\tablefirsthead{ \toprule \toprule 
\multicolumn{1}{c}{Star}&\multicolumn{1}{c}{$V_R$}&\multicolumn{1}{c}{$V_T$}&\multicolumn{1}{c}{$V_Z$}&\multicolumn{1}{c}{$r_{ap}$}&\multicolumn{1}{c}{$r_{peri}$}&\multicolumn{1}{c}{ecc.}&\multicolumn{1}{c}{$Z_{max}$}\\ 

\multicolumn{1}{c}{}&\multicolumn{1}{c}{[\kms]}&\multicolumn{1}{c}{[\kms]}&\multicolumn{1}{c}{[\kms]}&\multicolumn{1}{c}{[kpc]}&\multicolumn{1}{c}{[kpc]}&\multicolumn{1}{c}{}&\multicolumn{1}{c}{[kpc]}\\
\midrule}
\tablehead{%
\multicolumn{8}{l}%
{\bfseries \tablename \thetable{--continued.}} \\
\toprule
\multicolumn{1}{c}{Star}&\multicolumn{1}{c}{$V_R$}&\multicolumn{1}{c}{$V_T$}&\multicolumn{1}{c}{$V_Z$}&\multicolumn{1}{c}{$r_{ap}$}&\multicolumn{1}{c}{$r_{peri}$}&\multicolumn{1}{c}{ecc.}&\multicolumn{1}{c}{$Z_{max}$}\\ 

\multicolumn{1}{c}{}&\multicolumn{1}{c}{[\kms]}&\multicolumn{1}{c}{[\kms]}&\multicolumn{1}{c}{[\kms]}&\multicolumn{1}{c}{[kpc]}&\multicolumn{1}{c}{[kpc]}&\multicolumn{1}{c}{}&\multicolumn{1}{c}{[kpc]}\\
\midrule}
\tabletail{%
\midrule \multicolumn{8}{r}{{Continue on the next page}} \\}
\tablelasttail{%
\\\midrule
\bottomrule
}
\begin{supertabular}{lrrrrrrr}
 BD-11        3235 &  45.62 $\pm$   5.25 & 10.01 $\pm$  6.00 & $-$15.69 $\pm$  5.37 &  8.13 $\pm$  0.05  &  0.19 $\pm$  0.11 &  0.95 $\pm$  0.02 &  2.30 $\pm$  0.26 \\ 
 BD-12        5650 &  $-$7.89 $\pm$   0.55 & 206.02 $\pm$  3.01 & 12.21 $\pm$  0.44 &  7.36 $\pm$  0.10  &  6.29 $\pm$  0.25 &  0.08 $\pm$  0.01 &  0.36 $\pm$  0.04 \\ 
 BD-13         934 &  58.93 $\pm$   0.81 & 69.00 $\pm$  4.03 & 77.91 $\pm$  3.70 &  9.87 $\pm$  0.07  &  2.07 $\pm$  0.06 &  0.65 $\pm$  0.01 &  4.70 $\pm$  0.43 \\ 
 BD-13        3195 &  $-$135.34 $\pm$   3.32 & 52.63 $\pm$  2.42 & 29.82 $\pm$  1.48 & 10.11 $\pm$  0.11  &  1.04 $\pm$  0.06 &  0.81 $\pm$  0.01 &  1.69 $\pm$  0.01 \\ 
 BD-13        3488 &  $-$6.05 $\pm$   0.31 & 181.66 $\pm$  0.11 & 65.61 $\pm$  0.14 &  7.90 $\pm$  0.00  &  5.96 $\pm$  0.01 &  0.14 $\pm$  0.00 &  1.78 $\pm$  0.01 \\ 
 BD-13        6352 &  24.03 $\pm$   0.59 & 59.32 $\pm$  9.21 & $-$85.12 $\pm$  4.50 &  8.12 $\pm$  0.06  &  1.58 $\pm$  0.21 &  0.67 $\pm$  0.04 &  4.32 $\pm$  0.40 \\ 
 BD-14          52 &  18.29 $\pm$   0.94 & 110.20 $\pm$  7.04 & $-$86.08 $\pm$  1.82 &  8.77 $\pm$  0.07  &  3.53 $\pm$  0.21 &  0.43 $\pm$  0.03 &  4.35 $\pm$  0.25 \\ 
 BD-15        4109 &  $-$15.38 $\pm$   0.61 & 11.33 $\pm$ 10.40 & $-$123.45 $\pm$  5.18 &  6.45 $\pm$  0.02  &  0.22 $\pm$  0.17 &  0.93 $\pm$  0.05 &  4.53 $\pm$  0.28 \\ 
 BD-15        5449 &  $-$251.76 $\pm$  10.15 & 52.63 $\pm$ 11.01 & $-$0.66 $\pm$  1.92 & 12.05 $\pm$  0.40  &  0.69 $\pm$  0.17 &  0.89 $\pm$  0.03 &  1.72 $\pm$  0.19 \\ 
 BD-16        2232 &  382.80 $\pm$   2.41 & $-$66.67 $\pm$  3.28 & 72.18 $\pm$  0.17 & 74.56 $\pm$  2.40  &  1.07 $\pm$  0.06 &  0.97 $\pm$  0.00 & 12.82 $\pm$  0.48 \\ 
 BD-17        3143 &  $-$171.95 $\pm$   6.88 & 51.97 $\pm$ 10.86 & $-$69.62 $\pm$  6.05 & 14.00 $\pm$  0.67  &  1.15 $\pm$  0.25 &  0.85 $\pm$  0.04 &  3.81 $\pm$  0.50 \\ 
 BD-17        4251 &  $-$334.14 $\pm$   1.73 &  4.35 $\pm$  8.05 & 98.02 $\pm$  5.58 & 26.33 $\pm$  0.31  &  0.07 $\pm$  0.09 &  0.99 $\pm$  0.01 & 19.43 $\pm$  0.37 \\ 
 BD-18        5150 &  $-$26.08 $\pm$   0.29 & 239.84 $\pm$  0.08 &  4.24 $\pm$  0.10 &  8.70 $\pm$  0.04  &  6.80 $\pm$  0.02 &  0.12 $\pm$  0.00 &  0.19 $\pm$  0.00 \\ 
 BD-19        6363 &  $-$129.71 $\pm$   9.04 & 23.59 $\pm$ 15.71 & $-$70.36 $\pm$  4.32 &  9.18 $\pm$  0.24  &  0.51 $\pm$  0.30 &  0.89 $\pm$  0.06 &  4.85 $\pm$  0.76 \\ 
 CD-23        1855 &  $-$112.81 $\pm$   2.79 &  2.36 $\pm$  5.40 & $-$2.71 $\pm$  0.69 & 10.55 $\pm$  0.11  &  0.04 $\pm$  0.07 &  0.99 $\pm$  0.01 &  1.61 $\pm$  0.05 \\ 
 CD-23       11064 &  25.52 $\pm$   0.33 & 258.65 $\pm$  0.21 & 10.42 $\pm$  1.39 & 11.21 $\pm$  0.02  &  7.12 $\pm$  0.03 &  0.22 $\pm$  0.00 &  1.56 $\pm$  0.08 \\ 
 CD-24         613 &  34.28 $\pm$   2.30 & 191.39 $\pm$  2.00 & 19.72 $\pm$  2.11 &  8.90 $\pm$  0.06  &  6.74 $\pm$  0.09 &  0.14 $\pm$  0.01 &  2.13 $\pm$  0.12 \\ 
 CD-24        1384 &  85.81 $\pm$   0.62 & 53.46 $\pm$  2.72 & $-$94.33 $\pm$  1.16 &  9.89 $\pm$  0.02  &  1.37 $\pm$  0.08 &  0.76 $\pm$  0.01 &  4.62 $\pm$  0.04 \\ 
 CD-26        7083 &  116.95 $\pm$   4.27 & 209.54 $\pm$  5.32 &  1.45 $\pm$  1.79 & 14.54 $\pm$  0.89  &  6.33 $\pm$  0.17 &  0.39 $\pm$  0.01 &  1.48 $\pm$  0.18 \\ 
 CD-27       14182 &   7.99 $\pm$   0.62 & 244.62 $\pm$  0.48 &  2.27 $\pm$  0.29 &  8.66 $\pm$  0.01  &  6.85 $\pm$  0.04 &  0.12 $\pm$  0.00 &  0.56 $\pm$  0.02 \\ 
 CD-27       15535 &  $-$43.93 $\pm$   0.20 & 179.78 $\pm$  1.32 & $-$32.34 $\pm$  0.25 &  7.68 $\pm$  0.02  &  4.94 $\pm$  0.06 &  0.22 $\pm$  0.00 &  1.25 $\pm$  0.02 \\ 
 CD-27       16505 &  $-$377.06 $\pm$  15.13 &  3.92 $\pm$  8.09 & 148.23 $\pm$  1.76 & 73.99 $\pm$ 20.61  &  0.05 $\pm$  0.08 &  1.00 $\pm$  0.00 & 26.32 $\pm$  6.05 \\ 
 CD-28       10039 &  $-$40.01 $\pm$   0.21 & 179.67 $\pm$  0.49 & 69.56 $\pm$  1.19 &  7.37 $\pm$  0.04  &  5.35 $\pm$  0.01 &  0.16 $\pm$  0.00 &  2.27 $\pm$  0.07 \\ 
 CD-28       10387 &  90.87 $\pm$   4.12 & 200.78 $\pm$  3.19 & 14.26 $\pm$  2.18 &  9.73 $\pm$  0.08  &  5.10 $\pm$  0.16 &  0.31 $\pm$  0.02 &  0.60 $\pm$  0.04 \\ 
 CD-28       16762 &  80.54 $\pm$   0.26 & 27.25 $\pm$  5.90 & $-$36.10 $\pm$  3.00 &  6.96 $\pm$  0.04  &  0.44 $\pm$  0.10 &  0.88 $\pm$  0.03 &  1.38 $\pm$  0.07 \\ 
 CD-28       17446 &  $-$369.09 $\pm$  17.82 & $-$29.91 $\pm$ 12.95 & 69.11 $\pm$  7.29 & 40.38 $\pm$ 11.89  &  0.38 $\pm$  0.16 &  0.98 $\pm$  0.00 & 13.60 $\pm$  5.56 \\ 
 CD-29        9391 &  $-$17.33 $\pm$   0.22 & 189.58 $\pm$  1.71 & $-$44.15 $\pm$  2.03 &  7.92 $\pm$  0.01  &  6.00 $\pm$  0.08 &  0.14 $\pm$  0.01 &  1.49 $\pm$  0.07 \\ 
 CD-29       15930 &   7.78 $\pm$   0.64 & 148.89 $\pm$  3.26 & $-$38.60 $\pm$  1.90 &  5.75 $\pm$  0.08  &  3.01 $\pm$  0.12 &  0.31 $\pm$  0.01 &  0.97 $\pm$  0.04 \\ 
 CD-31       16658 &  $-$6.67 $\pm$   0.62 & 185.77 $\pm$  1.35 & $-$54.51 $\pm$  1.52 &  6.26 $\pm$  0.05  &  4.72 $\pm$  0.07 &  0.14 $\pm$  0.00 &  1.15 $\pm$  0.03 \\ 
 CD-31       16922 &  $-$34.03 $\pm$   8.31 & 218.84 $\pm$  1.40 &  0.71 $\pm$  3.65 &  8.05 $\pm$  0.29  &  6.40 $\pm$  0.11 &  0.11 $\pm$  0.03 &  0.35 $\pm$  0.02 \\ 
 CD-31       17277 &  $-$138.60 $\pm$   1.60 & 14.28 $\pm$  8.60 & $-$66.96 $\pm$  0.86 &  7.89 $\pm$  0.04  &  0.23 $\pm$  0.13 &  0.94 $\pm$  0.03 &  2.81 $\pm$  0.09 \\ 
 CD-32        4154 &  36.70 $\pm$   0.73 & 213.91 $\pm$  1.22 &  3.94 $\pm$  0.18 &  9.74 $\pm$  0.11  &  7.55 $\pm$  0.06 &  0.13 $\pm$  0.00 &  0.17 $\pm$  0.01 \\ 
 CD-32       13158 &   2.91 $\pm$   0.15 & 221.09 $\pm$  0.39 &  2.63 $\pm$  0.18 &  6.71 $\pm$  0.04  &  6.51 $\pm$  0.06 &  0.01 $\pm$  0.00 &  0.03 $\pm$  0.00 \\ 
 CD-32       14894 &  $-$109.30 $\pm$   0.33 & 281.31 $\pm$  2.63 & 48.37 $\pm$  4.18 & 13.56 $\pm$  0.07  &  4.96 $\pm$  0.10 &  0.46 $\pm$  0.01 &  1.87 $\pm$  0.18 \\ 
 CD-33        2721 &  15.36 $\pm$   1.76 & 36.47 $\pm$  0.90 & $-$111.53 $\pm$  0.82 &  9.37 $\pm$  0.03  &  1.07 $\pm$  0.01 &  0.79 $\pm$  0.00 &  6.60 $\pm$  0.16 \\ 
 CD-33       15063 &  49.34 $\pm$   0.95 & $-$76.52 $\pm$ 12.58 & $-$9.60 $\pm$  2.83 &  6.47 $\pm$  0.06  &  1.38 $\pm$  0.27 &  0.65 $\pm$  0.06 &  1.38 $\pm$  0.06 \\ 
 CD-34         242 &  62.55 $\pm$   2.67 & 217.59 $\pm$  0.46 & $-$10.11 $\pm$  1.43 & 10.43 $\pm$  0.12  &  6.70 $\pm$  0.05 &  0.22 $\pm$  0.01 &  2.09 $\pm$  0.10 \\ 
 CD-34        3626 &  $-$53.55 $\pm$   1.13 & 179.27 $\pm$  0.54 &  7.90 $\pm$  0.16 &  9.63 $\pm$  0.05  &  5.62 $\pm$  0.02 &  0.26 $\pm$  0.00 &  0.28 $\pm$  0.01 \\ 
 CD-34        5597 &  14.31 $\pm$   0.33 & 195.38 $\pm$  0.26 & $-$21.56 $\pm$  0.68 &  8.41 $\pm$  0.01  &  6.55 $\pm$  0.02 &  0.12 $\pm$  0.00 &  0.43 $\pm$  0.02 \\ 
 CD-35        4882 &  36.00 $\pm$   0.39 & 191.43 $\pm$  0.28 & $-$18.75 $\pm$  0.35 &  8.68 $\pm$  0.02  &  6.01 $\pm$  0.01 &  0.18 $\pm$  0.00 &  0.31 $\pm$  0.01 \\ 
 CD-35       13334 &  76.62 $\pm$   0.40 & 209.65 $\pm$  0.61 & 19.43 $\pm$  0.59 &  7.88 $\pm$  0.08  &  4.77 $\pm$  0.06 &  0.25 $\pm$  0.00 &  0.86 $\pm$  0.02 \\ 
 CD-35       13661 &  $-$157.80 $\pm$   5.49 & $-$66.73 $\pm$ 15.57 & 121.72 $\pm$  7.38 &  8.66 $\pm$  0.35  &  1.10 $\pm$  0.27 &  0.77 $\pm$  0.04 &  3.58 $\pm$  0.29 \\ 
 CD-35       14807 &  $-$26.36 $\pm$   1.21 & 227.47 $\pm$  1.23 &  6.35 $\pm$  1.20 &  8.26 $\pm$  0.10  &  6.68 $\pm$  0.09 &  0.11 $\pm$  0.00 &  1.29 $\pm$  0.13 \\ 
 CD-36         518 &  142.02 $\pm$   3.68 & 95.82 $\pm$  2.41 & $-$121.52 $\pm$  0.55 & 12.32 $\pm$  0.17  &  2.49 $\pm$  0.10 &  0.66 $\pm$  0.01 &  6.10 $\pm$  0.04 \\ 
 CD-36       11584 &  $-$2.20 $\pm$   0.39 & 227.27 $\pm$  0.33 & 10.83 $\pm$  0.18 &  6.74 $\pm$  0.08  &  6.49 $\pm$  0.06 &  0.02 $\pm$  0.00 &  0.12 $\pm$  0.00 \\ 
 CD-38       13823 &  251.76 $\pm$   4.14 & $-$166.72 $\pm$ 24.55 & $-$41.86 $\pm$  9.94 & 15.83 $\pm$  2.48  &  2.33 $\pm$  0.27 &  0.74 $\pm$  0.01 &  3.55 $\pm$  0.92 \\ 
 CD-39        6037 &  $-$223.93 $\pm$   3.58 & $-$28.97 $\pm$  1.77 & $-$74.37 $\pm$  2.22 & 14.94 $\pm$  0.37  &  0.50 $\pm$  0.03 &  0.94 $\pm$  0.00 &  3.20 $\pm$  0.22 \\ 
 CD-39        9313 &  $-$205.60 $\pm$   3.40 & 146.13 $\pm$  1.38 & $-$79.07 $\pm$  4.69 & 12.81 $\pm$  0.20  &  2.46 $\pm$  0.05 &  0.68 $\pm$  0.01 &  2.44 $\pm$  0.21 \\ 
 CD-40         471 &  $-$98.37 $\pm$   2.34 & 12.43 $\pm$  6.59 & $-$16.03 $\pm$  1.81 &  9.45 $\pm$  0.08  &  0.26 $\pm$  0.14 &  0.95 $\pm$  0.03 &  3.67 $\pm$  0.11 \\ 
 CD-41        4744 &  $-$8.88 $\pm$   0.07 & 215.50 $\pm$  0.17 & 10.60 $\pm$  0.07 &  8.26 $\pm$  0.01  &  7.69 $\pm$  0.01 &  0.04 $\pm$  0.00 &  0.17 $\pm$  0.00 \\ 
 CD-41        7816 &  $-$19.02 $\pm$   0.09 & 207.91 $\pm$  0.15 & 17.03 $\pm$  0.06 &  7.89 $\pm$  0.00  &  6.63 $\pm$  0.01 &  0.09 $\pm$  0.00 &  0.35 $\pm$  0.00 \\ 
 CD-41       14843 &  $-$33.23 $\pm$   0.52 & 265.83 $\pm$  0.67 & 31.63 $\pm$  0.35 & 12.61 $\pm$  0.09  &  7.16 $\pm$  0.01 &  0.28 $\pm$  0.00 &  1.92 $\pm$  0.05 \\ 
 CD-43        7161 &  $-$294.32 $\pm$   4.87 & 20.29 $\pm$  2.16 & $-$73.08 $\pm$  3.05 & 22.30 $\pm$  0.97  &  0.30 $\pm$  0.03 &  0.97 $\pm$  0.00 &  3.91 $\pm$  0.38 \\ 
 CD-43        8736 &  71.48 $\pm$   3.14 & 160.19 $\pm$  1.70 & 64.40 $\pm$  1.99 &  7.65 $\pm$  0.03  &  3.68 $\pm$  0.08 &  0.35 $\pm$  0.01 &  1.48 $\pm$  0.06 \\ 
 CD-44       12644 &  $-$2.71 $\pm$   1.28 & 54.26 $\pm$  4.34 & $-$59.02 $\pm$  1.48 &  6.49 $\pm$  0.04  &  0.94 $\pm$  0.09 &  0.75 $\pm$  0.02 &  1.20 $\pm$  0.04 \\ 
 CD-44       13783 &  $-$66.82 $\pm$   0.78 & 192.02 $\pm$  1.15 & $-$77.22 $\pm$  1.21 &  7.86 $\pm$  0.10  &  5.17 $\pm$  0.03 &  0.21 $\pm$  0.00 &  2.30 $\pm$  0.04 \\ 
 CD-44       13981 &   3.57 $\pm$   0.19 & 247.29 $\pm$  0.58 & 20.62 $\pm$  0.19 &  9.11 $\pm$  0.01  &  6.83 $\pm$  0.04 &  0.14 $\pm$  0.00 &  1.19 $\pm$  0.05 \\ 
 CD-44       15269 &  59.66 $\pm$   3.24 & 157.89 $\pm$  2.58 & $-$41.21 $\pm$  1.00 &  8.30 $\pm$  0.05  &  4.16 $\pm$  0.11 &  0.33 $\pm$  0.01 &  2.39 $\pm$  0.10 \\ 
 CD-46        8357 &  73.59 $\pm$   2.62 & 208.03 $\pm$  4.27 &  3.96 $\pm$  1.40 &  9.46 $\pm$  0.25  &  5.67 $\pm$  0.11 &  0.25 $\pm$  0.00 &  0.27 $\pm$  0.00 \\ 
 CD-46       13658 &  46.25 $\pm$   0.84 & 166.32 $\pm$  1.87 &  2.78 $\pm$  0.87 &  7.07 $\pm$  0.03  &  3.97 $\pm$  0.08 &  0.28 $\pm$  0.01 &  1.10 $\pm$  0.03 \\ 
 CD-47       14271 &  10.40 $\pm$   0.25 & 132.78 $\pm$  3.10 & 47.95 $\pm$  0.37 &  7.16 $\pm$  0.02  &  3.38 $\pm$  0.10 &  0.36 $\pm$  0.01 &  2.22 $\pm$  0.06 \\ 
 CD-48       12928 &  $-$53.31 $\pm$   1.42 & 216.62 $\pm$  0.67 & 71.14 $\pm$  2.51 &  8.41 $\pm$  0.01  &  5.65 $\pm$  0.07 &  0.20 $\pm$  0.01 &  1.64 $\pm$  0.07 \\ 
 CD-50         823 &  $-$363.77 $\pm$   7.60 & $-$51.63 $\pm$  7.13 & 86.36 $\pm$  2.76 & 53.99 $\pm$  6.54  &  0.79 $\pm$  0.11 &  0.97 $\pm$  0.00 & 10.87 $\pm$  1.52 \\ 
 CD-50         877 &  67.65 $\pm$   1.63 & 147.77 $\pm$  0.47 & $-$57.55 $\pm$  0.75 &  9.31 $\pm$  0.05  &  4.12 $\pm$  0.02 &  0.39 $\pm$  0.00 &  2.26 $\pm$  0.02 \\ 
 CD-52         976 &  27.53 $\pm$   0.42 & $-$12.27 $\pm$  1.50 &  2.22 $\pm$  1.85 &  8.42 $\pm$  0.01  &  0.23 $\pm$  0.03 &  0.95 $\pm$  0.01 &  1.01 $\pm$  0.02 \\ 
 CD-52        2441 &  14.47 $\pm$   0.48 & 221.52 $\pm$  0.57 & 10.87 $\pm$  0.17 &  8.67 $\pm$  0.04  &  7.84 $\pm$  0.02 &  0.05 $\pm$  0.00 &  0.24 $\pm$  0.01 \\ 
 CD-52        4849 &  235.24 $\pm$  13.19 & 17.85 $\pm$  2.02 & $-$67.30 $\pm$  3.91 & 13.66 $\pm$  1.14  &  0.28 $\pm$  0.03 &  0.96 $\pm$  0.00 &  3.81 $\pm$  0.67 \\ 
 CD-57        1959 &  248.51 $\pm$   4.41 & 119.11 $\pm$  1.28 & 51.91 $\pm$  1.36 & 19.32 $\pm$  0.70  &  2.24 $\pm$  0.02 &  0.79 $\pm$  0.01 &  3.34 $\pm$  0.24 \\ 
 CD-58         294 &  125.12 $\pm$   5.28 & 110.83 $\pm$  1.45 & $-$74.56 $\pm$  0.32 & 10.47 $\pm$  0.21  &  2.53 $\pm$  0.06 &  0.61 $\pm$  0.01 &  3.26 $\pm$  0.06 \\ 
 CD-59        6913 &  90.82 $\pm$   6.03 & $-$95.46 $\pm$  9.36 & 80.60 $\pm$  2.73 &  6.73 $\pm$  0.07  &  1.90 $\pm$  0.23 &  0.56 $\pm$  0.04 &  2.51 $\pm$  0.12 \\ 
 CD-62          15 &  $-$112.33 $\pm$   0.99 & -43.32 $\pm$  5.45 & 79.93 $\pm$  3.44 &  9.14 $\pm$  0.07  &  0.87 $\pm$  0.13 &  0.83 $\pm$  0.02 &  3.02 $\pm$  0.17 \\ 
CPD-62        1126 &  $-$18.99 $\pm$   0.13 & 236.21 $\pm$  0.14 & $-$12.19 $\pm$  0.31 &  9.55 $\pm$  0.01  &  7.74 $\pm$  0.00 &  0.10 $\pm$  0.00 &  0.30 $\pm$  0.01 \\ 
    HD        6518 &  140.73 $\pm$   4.26 & $-$21.28 $\pm$  7.28 & $-$106.87 $\pm$  0.44 & 11.15 $\pm$  0.21  &  0.46 $\pm$  0.16 &  0.92 $\pm$  0.03 &  5.49 $\pm$  0.04 \\ 
    HD       19367 &  36.37 $\pm$   0.75 & 18.67 $\pm$  7.30 & $-$86.62 $\pm$  0.95 &  9.93 $\pm$  0.06  &  0.54 $\pm$  0.22 &  0.90 $\pm$  0.04 &  5.91 $\pm$  0.11 \\ 
    HD       41020 &  $-$69.63 $\pm$   1.89 &  1.24 $\pm$  2.04 &  2.20 $\pm$  1.26 &  9.07 $\pm$  0.04  &  0.02 $\pm$  0.02 &  1.00 $\pm$  0.01 &  0.65 $\pm$  0.01 \\ 
    HD      114621 &  $-$22.34 $\pm$   0.18 & 217.21 $\pm$  0.23 & $-$2.43 $\pm$  0.17 &  7.80 $\pm$  0.01  &  6.68 $\pm$  0.02 &  0.08 $\pm$  0.00 &  0.03 $\pm$  0.00 \\ 
    HD      298008 &   5.77 $\pm$   0.24 & 214.89 $\pm$  0.51 & $-$36.99 $\pm$  0.58 &  7.98 $\pm$  0.01  &  7.59 $\pm$  0.03 &  0.02 $\pm$  0.00 &  0.64 $\pm$  0.01 \\ 
    HD      298296 &   9.21 $\pm$   0.34 & 208.54 $\pm$  0.20 & $-$11.04 $\pm$  0.28 &  8.13 $\pm$  0.01  &  7.14 $\pm$  0.01 &  0.06 $\pm$  0.00 &  0.16 $\pm$  0.00 \\ 
   TYC 5340-1656-1 &   0.46 $\pm$   4.50 & 187.57 $\pm$  2.79 & $-$13.05 $\pm$  2.28 &  8.60 $\pm$  0.01  &  6.31 $\pm$  0.18 &  0.15 $\pm$  0.01 &  0.37 $\pm$  0.03 \\ 
   TYC 5422-1192-1 &  13.07 $\pm$   0.78 & 223.97 $\pm$  2.48 & $-$10.97 $\pm$  1.45 & 11.66 $\pm$  0.49  & 10.09 $\pm$  0.20 &  0.07 $\pm$  0.01 &  0.41 $\pm$  0.05 \\ 
   TYC 5763-1084-1 &  66.76 $\pm$   1.81 & 173.94 $\pm$  1.64 &  0.14 $\pm$  1.52 &  7.27 $\pm$  0.03  &  3.99 $\pm$  0.09 &  0.29 $\pm$  0.01 &  1.20 $\pm$  0.05 \\ 
   TYC 6002-1724-1 &  $-$84.07 $\pm$   4.10 & $-$96.86 $\pm$  5.25 & $-$48.44 $\pm$  2.77 & 10.31 $\pm$  0.19  &  2.43 $\pm$  0.18 &  0.62 $\pm$  0.02 &  1.60 $\pm$  0.15 \\ 
   TYC  6108-150-1 &  80.17 $\pm$   5.50 & 36.58 $\pm$  4.33 & $-$4.27 $\pm$  3.64 &  8.22 $\pm$  0.07  &  0.68 $\pm$  0.09 &  0.85 $\pm$  0.02 &  1.18 $\pm$  0.10 \\ 
   TYC  6195-815-1 &  24.35 $\pm$   3.03 & 93.71 $\pm$  7.29 & 83.42 $\pm$  3.95 &  5.66 $\pm$  0.09  &  1.79 $\pm$  0.17 &  0.52 $\pm$  0.03 &  2.19 $\pm$  0.12 \\ 
   TYC  6199-351-1 &  $-$82.27 $\pm$   0.37 & 52.66 $\pm$  7.25 & $-$20.07 $\pm$  2.49 &  6.89 $\pm$  0.08  &  0.86 $\pm$  0.14 &  0.78 $\pm$  0.03 &  0.83 $\pm$  0.04 \\ 
   TYC 6310-1651-2 &  $-$15.62 $\pm$   1.13 & 242.76 $\pm$  0.17 & 15.11 $\pm$  0.59 &  7.58 $\pm$  0.09  &  5.98 $\pm$  0.06 &  0.12 $\pm$  0.00 &  0.80 $\pm$  0.03 \\ 
   TYC 7144-1122-1 &  $-$6.12 $\pm$   0.10 & 221.52 $\pm$  0.36 & $-$11.44 $\pm$  0.49 &  8.92 $\pm$  0.05  &  8.44 $\pm$  0.02 &  0.03 $\pm$  0.00 &  0.23 $\pm$  0.01 \\ 
   TYC  7351-734-1 &  25.94 $\pm$   0.21 & 228.15 $\pm$  0.33 & $-$6.50 $\pm$  0.14 &  7.79 $\pm$  0.05  &  6.50 $\pm$  0.03 &  0.09 $\pm$  0.00 &  0.31 $\pm$  0.01 \\ 
   TYC 7354-1196-1 &  $-$28.18 $\pm$   0.65 & 261.22 $\pm$  1.35 & $-$2.53 $\pm$  0.52 & 11.11 $\pm$  0.10  &  7.16 $\pm$  0.03 &  0.22 $\pm$  0.01 &  0.16 $\pm$  0.01 \\ 
   TYC   7427-35-1 &  39.30 $\pm$   0.32 & 253.85 $\pm$  0.43 & $-$8.41 $\pm$  0.50 & 10.31 $\pm$  0.02  &  6.78 $\pm$  0.02 &  0.21 $\pm$  0.00 &  0.44 $\pm$  0.01 \\ 
   TYC  8025-237-1 &  101.42 $\pm$   6.33 & 48.21 $\pm$  7.00 & $-$48.99 $\pm$  0.22 &  8.76 $\pm$  0.13  &  0.95 $\pm$  0.16 &  0.80 $\pm$  0.03 &  2.46 $\pm$  0.09 \\ 
   TYC  8161-753-1 &  $-$68.35 $\pm$   0.77 &  8.79 $\pm$  0.34 &  6.48 $\pm$  0.34 &  8.53 $\pm$  0.01  &  0.15 $\pm$  0.01 &  0.96 $\pm$  0.00 &  0.16 $\pm$  0.00 \\ 
   TYC  8178-334-1 &  19.70 $\pm$   0.49 & 217.94 $\pm$  0.48 & 10.88 $\pm$  0.11 &  8.53 $\pm$  0.03  &  7.46 $\pm$  0.02 &  0.07 $\pm$  0.00 &  0.16 $\pm$  0.00 \\ 
   TYC 8263-2160-1 &  $-$45.69 $\pm$   1.72 & 197.72 $\pm$  1.62 &  7.87 $\pm$  0.45 &  7.21 $\pm$  0.05  &  5.01 $\pm$  0.11 &  0.18 $\pm$  0.01 &  0.71 $\pm$  0.04 \\ 
   TYC 8325-6844-1 &  82.12 $\pm$   0.53 & 168.58 $\pm$  0.78 & $-$11.76 $\pm$  0.16 &  8.66 $\pm$  0.02  &  4.07 $\pm$  0.03 &  0.36 $\pm$  0.00 &  0.17 $\pm$  0.00 \\ 
   TYC   8394-14-1 &  -23.62 $\pm$   1.01 & 234.72 $\pm$  0.25 & $-$2.79 $\pm$  0.64 &  7.31 $\pm$  0.07  &  5.78 $\pm$  0.11 &  0.12 $\pm$  0.00 &  1.46 $\pm$  0.09 \\ 
   TYC 8400-1610-1 &  $-$16.58 $\pm$   0.22 & 216.65 $\pm$  0.55 & $-$9.91 $\pm$  0.61 &  6.79 $\pm$  0.07  &  6.12 $\pm$  0.06 &  0.05 $\pm$  0.00 &  1.06 $\pm$  0.04 \\ 
   TYC  8584-127-1 &  -8.17 $\pm$   0.22 & 212.39 $\pm$  0.50 &  2.67 $\pm$  0.20 &  8.27 $\pm$  0.02  &  7.56 $\pm$  0.06 &  0.05 $\pm$  0.00 &  0.11 $\pm$  0.00 \\ 
   TYC 8633-2281-1 &  $-$271.23 $\pm$   3.24 & $-$35.06 $\pm$  0.93 & 26.48 $\pm$  0.77 & 17.04 $\pm$  0.35  &  0.52 $\pm$  0.01 &  0.94 $\pm$  0.00 &  1.44 $\pm$  0.02  \\
   
\end{supertabular}

    
\newpage
\topcaption{\label{dynamics} Total orbital energy (E), angular momentum (L$_Z$), radial and vertical actions (J$_R$, J$_Z$) and classification (class, see bottom of the table) of the program stars.}
\tablefirsthead{ \toprule \toprule 
\multicolumn{1}{c}{Star}&\multicolumn{1}{c}{$E$}&\multicolumn{1}{c}{$L_Z$}&\multicolumn{1}{c}{$J_R$}&\multicolumn{1}{c}{$J_Z$}&\multicolumn{1}{c}{Class}\\ 

\multicolumn{1}{c}{}&\multicolumn{1}{c}{[$\rm km^2 s ^{-2}$]}&\multicolumn{1}{c}{[kpc \kms]}&\multicolumn{1}{c}{[kpc \kms]}&\multicolumn{1}{c}{[kpc \kms]}&\multicolumn{1}{c}{}\\
\midrule}
\tablehead{%
\multicolumn{6}{l}%
{\bfseries \tablename \thetable{--continued.}} \\
\toprule
\multicolumn{1}{c}{Star}&\multicolumn{1}{c}{$E$}&\multicolumn{1}{c}{$L_Z$}&\multicolumn{1}{c}{$J_R$}&\multicolumn{1}{c}{$J_Z$}&\multicolumn{1}{c}{Class}\\ 

\multicolumn{1}{c}{}&\multicolumn{1}{c}{[$\rm km^2 s ^{-2}$]}&\multicolumn{1}{c}{[kpc \kms]}&\multicolumn{1}{c}{[kpc \kms]}&\multicolumn{1}{c}{[kpc \kms]}&\multicolumn{1}{c}{}\\
\midrule}
\tabletail{%
\midrule \multicolumn{6}{r}{{Continue on the next page}} \\}
\tablelasttail{%
\\\midrule
\bottomrule
\\
\multicolumn{6}{r}{\textbf{Class:} 1: Thin disk; 2: Thick disk; 3: Thin-Thick transition; 4: Halo; 5: GSE candidate; 6: Seq candidate. }\\}
\begin{supertabular}{lrrrrc}
 BD-11        3235 &  $-$63257 $\pm$    498 &    77 $\pm$    46 &   573 $\pm$    18 &    73 $\pm$    12  & 4\\ 
 BD-12        5650 &  $-$49252 $\pm$   1160 &  1506 $\pm$    41 &     6 $\pm$     2 &     5 $\pm$     1  & 1\\ 
 BD-13         934 &  $-$50818 $\pm$    316 &   646 $\pm$    35 &   365 $\pm$     7 &   192 $\pm$    27  & 2\\ 
 BD-13        3195 &  $-$53493 $\pm$    350 &   431 $\pm$    19 &   593 $\pm$    19 &    36 $\pm$     0  & 2\\ 
 BD-13        3488 &  $-$47972 $\pm$     17 &  1426 $\pm$     1 &    20 $\pm$     0 &    63 $\pm$     1  & 2\\ 
 BD-13        6352 &  $-$59376 $\pm$    117 &   443 $\pm$    70 &   304 $\pm$    22 &   206 $\pm$    26  & 2\\ 
 BD-14          52 &  $-$51604 $\pm$    263 &   890 $\pm$    56 &   158 $\pm$    18 &   218 $\pm$    15  & 2\\ 
 BD-15        4109 &  $-$69673 $\pm$    153 &    67 $\pm$    62 &   328 $\pm$    15 &   281 $\pm$    29  & 4\\ 
 BD-15        5449 &  $-$46630 $\pm$   1287 &   310 $\pm$    71 &   821 $\pm$    66 &    28 $\pm$     3  & 4\\ 
 BD-16        2232 &  17298 $\pm$    859 &  $-$600 $\pm$    28 &  5048 $\pm$   167 &    55 $\pm$     0  & 4\\ 
 BD-17        3143 &  $-$39506 $\pm$   1714 &   469 $\pm$    94 &   845 $\pm$    94 &    78 $\pm$     9  & 2\\ 
 BD-17        4251 &  $-$15189 $\pm$    425 &    27 $\pm$    50 &  1807 $\pm$    16 &   430 $\pm$     9  & 5\\ 
 BD-18        5150 &  $-$43547 $\pm$    157 &  1688 $\pm$     5 &    17 $\pm$     0 &     1 $\pm$     0  & 1\\ 
 BD-19        6363 &  $-$56061 $\pm$   1232 &   177 $\pm$   118 &   535 $\pm$    49 &   200 $\pm$    42  & 4\\ 
 CD-23        1855 &  $-$52653 $\pm$    492 &    21 $\pm$    49 &   843 $\pm$    21 &    30 $\pm$     1  & 4\\ 
 CD-23       11064 &  $-$36132 $\pm$     63 &  1855 $\pm$     8 &    66 $\pm$     1 &    32 $\pm$     3  & 1\\ 
 CD-24         613 &  $-$42832 $\pm$     71 &  1583 $\pm$    14 &    22 $\pm$     3 &    74 $\pm$     6  & 1\\ 
 CD-24        1384 &  $-$52171 $\pm$     96 &   458 $\pm$    23 &   465 $\pm$    15 &   177 $\pm$     4  & 4\\ 
 CD-26        7083 &  $-$30348 $\pm$   2052 &  1919 $\pm$    65 &   233 $\pm$    28 &    20 $\pm$     2  & 2\\ 
 CD-27       14182 &  $-$43472 $\pm$    125 &  1677 $\pm$     6 &    16 $\pm$     1 &     8 $\pm$     1  & 1\\ 
 CD-27       15535 &  $-$52353 $\pm$    284 &  1299 $\pm$    13 &    45 $\pm$     2 &    35 $\pm$     1  & 1\\ 
 CD-27       16505 &  17075 $\pm$   6102 &    30 $\pm$    62 &  5246 $\pm$  1254 &   132 $\pm$     7  & 4\\ 
 CD-28       10039 &  $-$51640 $\pm$    114 &  1249 $\pm$    10 &    23 $\pm$     1 &   101 $\pm$     6  & 2\\ 
 CD-28       10387 &  $-$45021 $\pm$    463 &  1497 $\pm$    29 &   109 $\pm$    10 &     7 $\pm$     1  & 1\\ 
 CD-28       16762 &  $-$71018 $\pm$    279 &   172 $\pm$    39 &   449 $\pm$    14 &    41 $\pm$     3  & 4\\ 
 CD-28       17446 &  $-$1048 $\pm$   7479 &  $-$188 $\pm$    79 &  2932 $\pm$   743 &   116 $\pm$    17  & 5\\ 
 CD-29        9391 &  $-$48010 $\pm$    236 &  1449 $\pm$    14 &    19 $\pm$     2 &    46 $\pm$     4  & 1\\ 
 CD-29       15930 & $-$69693 $\pm$   1037 &   845 $\pm$    32 &    68 $\pm$     3 &    32 $\pm$     3  & 2\\ 
 CD-31       16658 &  $-$58964 $\pm$    495 &  1153 $\pm$    18 &    16 $\pm$     1 &    40 $\pm$     2  & 2\\ 
 CD-31       16922 &  $-$46652 $\pm$    602 &  1580 $\pm$    12 &    14 $\pm$     7 &     4 $\pm$     0  & 1\\ 
 CD-31       17277 &  $-$63962 $\pm$    206 &    91 $\pm$    55 &   533 $\pm$    20 &   105 $\pm$     6  & 4\\ 
 CD-32        4154 &  $-$38848 $\pm$    424 &  1861 $\pm$    16 &    20 $\pm$     1 &     1 $\pm$     0  & 1\\ 
 CD-32       13158 &  $-$50870 $\pm$    349 &  1478 $\pm$    10 &     0 $\pm$     0 &     0 $\pm$     0  & 1\\ 
 CD-32       14894 &  $-$34827 $\pm$     72 &  1603 $\pm$    24 &   294 $\pm$    12 &    31 $\pm$     5  & 2\\ 
 CD-33        2721 &  $-$53295 $\pm$    194 &   328 $\pm$     7 &   420 $\pm$     1 &   349 $\pm$    13  & 4\\ 
 CD-33       15063 &  $-$71791 $\pm$    660 &  $-$469 $\pm$    71 &   259 $\pm$    40 &    46 $\pm$     4  & 6\\ 
 CD-34         242 &  $-$38883 $\pm$    192 &  1710 $\pm$     5 &    59 $\pm$     5 &    57 $\pm$     3  & 1\\ 
 CD-34        3626 &  $-$43989 $\pm$     93 &  1587 $\pm$     1 &    79 $\pm$     3 &     2 $\pm$     0  & 1\\ 
 CD-34        5597 &  $-$45056 $\pm$    112 &  1626 $\pm$     4 &    17 $\pm$     0 &     5 $\pm$     0  & 1\\ 
 CD-35        4882 &  $-$45803 $\pm$     90 &  1580 $\pm$     3 &    37 $\pm$     0 &     3 $\pm$     0  & 1\\ 
 CD-35       13334 &  $-$52334 $\pm$    491 &  1312 $\pm$    16 &    59 $\pm$     0 &    18 $\pm$     1  & 1\\ 
 CD-35       13661 &  $-$58418 $\pm$   2162 &  $-$376 $\pm$    79 &   428 $\pm$    21 &   139 $\pm$    12  & 4\\ 
 CD-35       14807 &  $-$44988 $\pm$    499 &  1580 $\pm$    27 &    12 $\pm$     1 &    35 $\pm$     7  & 1\\ 
 CD-36         518 &  $-$41841 $\pm$    314 &   768 $\pm$    19 &   471 $\pm$    24 &   222 $\pm$     4  & 4\\ 
 CD-36       11584 &  $-$50787 $\pm$    486 &  1478 $\pm$    14 &     0 $\pm$     0 &     1 $\pm$     0  & 1\\ 
 CD-38       13823 &  $-$33521 $\pm$   5134 &  $-$926 $\pm$   108 &   767 $\pm$   127 &    62 $\pm$    10  & 4\\ 
 CD-39        6037 &  $-$37384 $\pm$   1049 &  $-$235 $\pm$    14 &  1077 $\pm$    19 &    61 $\pm$     4  & 5\\ 
 CD-39        9313 &  $-$41310 $\pm$    530 &   937 $\pm$    17 &   541 $\pm$    19 &    48 $\pm$     6  & 2\\ 
 CD-40         471 &  $-$55925 $\pm$    356 &   102 $\pm$    54 &   633 $\pm$    30 &   123 $\pm$     4  & 4\\ 
 CD-41        4744 &  $-$42378 $\pm$     44 &  1752 $\pm$     2 &     2 $\pm$     0 &     1 $\pm$     0  & 1\\ 
 CD-41        7816 &  $-$46474 $\pm$     43 &  1594 $\pm$     2 &     8 $\pm$     0 &     4 $\pm$     0  & 1\\ 
 CD-41       14843 &  $-$32872 $\pm$    169 &  1936 $\pm$     1 &   108 $\pm$     4 &    38 $\pm$     1  & 1\\ 
 CD-43        7161 &  $-$22027 $\pm$   1620 &   154 $\pm$    16 &  1697 $\pm$    79 &    43 $\pm$     4  & 5\\ 
 CD-43        8736 &  $-$56701 $\pm$    150 &  1072 $\pm$    18 &   106 $\pm$     6 &    45 $\pm$     3  & 2\\ 
 CD-44       12644 &  $-$73263 $\pm$    420 &   350 $\pm$    30 &   330 $\pm$    10 &    37 $\pm$     2  & 2\\ 
 CD-44       13783 &  $-$50456 $\pm$    418 &  1271 $\pm$    16 &    40 $\pm$     2 &    95 $\pm$     4  & 2\\ 
 CD-44       13981 &  $-$42152 $\pm$     94 &  1679 $\pm$     8 &    24 $\pm$     1 &    27 $\pm$     2  & 1\\ 
 CD-44       15269 &  $-$52225 $\pm$    120 &  1132 $\pm$    23 &   102 $\pm$     9 &    90 $\pm$     5  & 2\\ 
 CD-46        8357 &  $-$44387 $\pm$   1065 &  1584 $\pm$    33 &    71 $\pm$     4 &     2 $\pm$     0  & 1\\ 
 CD-46       13658 &  $-$58493 $\pm$    449 &  1106 $\pm$    19 &    67 $\pm$     3 &    31 $\pm$     1  & 1\\ 
 CD-47       14271 &  $-$59639 $\pm$    434 &   921 $\pm$    26 &   101 $\pm$     6 &    93 $\pm$     4  & 2\\ 
 CD-48       12928 &  $-$47328 $\pm$    209 &  1438 $\pm$    16 &    40 $\pm$     2 &    51 $\pm$     4  & 2\\ 
 CD-50         823 &   8043 $\pm$   3487 &  $-$420 $\pm$    58 &  3795 $\pm$   403 &    53 $\pm$     2  & 4\\ 
 CD-50         877 &  $-$48696 $\pm$    114 &  1204 $\pm$     3 &   150 $\pm$     4 &    70 $\pm$     1  & 2\\ 
 CD-52         976 &  $-$63024 $\pm$     92 &  $-$102 $\pm$    13 &   631 $\pm$     6 &    20 $\pm$     1  & 4\\ 
 CD-52        2441 &  $-$40843 $\pm$    150 &  1804 $\pm$     5 &     3 $\pm$     0 &     2 $\pm$     0  & 1\\ 
 CD-52        4849 &  $-$40935 $\pm$   3397 &   128 $\pm$    14 &  1008 $\pm$    74 &    86 $\pm$    13  & 5\\ 
 CD-57        1959 &  $-$26202 $\pm$   1354 &   966 $\pm$    11 &  1029 $\pm$    45 &    46 $\pm$     3  & 4\\ 
 CD-58         294 &  $-$48498 $\pm$    593 &   842 $\pm$    12 &   367 $\pm$    21 &    99 $\pm$     0  & 2\\ 
 CD-59        6913 &  $-$66800 $\pm$   1307 &  $-$570 $\pm$    50 &   198 $\pm$    21 &   114 $\pm$     9  & 6\\ 
 CD-62          15 &  $-$57119 $\pm$    632 &  $-$322 $\pm$    40 &   518 $\pm$    20 &    98 $\pm$     8  & 4\\ 
CPD-62        1126 &  $-$38841 $\pm$     34 &  1866 $\pm$     1 &    14 $\pm$     0 &     2 $\pm$     0  & 1\\ 
    HD        6518 &  $-$48305 $\pm$    808 &  $-$172 $\pm$    59 &   706 $\pm$    11 &   187 $\pm$     6  & 4\\ 
    HD       19367 &  $-$52477 $\pm$    126 &   173 $\pm$    67 &   566 $\pm$    37 &   243 $\pm$     3  & 4\\ 
    HD       41020 &  $-$60065 $\pm$    214 &    11 $\pm$    18 &   746 $\pm$     8 &     9 $\pm$     0  & 4\\ 
    HD      114621 &  $-$46641 $\pm$     97 &  1600 $\pm$     3 &     7 $\pm$     0 &     0 $\pm$     0  & 1\\ 
    HD      298008 &  $-$43253 $\pm$    111 &  1698 $\pm$     4 &     1 $\pm$     0 &    11 $\pm$     0  & 1\\ 
    HD      298296 &  $-$44276 $\pm$     62 &  1680 $\pm$     2 &     5 $\pm$     0 &     1 $\pm$     0  & 1\\ 
   TYC 5340-1656-1 &  $-$45200 $\pm$    493 &  1612 $\pm$    24 &    26 $\pm$     4 &     4 $\pm$     1  & 1\\ 
   TYC 5422-1192-1 &  $-$29474 $\pm$   1248 &  2295 $\pm$    60 &     8 $\pm$     3 &     3 $\pm$     1  & 1\\ 
   TYC 5763-1084-1 &  $-$57502 $\pm$    459 &  1117 $\pm$    22 &    74 $\pm$     3 &    35 $\pm$     3  & 1\\ 
   TYC 6002-1724-1 &  $-$49743 $\pm$   1213 &  $-$898 $\pm$    53 &   382 $\pm$    14 &    34 $\pm$     4  & 6\\ 
   TYC  6108-150-1 &  $-$63492 $\pm$    355 &   278 $\pm$    33 &   517 $\pm$    20 &    26 $\pm$     3  & 4\\ 
   TYC  6195-815-1 &  $-$74099 $\pm$   1102 &   498 $\pm$    52 &   144 $\pm$     8 &   112 $\pm$    10  & 2\\ 
   TYC  6199-351-1 &  $-$71505 $\pm$    793 &   334 $\pm$    50 &   381 $\pm$    16 &    19 $\pm$     2  & 2\\ 
   TYC 6310-1651-2 &  $-$49417 $\pm$    490 &  1464 $\pm$    17 &    14 $\pm$     0 &    17 $\pm$     1  & 1\\ 
   TYC 7144-1122-1 &  $-$38736 $\pm$    170 &  1889 $\pm$     6 &     1 $\pm$     0 &     2 $\pm$     0  & 1\\ 
   TYC  7351-734-1 &  $-$47178 $\pm$    234 &  1572 $\pm$     8 &     9 $\pm$     0 &     3 $\pm$     0  & 1\\ 
   TYC 7354-1196-1 &  $-$36398 $\pm$    164 &  1906 $\pm$     2 &    62 $\pm$     4 &     1 $\pm$     0  & 1\\ 
   TYC   7427-35-1 &  $-$39188 $\pm$     28 &  1795 $\pm$     1 &    54 $\pm$     1 &     4 $\pm$     0  & 1\\ 
   TYC  8025-237-1 &  $-$59167 $\pm$    413 &   353 $\pm$    53 &   483 $\pm$    35 &    75 $\pm$     2  & 4\\ 
   TYC  8161-753-1 &  $-$63294 $\pm$     67 &    71 $\pm$     3 &   677 $\pm$     2 &     1 $\pm$     0  & 4\\ 
   TYC  8178-334-1 &  $-$42253 $\pm$    112 &  1751 $\pm$     4 &     5 $\pm$     0 &     1 $\pm$     0  & 1\\ 
   TYC 8263-2160-1 &  $-$54152 $\pm$    600 &  1307 $\pm$    23 &    31 $\pm$     2 &    15 $\pm$     1  & 1\\ 
   TYC 8325-6844-1 &  $-$51958 $\pm$    143 &  1264 $\pm$     7 &   129 $\pm$     2 &     1 $\pm$     0  & 3\\ 
   TYC   8394-14-1 &  $-$50821 $\pm$    560 &  1367 $\pm$    27 &    13 $\pm$     1 &    49 $\pm$     6  & 1\\ 
   TYC 8400-1610-1 &  $-$51590 $\pm$    444 &  1387 $\pm$    18 &     2 $\pm$     0 &    32 $\pm$     3  & 1\\ 
   TYC  8584-127-1 &  $-$42710 $\pm$    208 &  1740 $\pm$     8 &     2 $\pm$     0 &     0 $\pm$     0  & 1\\ 
   TYC 8633-2281-1 &  $-$32584 $\pm$    826 &  $-$258 $\pm$     6 &  1265 $\pm$    24 &    14 $\pm$     0  & 5 \\
\end{supertabular}

\begin{table} 
\caption{Line list adopted for the star BD-11\,3235.}
\label{linelist}
\begin{minipage}{0.5\textwidth}
\begin{tabular}{lrlrl}
\hline \hline
Elt & Ion& Wavelength &log \textit{gf}&$\chi_{exc}$\\
&&[nm]&[cm$^{-1}$]&\\
\hline
    Na&I&568.2632&$-$0.7060&16956.170\\
    Na&I&568.8193&$-$1.0406&16973.365\\
    Na&I&616.0746&$-$1.246&16973.365\\

    Mg&I&571.1088&$-$1.724&35051.266\\
    Mg&I&706.0414&$-$1.351&46403.06\\
    Mg&I&781.1122&$-$1.717&47957.027\\
    Mg&I&809.8707&$-$2.229&47957.027\\
    Mg&I&821.3013&$-$2.661&46403.065\\
    Mg&I&871.2676&$-$1.690&59318.793\\
    Mg&I&871.7803&$-$2.866&59318.793\\
    Mg&I&873.6006&$-$1.758&47957.027\\
    Mg&I&892.3569&$-$1.678&43503.33\\

    Al&I&669.6012&$-$1.569&25347.756\\
    Al&I&877.3888&$-$0.192&43831.105\\
    
    Si&I&564.5613&$-$2.140&39760.285\\
    Si&I&566.5555&$-$2.040&39683.164\\
    Si&I&568.4484&$-$1.773&39955.05\\
    Si&I&569.0425&$-$1.870&39760.285\\
    Si&I&570.1104&$-$1.953&39760.285\\
    Si&I&577.2027&$-$3.107&45276.188\\
    Si&I&579.3073&$-$1.963&39760.285\\
    Si&I&594.8541&$-$1.231&40991.883\\
    Si&I&613.1573&$-$1.557&45293.629\\
    Si&I&615.5134&$-$0.855&45321.848\\
    Si&I&623.7319&$-$1.070&45276.188\\
    Si&I&624.4114&$-$2.288&45293.629\\
    Si&I&655.5463&$-$1.163&48264.293\\
    Si&I&672.1848&$-$1.140&47284.062\\
    Si&I&700.3569&$-$0.939&48102.32\\
    Si&I&793.2348&$-$0.469&48102.323\\
    Si&I&855.6777&$-$0.151&59034.988\\
    Si&I&855.6805&$-$0.407&61617.170\\
    Si&I&874.2446&$-$0.071&58786.860\\
    Si&I&875.2007&0.078&58774.368\\
    Si&I&889.2720&$-$0.758&48264.292\\

    S&I&921.2863&0.396&63475.051\\

    Ca&I&468.5268&$-$0.880&23652.305\\
    Ca&I&526.0387&$-$1.900&20335.359\\
    Ca&I&534.9465&$-$0.428&21849.635\\
    Ca&I&551.2980&$-$0.300&23652.305\\
    Ca&I&558.1965&$-$0.517&20349.260\\
    Ca&I&559.0114&$-$0.547&20335.359\\
    Ca&I&559.4462&$-$0.050&20349.260\\
    Ca&I&560.1277&$-$0.496&20371.000\\
    Ca&I&585.7451&0.230&23652.305\\
    Ca&I&610.2439&$-$3.373&20349.260\\
    Ca&I&616.1297&$-$1.266&20349.260\\
    Ca&I&616.6439&$-$1.142&20335.359\\
    Ca&I&616.9042&$-$0.540&20349.260\\
    Ca&I&616.9563&$-$0.270&20371.00\\
    Ca&I&644.9808&$-$0.550&20335.359\\
    Ca&I&645.5598&$-$1.350&20349.260\\
    Ca&I&647.1662&$-$0.590&20371.000\\
    Ca&I&649.9650&$-$0.590&20349.260\\
    Ca&I&671.7681&$-$0.610&21849.635\\

    Sc&I&568.6847&0.370&11610.280\\

    Ti&I&474.2789&0.010&18037.213\\
    Ti&I&475.9141&$-$3.002&22404.738\\
    Ti&I&475.9270&0.570&18192.570 \\

\hline
\end{tabular}

\end{minipage} \hfill
\begin{minipage}{0.5\textwidth}
\begin{tabular}{lrlrl}
\hline 
Elt & Ion& Wavelength &log \textit{gf} &$\chi_{exc}$\\
&&[nm]&[cm$^{-1}$]&\\
\hline

    Ti&I&   487.0125 &  0.440 &   18141.266 \\
    Ti&I&   497.5291 & $-$2.930 &   28951.973 \\
    Ti&I&   497.5343 &  0.150 &   20209.447 \\
    Ti&I&   497.8188 & $-$0.303 &   15877.081 \\
    Ti&I&   500.0990 &  0.022 &   16106.076 \\
    Ti&I&   500.1018 & $-$2.448 &   25102.875 \\
    Ti&I&   500.1128 & $-$1.699 &   25107.410 \\
    Ti&I&   500.9645 & $-$2.203 &     170.133 \\
    Ti&I&   502.5565 & $-$1.728 &   21588.494 \\
    Ti&I&   502.5570 &  0.250 &   16458.672 \\
    Ti&I&   503.8349 &  0.069 &   31206.008 \\
    Ti&I&   503.8398 &  0.020 &   11531.761 \\
    Ti&I&   504.3584 & $-$1.677 &    6742.756 \\
    Ti&I&   506.2102 & $-$0.408 &   17423.855 \\
    Ti&I&   507.1467 & $-$1.007 &   11776.812 \\
    Ti&I&   508.6834 & $-$3.285 &   18037.213 \\
    Ti&I&   508.7058 & $-$0.880 &   11531.761 \\
    Ti&I&   511.3440 & $-$0.727 &   11639.811 \\
    Ti&I&   514.5460 & $-$0.718 &   11776.812 \\
    Ti&I&   514.7762 & $-$1.055 &   28788.381 \\
    Ti&I&   521.9489 & $-$3.481 &   19322.984 \\
    Ti&I&   521.9503 & $-$2.166 &   25068.842 \\
    Ti&I&   521.9634 & $-$1.846 &   19322.984 \\
    Ti&I&   521.9702 & $-$2.236 &     170.133 \\
    Ti&I&   523.8537 & $-$0.916 &   16875.123 \\
    Ti&I&   523.8586 & $-$2.074 &    6842.962 \\
    Ti&I&   528.2376 & $-$1.810 &    8492.422 \\
    Ti&I&   528.2554 & $-$2.838 &   29914.736 \\
    Ti&I&   529.5776 & $-$1.577 &    8602.344 \\
    Ti&I&   529.5839 & $-$2.453 &   25102.875 \\
    Ti&I&   542.6249 & $-$2.950 &     170.133 \\
    Ti&I&   542.6549 & $-$2.714 &   30039.170 \\
    Ti&I&   548.9873 & $-$0.647 &   27750.135 \\
    Ti&I&   549.0130 & $-$1.001 &   26892.936 \\
    Ti&I&   549.0148 & $-$0.877 &   11776.812 \\
    Ti&I&   549.0198 & $-$2.504 &   29661.250 \\
    Ti&I&   568.9461 & $-$0.413 &   18525.059 \\
    Ti&I&   586.6367 & $-$1.055 &   26657.416 \\
    Ti&I&   586.6451 & $-$0.784 &    8602.344 \\
    Ti&I&   586.6808 & $-$2.213 &   26803.420 \\
    Ti&I&   591.8536 & $-$1.640 &    8602.344\\
    Ti&I&   595.3159 & $-$0.273 &   15220.393 \\
    Ti&I&   594.1382 & $-$3.395 &   25388.330 \\
    Ti&I&   594.1696 & $-$1.288 &   25227.223 \\
    Ti&I&   594.1752 & $-$1.510 &    8492.422 \\
    Ti&I&   597.8541 & $-$0.440 &   15108.111 \\
    Ti&I&   597.8878 & $-$3.302 &   32514.566 \\
    Ti&I&   606.4626 & $-$1.888 &    8436.618 \\
    Ti&I&   609.1171 & $-$0.367 &   18287.555 \\
    Ti&I&   612.6216 & $-$1.369 &    8602.344 \\
    Ti&I&   612.6273 & $-$2.845 &   25438.908 \\
    Ti&I&   625.8102 & $-$0.299 &   11639.811 \\
    Ti&I&   626.1099 & $-$0.423 &   11531.761 \\
    Ti&I&   626.1124 & $-$1.571 &   25227.223 \\
    Ti&I&   633.6099 & $-$1.687 &   11639.811 \\
    Ti&I&   655.3671 & $-$2.048 &   26564.400 \\
    Ti&I&   655.3682 & $-$2.915 &   33660.652 \\
    ...&&&&\\  
\hline\hline
\end{tabular}

\end{minipage}
\end{table}

\begin{table}
\footnotesize
\caption{Abundances and line-to-line dispersion ($\sigma$) of elements in their ionisation state from \ion{Na}{i} to \ion{Sc}{ii} in [X/H].}
\label{abu1_mince1}
\centering
\begin{tabular}{lcccccccccccccc}
    \hline \hline
    Star & \ion{Na}{i} & $\sigma$ & \ion{Mg}{i} & $\sigma$ & \ion{Al}{i} & $\sigma$ & \ion{Si}{i} & $\sigma$ & \ion{Ca}{i} & $\sigma$ & \ion{Sc}{i} & $\sigma$ & \ion{Sc}{ii} & $\sigma$\\
    \hline
BD-11 3235 & -2.28 & 0.03 & -1.53 & 0.04 & -2.06 & 0.19 & -1.70 & 0.12 & -1.73 & 0.05 & -2.05 & 0.00 & -1.59 & 0.13 \\
BD-13 934 & -2.00 & 0.09 & -1.39 & 0.06 & -1.82 & 0.12 & -1.51 & 0.08 & -1.53 & 0.06 & -1.89 & 0.00 & -1.56 & 0.12 \\
BD-13 3195 & -1.31 & 0.01 & -0.72 & 0.10 & -0.88 & 0.07 & -0.82 & 0.12 & -0.96 & 0.07 & -- & -- & -1.03 & 0.14 \\
BD-15 5449 & -1.89 & 0.02 & -1.73 & 0.09 & -1.28 & 0.05 & -1.74 & 0.13 & -1.79 & 0.06 & -2.19 & 0.00 & -1.87 & 0.06 \\
BD-16 2232 & -1.87 & 0.04 & -1.28 & 0.07 & -1.78 & 0.10 & -1.41 & 0.12 & -1.37 & 0.05 & -1.65 & 0.00 & -1.51 & 0.09 \\
BD-17 3143 & -1.76 & 0.05 & -1.21 & 0.08 & -1.69 & 0.12 & -1.43 & 0.11 & -1.27 & 0.04 & -1.47 & 0.00 & -1.68 & 0.09 \\
BD-19 6363 & -2.52 & 0.04 & -1.96 & 0.03 & -- & -- & -1.99 & 0.15 & -2.03 & 0.06 & -2.58 & 0.00 & -2.27 & 0.02 \\
CD-23 1855 & -2.01 & 0.11 & -1.51 & 0.07 & -1.92 & 0.11 & -1.55 & 0.11 & -1.55 & 0.05 & -1.80 & 0.00 & -1.65 & 0.11 \\
CD-27 16505 & -1.39 & 0.05 & -0.86 & 0.07 & -1.25 & 0.07 & -0.96 & 0.14 & -0.99 & 0.05 & -- & -- & -0.93 & 0.14 \\
CD-28 10039 & -1.09 & 0.06 & -0.74 & 0.06 & -0.95 & 0.07 & -0.91 & 0.13 & -0.87 & 0.04 & -- & -- & -0.96 & 0.13 \\
CD-28 16762 & -1.49 & 0.04 & -1.21 & 0.09 & -1.01 & 0.09 & -1.29 & 0.12 & -1.25 & 0.05 & -1.63 & 0.00 & -1.33 & 0.13 \\
CD-28 17446 & -2.02 & 0.03 & -1.41 & 0.07 & -1.84 & 0.11 & -1.48 & 0.08 & -1.51 & 0.05 & -1.87 & 0.00 & -1.41 & 0.11 \\
CD-29 15930 & -1.82 & 0.03 & -1.24 & 0.05 & -1.64 & 0.13 & -1.43 & 0.12 & -1.39 & 0.06 & -1.72 & 0.00 & -1.42 & 0.12 \\
CD-33 2721 & -1.66 & 0.041 & -1.48 & 0.07 & -1.19 & 0.08 & -1.54 & 0.08 & -1.49 & 0.06 & -1.85 & 0.00 & -1.72 & 0.09 \\
CD-33 15063 & -1.65 & 0.05 & -1.19 & 0.08 & -1.31 & 0.06 & -1.34 & 0.10 & -1.39 & 0.05 & -1.76 & 0.00 & -1.37 & 0.15 \\
CD-36 518 & -1.92 & 0.04 & -1.32 & 0.05 & -1.75 & 0.11 & -1.44 & 0.11 & -1.40 & 0.05 & -1.83 & 0.00 & -1.44 & 0.09 \\
CD-38 13823 & -2.84 & 0.00 & -1.97 & 0.09 & -- & -- & -2.19 & 0.16 & -2.40 & 0.09 & -2.86 & 0.00 & -2.63 & 0.12 \\
CD-39 6037 & -1.21 & 0.09 & -0.64 & 0.08 & -1.01 & 0.07 & -0.75 & 0.07 & -0.85 & 0.05 & -- & -- & -0.85 & 0.15 \\
CD-39 9313 & -1.85 & 0.04 & -1.29 & 0.07 & -1.57 & 0.00 & -1.35 & 0.05 & -1.41 & 0.04 & -- & -- & -1.45 & 0.02 \\
CD-40 471 & -2.73 & 0.05 & -1.93 & 0.00 & -- & -- & -2.02 & 0.09 & -2.11 & 0.09 & -2.71 & 0.00 & -2.23 & 0.09 \\
CD-43 7161 & -1.68 & 0.06 & -1.06 & 0.08 & -1.50 & 0.09 & -1.17 & 0.12 & -1.21 & 0.06 & -- & -- & -1.15 & 0.10 \\
CD-44 12644 & -1.54 & 0.04 & -0.98 & 0.06 & -1.23 & 0.13 & -1.16 & 0.07 & -1.03 & 0.03 & -1.32 & 0.00 & -1.32 & 0.10 \\
CD-44 15269 & -1.63 & 0.06 & -0.88 & 0.08 & -1.43 & 0.09 & -0.97 & 0.09 & -1.09 & 0.09 & -1.48 & 0.00 & -1.17 & 0.15 \\
CD-52 976 & -2.18 & 0.02 & -1.50 & 0.05 & -2.11 & 0.00 & -1.55 & 0.05 & -1.59 & 0.07 & -2.03 & 0.00 & -1.72 & 0.12 \\
CD-57 1959 & -2.19 & 0.03 & -1.51 & 0.04 & -2.09 & 0.00 & -1.54 & 0.07 & -1.64 & 0.05 & -- & -- & -1.63 & 0.11 \\
CD-62 15 & -2.23 & 0.04 & -1.58 & 0.08 & -1.99 & 0.00 & -1.73 & 0.14 & -1.74 & 0.05 & -2.27 & 0.00 & -1.89 & 0.03 \\
HD 6518 & -1.52 & 0.04 & -0.97 & 0.07 & -1.36 & 0.08 & -1.05 & 0.08 & -1.05 & 0.06 & -1.34 & 0.00 & -0.97 & 0.12 \\
HD 19367 & -2.32 & 0.09 & -1.66 & 0.08 & -2.24 & 0.00 & -1.79 & 0.09 & -1.77 & 0.05 & -1.95 & 0.00 & -1.91 & 0.08 \\
TYC 6002-1724-1 & -2.54 & 0.06 & -1.84 & 0.09 & -- & -- & -1.95 & 0.07 & -1.96 & 0.05 & -2.40 & 0.00 & -2.19 & 0.04 \\
TYC 6199-351-1 & -1.77 & 0.05 & -1.09 & 0.09 & -1.51 & 0.09 & -1.15 & 0.07 & -1.30 & 0.05 & -1.71 & 0.00 & -1.25 & 0.01 \\
TYC 8161-753-1 & -1.78 & 0.04 & -1.11 & 0.07 & -1.55 & 0.07 & -1.18 & 0.12 & -1.28 & 0.05 & -1.78 & 0.00 & -1.22 & 0.14 \\
TYC 8633-2281-1 & -2.13 & 0.13 & -1.49 & 0.06 & -1.88 & 0.11 & -1.61 & 0.07 & -1.67 & 0.06 & -- & -- & -1.69 & 0.15 \\
    \hline \hline
\end{tabular}
\end{table}
\begin{table}
\footnotesize
\centering
\caption{Abundances and line-to-line dispersion ($\sigma$) of elements in their ionisation state from \ion{Ti}{i} to \ion{Mn}{ii} in [X/H].}
\label{abu2_mince1}
\begin{tabular}{lcccccccccccccccc}
\hline \hline
Star & \ion{Ti}{i} & $\sigma$ & \ion{Ti}{ii} & $\sigma$ & \ion{V}{i} & $\sigma$ & \ion{V}{ii} &$\sigma$ & \ion{Cr}{i} & $\sigma$ & \ion{Cr}{ii} & $\sigma$ & 
\ion{Mn}{i} & $\sigma$ &\ion{Mn}{ii} &$\sigma$\\
\hline
BD-11 3235 & -1.75 & 0.07 & -1.51 & 0.08 & -2.10 & 0.07 & -1.94 & 0.14 & -2.10 & 0.10 & -1.89 & 0.07 & -2.39 & 0.08 & -2.46 & 0.00 \\
BD-13 934 & -1.56 & 0.06 & -1.44 & 0.08 & -1.92 & 0.06 & -1.94 & 0.05 & -1.89 & 0.09 & -1.76 & 0.03 & -2.21 & 0.09 & -- & -- \\
BD-13 3195 & -0.92 & 0.08 & -0.99 & 0.08 & -1.25 & 0.09 & -1.34 & 0.20 & -1.29 & 0.11 & -1.32 & 0.00 & -1.55 & 0.09 & -- & -- \\
BD-15 5449 & -1.87 & 0.05 & -1.71 & 0.04 & -2.20 & 0.07 & -2.25 & 0.04 & -2.15 & 0.14 & -2.05 & 0.08 & -2.49 & 0.11 & -- & -- \\
BD-16 2232 & -1.41 & 0.06 & -1.38 & 0.09 & -1.76 & 0.05 & -1.76 & 0.02 & -1.75 & 0.09 & -1.69 & 0.03 & -2.01 & 0.07 & -2.01 & 0.00 \\
BD-17 3143 & -1.19 & 0.09 & -1.48 & 0.13 & -1.54 & 0.07 & -1.95 & 0.00 & -1.58 & 0.11 & -1.81 & 0.06 & -1.91 & 0.09 & -- & -- \\
BD-19 6363 & -2.14 & 0.07 & -1.92 & 0.09 & -2.63 & 0.14 & -2.27 & 0.09 & -2.58 & 0.12 & -2.42 & 0.14 & -2.89 & 0.04 & -2.65 & 0.00 \\
CD-23 1855 & -1.57 & 0.07 & -1.57 & 0.11 & -1.93 & 0.07 & -1.86 & 0.07 & -1.98 & 0.07 & -1.96 & 0.09 & -2.23 & 0.05 & -2.13 & 0.00 \\
CD-27 16505 & -0.95 & 0.08 & -0.92 & 0.10 & -1.24 & 0.07 & -0.91 & 0.00 & -1.25 & 0.13 & -1.09 & 0.00 & -1.52 & 0.13 & -- & -- \\
CD-28 10039 & -0.81 & 0.08 & -0.92 & 0.09 & -1.09 & 0.08 & -0.82 & 0.00 & -1.06 & 0.14 & -0.98 & 0.00 & -1.46 & 0.12 & -- & -- \\
CD-28 16762 & -1.29 & 0.07 & -1.25 & 0.08 & -1.60 & 0.07 & -1.61 & 0.13 & -1.58 & 0.13 & -1.55 & 0.10 & -1.84 & 0.05 & -- & -- \\
CD-28 17446 & -1.53 & 0.06 & -1.27 & 0.08 & -1.89 & 0.04 & -1.99 & 0.00 & -1.87 & 0.11 & -1.52 & 0.00 & -2.16 & 0.09 & -- & -- \\
CD-29 15930 & -1.38 & 0.06 & -1.29 & 0.06 & -1.75 & 0.06 & -1.99 & 0.05 & -1.77 & 0.09 & -1.66 & 0.04 & -2.08 & 0.12 & -- & -- \\
CD-33 2721 & -1.49 & 0.07 & -1.56 & 0.07 & -1.85 & 0.04 & -1.95 & 0.04 & -1.84 & 0.08 & -1.77 & 0.25 & -2.12 & 0.06 & -2.09 & 0.00 \\
CD-33 15063 & -1.36 & 0.08 & -1.25 & 0.13 & -1.75 & 0.06 & -1.84 & 0.09 & -1.74 & 0.09 & -1.54 & 0.00 & -2.07 & 0.10 & -- & -- \\
CD-36 518 & -1.44 & 0.06 & -1.34 & 0.05 & -1.80 & 0.05 & -1.70 & 0.03 & -1.74 & 0.09 & -1.59 & 0.01 & -2.03 & 0.06 & -2.01 & 0.00 \\
CD-38 13823 & -2.42 & 0.06 & -2.44 & 0.11 & -2.72 & 0.15 & -2.72 & 0.06 & -2.52 & 0.08 & -2.53 & 0.06 & -2.56 & 0.09 & -2.63 & 0.00 \\
CD-39 6037 & -0.87 & 0.07 & -0.77 & 0.08 & -1.15 & 0.07 & -0.90 & 0.00 & -1.06 & 0.16 & -- & -- & -1.20 & 0.11 & -- & -- \\
CD-39 9313 & -1.36 & 0.06 & -1.25 & 0.06 & -1.69 & 0.05 & -- & -- & -1.66 & 0.06 & -1.62 & 0.07 & -1.97 & 0.12 & -- & -- \\
CD-40 471 & -2.12 & 0.09 & -2.08 & 0.09 & -2.51 & 0.09 & -2.58 & 0.11 & -2.51 & 0.06 & -2.55 & 0.14 & -2.85 & 0.07 & -2.73 & 0.00 \\
CD-43 7161 & -1.17 & 0.07 & -1.07 & 0.07 & -1.53 & 0.09 & -- & -- & -1.51 & 0.11 & -1.41 & 0.05 & -1.81 & 0.13 & -- & -- \\
CD-44 12644 & -0.98 & 0.06 & -1.22 & 0.10 & -1.31 & 0.04 & -- & -- & -1.39 & 0.07 & -1.10 & 0.00 & -1.61 & 0.05 & -- & -- \\
CD-44 15269 & -1.10 & 0.07 & -0.94 & 0.07 & -1.47 & 0.05 & -- & -- & -1.43 & 0.07 & -1.39 & 0.08 & -1.76 & 0.10 & -- & -- \\
CD-52 976 & -1.64 & 0.07 & -1.53 & 0.13 & -1.97 & 0.06 & -1.78 & 0.00 & -1.97 & 0.08 & -1.88 & 0.03 & -2.23 & 0.08 & -2.14 & 0.00 \\
CD-57 1959 & -1.73 & 0.07 & -1.51 & 0.05 & -2.07 & 0.07 & -1.86 & 0.08 & -2.03 & 0.11 & -1.76 & 0.03 & -2.28 & 0.05 & -2.28 & 0.00 \\
CD-62 15 & -1.73 & 0.13 & -1.72 & 0.09 & -2.09 & 0.13 & -2.06 & 0.09 & -2.17 & 0.09 & -2.11 & 0.09 & -2.41 & 0.04 & -2.33 & 0.00 \\
HD 6518 & -1.08 & 0.08 & -0.99 & 0.07 & -1.41 & 0.09 & -1.51 & 0.09 & -1.41 & 0.11 & -1.38 & 0.08 & -1.65 & 0.06 & -- & -- \\
HD 19367 & -1.75 & 0.08 & -1.79 & 0.08 & -2.09 & 0.07 & -2.08 & 0.00 & -2.10 & 0.09 & -2.11 & 0.04 & -2.37 & 0.05 & -2.31 & 0.00 \\
TYC 6002-1724-1 & -2.01 & 0.06 & -1.95 & 0.07 & -2.39 & 0.03 & -2.31 & 0.07 & -2.44 & 0.07 & -2.41 & 0.09 & -2.79 & 0.07 & -2.59 & 0.00 \\
TYC 6199-351-1 & -1.36 & 0.07 & -1.08 & 0.07 & -1.72 & 0.06 & -1.74 & 0.12 & -1.76 & 0.08 & -1.48 & 0.11 & -1.97 & 0.07 & -- & -- \\
TYC 8161-753-1 & -1.43 & 0.07 & -1.11 & 0.09 & -1.81 & 0.06 & -1.67 & 0.11 & -1.80 & 0.12 & -1.49 & 0.09 & -2.11 & 0.09 & -- & -- \\
TYC 8633-2281-1 & -1.68 & 0.08 & -1.46 & 0.12 & -1.99 & 0.04 & -2.05 & 0.04 & -1.99 & 0.08 & -1.85 & 0.06 & -2.27 & 0.08 & -- & -- \\
\hline \hline
\end{tabular}
\end{table}

\begin{table}
\footnotesize
\centering
\caption{Abundances and line-to-line dispersion ($\sigma$) of elements in their ionisation state from \ion{Fe}{i} to \ion{Zn}{i} in [X/H].}
\label{abu3_mince1}
\begin{tabular}{lcccccccccccc}
\hline \hline
Star & \ion{Fe}{i} & $\sigma$ & \ion{Fe}{ii} & $\sigma$ & \ion{Co}{i} & $\sigma$ & \ion{Ni}{i} & $\sigma$ & \ion{Cu}{i} & $\sigma$ & \ion{Zn}{i} & $\sigma$ \\
\hline
BD-11 3235 & $-$2.11 & 0.14 & $-$1.75 & 0.13 & $-$1.97 & 0.13 & $-$2.09 & 0.15 & $-$2.59 & 0.12 & $-$1.98 & 0.00 \\
BD-13 934 & $-$1.93 & 0.12 & $-$1.73 & 0.12 & $-$1.81 & 0.13 & $-$1.91 & 0.13 & $-$2.33 & 0.14 & $-$1.84 & 0.00 \\
BD-13 3195 & $-$1.41 & 0.15 & $-$1.43 & 0.12 & $-$1.19 & 0.15 & $-$1.34 & 0.13 & $-$1.61 & 0.11 & $-$1.46 & 0.03 \\
BD-15 5449 & $-$2.19 & 0.12 & $-$1.94 & 0.09 & $-$2.05 & 0.14 & $-$2.17 & 0.14 & $-$2.68 & 0.11 & -- & -- \\
BD-16 2232 & $-$1.77 & 0.12 & $-$1.71 & 0.10 & $-$1.63 & 0.15 & $-$1.76 & 0.15 & $-$2.20 & 0.14 & $-$1.83 & 0.00 \\
BD-17 3143 & $-$1.68 & 0.14 & $-$1.91 & 0.11 & $-$1.54 & 0.13 & $-$1.71 & 0.15 & $-$2.06 & 0.06 & $-$1.94 & 0.00 \\
BD-19 6363 & $-$2.51 & 0.12 & $-$2.37 & 0.13 & $-$2.49 & 0.08 & $-$2.44 & 0.13 & -- & -- & $-$2.25 & 0.02 \\
CD-23 1855 & $-$1.96 & 0.11 & $-$1.89 & 0.09 & $-$1.84 & 0.11 & $-$1.93 & 0.11 & $-$2.41 & 0.09 & $-$1.95 & 0.02 \\
CD-27 16505 & $-$1.33 & 0.16 & $-$1.09 & 0.14 & $-$1.18 & 0.14 & $-$1.29 & 0.13 & $-$1.73 & 0.09 & $-$1.49 & 0.00 \\
CD-28 10039 & $-$1.12 & 0.20 & $-$1.02 & 0.14 & $-$1.10 & 0.15 & $-$1.16 & 0.11 & $-$1.46 & 0.12 & $-$1.45 & 0.01 \\
CD-28 16762 & $-$1.63 & 0.13 & $-$1.52 & 0.12 & $-$1.52 & 0.13 & $-$1.66 & 0.13 & $-$2.09 & 0.13 & $-$1.69 & 0.00 \\
CD-28 17446 & $-$1.84 & 0.17 & $-$1.43 & 0.14 & $-$1.79 & 0.14 & $-$1.85 & 0.14 & $-$2.35 & 0.10 & $-$1.85 & 0.01 \\
CD-29 15930 & $-$1.79 & 0.11 & $-$1.66 & 0.12 & $-$1.67 & 0.15 & $-$1.79 & 0.14 & $-$2.16 & 0.13 & -- & -- \\
CD-33 2721 & $-$1.88 & 0.12 & $-$1.91 & 0.13 & $-$1.81 & 0.13 & $-$1.91 & 0.14 & $-$2.34 & 0.08 & $-$2.02 & 0.00 \\
CD-33 15063 & $-$1.80 & 0.16 & $-$1.61 & 0.15 & $-$1.65 & 0.15 & $-$1.81 & 0.15 & $-$2.20 & 0.12 & $-$1.75 & 0.01 \\
CD-36 518 & $-$1.74 & 0.11 & $-$1.59 & 0.12 & $-$1.73 & 0.13 & $-$1.81 & 0.11 & $-$2.28 & 0.12 & $-$1.81 & 0.03 \\
CD-38 13823 & $-$2.46 & 0.14 & $-$2.45 & 0.14 & $-$2.35 & 0.12 & $-$2.38 & 0.11 & $-$3.24 & 0.00 & $-$2.59 & 0.02 \\
CD-39 6037 & $-$0.99 & 0.16 & $-$0.78 & 0.11 & $-$1.03 & 0.13 & $-$1.04 & 0.12 & $-$1.49 & 0.17 & -- & -- \\
CD-39 9313 & $-$1.68 & 0.15 & $-$1.47 & 0.15 & $-$1.57 & 0.15 & $-$1.67 & 0.13 & $-$2.16 & 0.12 & $-$1.63 & 0.00 \\
CD-40 471 & $-$2.48 & 0.14 & $-$2.47 & 0.13 & $-$2.36 & 0.11 & $-$2.49 & 0.13 & $-$2.95 & 0.11 & $-$2.529 & 0.03 \\
CD-43 7161 & $-$1.50 & 0.13 & $-$1.34 & 0.16 & $-$1.41 & 0.14 & $-$1.50 & 0.13 & $-$1.99 & 0.09 & $-$1.59 & 0.00 \\
CD-44 12644 & $-$1.44 & 0.12 & $-$1.66 & 0.07 & $-$1.26 & 0.12 & $-$1.46 & 0.14 & $-$1.84 & 0.07 & $-$1.65 & 0.00 \\
CD-44 15269 & $-$1.39 & 0.15 & $-$1.19 & 0.17 & $-$1.31 & 0.12 & $-$1.37 & 0.14 & $-$1.85 & 0.09 & -- & -- \\
CD-52 976 & $-$1.90 & 0.09& $-$1.79 & 0.09 & $-$1.86 & 0.09 & $-$1.92 & 0.09 & $-$2.44 & 0.07 & $-$1.86 & 0.02 \\
CD-57 1959 & $-$2.00 & 0.12 & $-$1.71 & 0.11 & $-$1.94 & 0.08 & $-$2.02 & 0.09 & $-$2.55 & 0.08 & $-$1.91 & 0.03 \\
CD-62 15 & $-$2.11 & 0.12 & $-$2.06 & 0.11 & $-$2.06 & 0.15 & $-$2.11 & 0.13 & $-$2.47 & 0.06 & $-$2.06 & 0.02 \\
HD 6518 & $-$1.44 & 0.14 & $-$1.29 & 0.14 & $-$1.31 & 0.13 & $-$1.43 & 0.12 & $-$1.88 & 0.07 & $-$1.47 & 0.04 \\
HD 19367 & $-$2.12 & 0.14 & $-$2.08 & 0.11 & $-$2.00 & 0.11 & $-$2.13& 0.14 & $-$2.59 & 0.05 & $-$2.23 & 0.03 \\
TYC 6002-1724-1 & $-$2.34 & 0.12 & $-$2.30 & 0.08 & $-$2.34 & 0.08 & $-$2.35 & 0.11 & $-$2.88 & 0.00 & $-$2.35 & 0.03 \\
TYC 6199-351-1 & $-$1.66 & 0.16 & $-$1.29 & 0.18 & $-$1.53 & 0.15 & $-$1.62 & 0.14 & $-$2.09 & 0.11 & $-$1.54 & 0.05 \\
TYC 8161-753-1 & $-$1.77 & 0.13 & $-$1.56 & 0.12 & $-$1.66 & 0.12 & $-$1.76 & 0.10 & $-$2.20 & 0.15 & $-$1.51 & 0.02 \\
TYC 8633-2281-1 & $-$2.05 & 0.19 & $-$1.80 & 0.12 & $-$1.97 & 0.09 & $-$2.02 & 0.12 & $-$2.45 & 0.14 & $-$2.02 & 0.00 \\
\hline \hline
\end{tabular}
\end{table}

\begin{table}
\footnotesize
\centering
\caption{Abundances and line-to-line dispersion ($\sigma$) of elements in their ionisation state from \ion{Rb}{i} to \ion{La}{ii} in [X/H].}
\label{abu4_mince1}
\begin{tabular}{lcccccccccccc}
\hline \hline
Star & \ion{Rb}{i} & $\sigma$& \ion{Sr}{i} & $\sigma$ & \ion{Y}{ii} & $\sigma$& \ion{Zr}{ii} & $\sigma$ & \ion{Ba}{ii} & $\sigma$ & \ion{La}{ii} & $\sigma$ \\
\hline
BD-11 3235 & $-$1.46& 0.00&$-$2.53 & 0.00 & $-$2.00 & 0.06 & $-$1.52 & 0.07 & $-$1.92 & 0.08 & $-$2.03 & 0.05 \\
BD-13 934 & --&--&$-$1.97 & 0.00 & $-$1.66 & 0.07 & $-$1.25 & 0.12 & $-$1.66 & 0.05 & $-$1.60 & 0.03 \\
BD-13 3195 & $-$0.82&0.00& -- & -- & $-$1.07 & 0.05 & $-$0.71 & 0.09 & $-$0.95 & 0.05 & $-$1.08 & 0.07 \\
BD-15 5449 & --&--&$-$2.46 & 0.00 & $-$2.18 & 0.06 & $-$1.75 & 0.18 & $-$1.94 & 0.08 & $-$2.06 & 0.03 \\
BD-16 2232 & --&--&$-$2.04 & 0.00 & $-$1.72 & 0.06 & $-$1.25 & 0.06 & $-$1.66 & 0.04 & $-$1.54 & 0.05 \\
BD-17 3143 & --&--&$-$1.92 & 0.00 & $-$1.88 & 0.08 & $-$1.35 & 0.07 & $-$1.68 & 0.03 & $-$1.61 & 0.06 \\
BD-19 6363 & --&--&$-$2.61 & 0.00 & $-$2.39 & 0.04 & $-$1.97 & 0.08 & $-$2.39 & 0.04 & $-$2.29 & 0.02 \\
CD-23 1855 & --&--&$-$2.18 & 0.00 & $-$1.70 & 0.07 & $-$1.20 & 0.05 & $-$1.68 & 0.06 & $-$1.55 & 0.06 \\
CD-27 16505 & $-$1.35&0.00&-- & -- & $-$1.19 & 0.10 & $-$0.76 & 0.01 & $-$1.04 & 0.00 & $-$1.08 & 0.01 \\
CD-28 10039 & $-$1.10&0.00&-- & -- & $-$1.04 & 0.00 & $-$0.67 & 0.03 & $-$0.84 & 0.09 & $-$0.69 & 0.07 \\
CD-28 16762 & $-$1.23&0.00&$-$1.96 & 0.00 & -$-$.64 & 0.09 & $-$1.12 & 0.03 & $-$1.35 & 0.05 & $-$1.35 & 0.09 \\
CD-28 17446 & --&--&$-$2.35 & 0.00 & $-$1.66 & 0.08 & $-$1.15 & 0.06 & $-$1.39 & 0.01 & $-$1.42 & 0.05 \\
CD-29 15930 & --&--&$-$2.15 & 0.00 & $-$1.59 & 0.06 & $-$1.09 & 0.08 & $-$1.54 & 0.07 & $-$1.61 & 0.12 \\
CD-33 2721 & --&--&$-$2.12 & 0.00 & $-$1.96 & 0.09 & $-$1.49 & 0.07 & $-$1.72 & 0.06 & $-$1.74 & 0.06 \\
CD-33 15063 & --&--&$-$1.95 & 0.00 & $-$1.45 & 0.07 & $-$1.01 & 0.12 & $-$0.89 & 0.07 & $-$1.18 & 0.09 \\
CD-36 518 & --&--&$-$1.99 & 0.00 & $-$1.65 & 0.06 & $-$1.19 & 0.05 & $-$1.40 & 0.05 & $-$1.41 & 0.06 \\
CD-38 13823 & --&--&-- & -- & $-$3.15 & 0.01 & $-$2.84 & 0.08 & $-$3.17 & 0.03 & $-$2.94 & 0.04 \\
CD-39 6037 & $-$1.22&0.00&-- & -- & $-$1.060 & 0.07 & -- & -- & $-$0.47 & 0.09 & $-$0.84 & 0.03 \\
CD-39 9313 & --&--&$-$2.06 & 0.00 & $-$1.56 & 0.09 & $-$1.24 & 0.12 & $-$1.08 & 0.06 & $-$1.39 & 0.03 \\
CD-40 471 & $-$1.35&0.00&$-$2.72 & 0.00 & $-$2.56 & 0.08 & $-$2.12 & 0.09 & $-$2.24 & 0.09 & $-$2.28 & 0.06 \\
CD-43 7161 & $-$1.52&0.00&$-$2.08 & 0.00 & $-$1.45 & 0.06 & $-$0.97 & 0.04 & $-$1.13 & 0.04 & $-$1.28 & 0.04 \\
CD-44 12644 & --&--&$-$1.51 & 0.00 & $-$1.482 & 0.09 & $-$0.93 & 0.07 & $-$1.05 & 0.05 & $-$1.05 & 0.07 \\
CD-44 15269 & $-$1.08&0.00&$-$1.79 & 0.00 & $-$1.29 & 0.05 & $-$0.87 & 0.05 & $-$0.79 & 0.09 & $-$1.34 & 0.07 \\
CD-52 976 & --&--&$-$2.29 & 0.00 & $-$2.13 & 0.06 & $-$1.59 & 0.03 & $-$1.76 & 0.11 & $-$1.89 & 0.06 \\
CD-57 1959 & --&--&$-$2.33 & 0.00 & $-$1.96 & 0.08 & $-$1.48 & 0.07 & $-$1.77 & 0.04 & $-$1.75 & 0.06 \\
CD-62 15 & --&--&$-$2.27 & 0.00 & $-$2.09 & 0.07 & $-$1.66 & 0.02 & $-$2.09 & 0.01 & $-$2.05 & 0.03 \\
HD 6518 & $-$1.61&0.00&$-$1.87 & 0.00 & $-$1.442 & 0.11 & $-$0.93 & 0.05 & $-$1.28 & 0.06 & $-$1.32 & 0.07 \\
HD 19367 & --&--&$-$2.38 & 0.00 & $-$2.17 & 0.09 & $-$1.72 & 0.12 & $-$1.89 & 0.04 & $-$1.84 & 0.07 \\
TYC 6002-1724-1 & --&--&$-$2.53 & 0.00 & $-$2.55 & 0.06 & $-$2.03 & 0.01 & $-$2.61 & 0.05 & $-$2.49 & 0.05 \\
TYC 6199-351-1 & $-$1.61&0.00&$-$2.17 & 0.00 & $-$1.59 & 0.11 & $-$1.05 & 0.04 & $-$1.43 & 0.07 & $-$1.53 & 0.05 \\
TYC 8161-753-1 & $-$1.39&0.00&$-$1.95 & 0.00 & $-$1.41 & 0.07 & $-$0.90 & 0.05 & $-$1.13 & 0.04 & $-$1.22 & 0.03 \\
TYC 8633-2281-1 & --&--&$-$2.44 & 0.00 & $-$2.05 & 0.10 & $-$1.47 & 0.05 & $-$1.67 & 0.06 & $-$1.79 & 0.05 \\
\hline \hline
\end{tabular}
\end{table}


\begin{table}
\footnotesize
\centering
\caption{Abundances and line-to-line dispersion ($\sigma$) of elements in their ionisation state from Ce II to Eu II in [X/H].}
\label{abu5_mince1}
\begin{tabular}{lcccccccccc}
\hline \hline
Star & Ce II & $\sigma$ & Pr II & $\sigma$ & Nd II & $\sigma$ & Sm II & $\sigma$ & Eu II & $\sigma$ \\
\hline
BD-11 3235 & $-$2.14 & 0.12 & $-$1.98 & 0.05 & $-$1.91 & 0.06 & $-$1.70 & 0.09 & $-$1.56 & 0.00 \\
BD-13 934 & $-$1.77 & 0.09 & $-$1.62 & 0.01 & $-$1.54 & 0.02 & $-$1.39 & 0.05 & $-$1.19 & 0.00 \\
BD-13 3195 & $-$1.11 & 0.08 & $-$1.16 & 0.02 & -- & -- & $-$0.89 & 0.02 & $-$1.00 & 0.00 \\
BD-15 5449 & $-$2.18 & 0.08 & $-$2.11 & 0.04 & $-$1.99 & 0.03 & $-$1.89 & 0.02 & $-$1.88 & 0.00 \\
BD-16 2232 & $-$1.66 & 0.05 & $-$1.52 & 0.04 & $-$1.45 & 0.01 & $-$1.23 & 0.03 & $-$1.02 & 0.00 \\
BD-17 3143 & $-$1.71 & 0.08 & $-$1.58 & 0.01 & $-$1.55 & 0.03 & $-$1.41 & 0.04 & $-$1.31 & 0.00 \\
BD-19 6363 & $-$2.37 & 0.04 & -- & -- & $-$2.22 & 0.04 & $-$1.97 & 0.05 & $-$2.15 & 0.00 \\
CD-23 1855 & $-$1.67 & 0.01 & -- & -- & $-$1.45 & 0.02 & $-$1.26 & 0.02 & $-$1.15 & 0.00 \\
CD-27 16505 & $-$1.25 & 0.06 & $-$0.93 & 0.00 & -- & -- & $-$0.69 & 0.03 & $-$0.48 & 0.00 \\
CD-28 10039 & $-$1.12 & 0.07 & $-$0.86 & 0.00 & -- & -- & -- & -- & $-$0.49 & 0.00 \\
CD-28 16762 & $-$1.26 & 0.26 & $-$1.32 & 0.00 & $-$1.28 & 0.04 & $-$1.14 & 0.07 & $-$0.93 & 0.00 \\
CD-28 17446 & $-$1.54 & 0.11 & $-$1.38 & 0.07 & $-$1.33 & 0.02&  $-$1.09 & 0.02 & $-$0.86 & 0.00 \\
CD-29 15930 & $-$1.74 & 0.10 & $-$1.58 & 0.03 & -- & -- & $-$1.41 & 0.01 & $-$1.24 & 0.00 \\
CD-33 2721 & $-$1.88& 0.06 & $-$1.73 & 0.00 & $-$1.69 & 0.03 & $-$1.54 & 0.01 & $-$1.44 & 0.00 \\
CD-33 15063 & $-$1.21 & 0.08 & $-$1.43 & 0.00 & $-$1.16 & 0.06 & $-$1.35 & 0.03 & $-$1.43 & 0.00 \\
CD-36 518 & $-$1.49 & 0.05 & $-$1.42 & 0.00 & $-$1.34 & 0.03 & $-$1.22 & 0.04 & $-$1.06 & 0.00 \\
CD-38 13823 & -- & -- & -- & -- & $-$2.79 & 0.04 & $-$2.69 & 0.03 & $-$2.48 & 0.00 \\
CD-39 6037 & $-$0.87 & 0.13 & $-$0.79 & 0.00 & -- & -- & $-$0.48 & 0.09 & $-$0.38 & 0.00 \\
CD-39 9313 & $-$1.46 & 0.0 & $-$1.49 & 0.03 & $-$1.34 & 0.03 & $-$1.31 & 0.07 & $-$1.14 & 0.00 \\
CD-40 471 & $-$2.414 & 0.03 & $-$2.17 & 0.00 & $-$2.17 & 0.06 & $-$1.91 & 0.04 & $-$1.75 & 0.00 \\
CD-43 7161 & $-$1.35 & 0.06 & $-$1.19 & 0.02 & $-$1.14 & 0.00 & $-$1.05 & 0.09 & $-$0.71 & 0.00 \\
CD-44 12644 & $-$1.12 & 0.07 & $-$1.13 & 0.00 & $-$1.01 & 0.02 & $-$0.93 & 0.05 & $-$0.89 & 0.00 \\
CD-44 15269 & $-$1.38 & 0.04 & $-$1.38 & 0.03 & $-$1.26 & 0.06 & $-$1.06 & 0.04 & $-$0.94 & 0.00 \\
CD-52 976 & $-$1.99 & 0.04 & $-$1.86 & 0.00 & $-$1.79 & 0.03 & $-$1.62 & 0.01 & $-$1.56 & 0.00 \\
CD-57 1959 & $-$1.86 & 0.05 & $-$1.71 & 0.00 & $-$1.69 & 0.02 & $-$1.46 & 0.05 & $-$1.23 & 0.00 \\
CD-62 15 & $-$2.16 & 0.02 & -- & -- & $-$1.988 & 0.03 & $-$1.85 & 0.05 & $-$1.76 & 0.00 \\
HD 6518 & $-$1.41 & 0.09 & $-$1.22 & 0.00 & -- & -- & $-$0.97 & 0.01 & $-$0.81 & 0.00 \\
HD 19367 & $-$1.95 & 0.05 & $-$1.72 & 0.00 & $-$1.80 & 0.04 & $-$1.56 & 0.07 & $-$1.35 & 0.00 \\
TYC 6002-1724-1 & -- & -- & -- & -- & $-$2.41 & 0.00 & $-$2.20 & 0.09 & $-$2.26 & 0.00 \\
TYC 6199-351-1 & $-$1.60 & 0.11 & $-$1.46 & 0.02 & --& -- & $-$1.19 & 0.02 & $-$1.07 & 0.00 \\
TYC 8161-753-1 & $-$1.28 & 0.08 & $-$1.35 & 0.00 & $-$1.19 & 0.04 & $-$1.19 & 0.09 & $-$1.07 & 0.00 \\
TYC 8633-2281-1 & $-$1.89 & 0.09 & $-$1.78 & 0.00 & $-$1.74 & 0.04 & $-$1.60 & 0.06 & $-$1.44 & 0.00 \\
\hline \hline
\end{tabular}
\end{table}

\begin{table}
    \centering
    \caption{LTE and NLTE sulfur abundances obtained line by line for the MINCE stars.}
    \label{tab_S_mince}
    \begin{tabular}{lcccccccc}
          \hline \hline
          Star&921.2&922.8&923.7&$\langle$A(S)$\rangle$&921.2&922.8&923.7&$\langle$A(S)$\rangle$\\
          &LTE&LTE&LTE&LTE&NLTE&NLTE&NLTE&NLTE\\
         \hline
            BD-11 3235  &	5.95 &	-- &	-- &	5.95 &	5.88 &	-- &	-- &	5.88\\
  BD-13 3195  &	6.38 &	-- &	-- &	6.38 &	6.09 &	-- &	-- &	6.09\\
  BD-15 5449   &	-- &	5.79 &	5.79 &	5.79 $\pm$ 0.00 &	-- &	5.72 &	5.72 &	5.72 $\pm$	0.00\\
  BD-16 2232   &	5.97 &	-- &	-- &	5.971  &	5.72 &	-- &	-- &	5.72\\
  BD-17 3143   &	5.96 &	-- &	-- &	5.96  &	5.65 &	-- &	-- &	5.65\\
  BD-19 6363   &	5.39 &	-- &	-- &	5.39 &	5.13 &	-- &	-- &	5.13\\
  CD-27 16505  &	6.55 &	6.38 &	6.44 &	6.46 $\pm$ 0.07 &	6.30 &	6.28 &	6.36 &	6.32 $\pm$	0.03\\
  CD-28 10039 &	6.61 &	-- &	-- &	6.61 &	6.28 &	-- &	-- &	6.28\\
  CD-28 16762  &	6.32 &	-- &	-- &	6.32 &	6.02 &	-- &	-- &	6.02\\
  CD-29 15930  &	-- &	5.97 &	-- &	5.97  &	-- &	5.95 &	-- &	5.95 \\
  CD-33 2721   &	5.94 &	-- &	5.69 &	5.82 $\pm$ 0.12 &	5.69 &	-- &	5.55 &	5.62 $\pm$	0.07\\
  CD-33 15063  &	6.10 &	6.25 &	-- &	6.18 $\pm$ 0.07 &	5.99 &	6.18 &	-- &	6.09 $\pm$	0.09\\
  CD-36 518    &	6.17 &	6.19 &	-- &	6.18 $\pm$ 0.02 &	5.86 &	5.92 &	-- &	5.89 $\pm$	0.03\\ 
  CD-38 13823  &	4.91 &	5.02 &	-- &	4.96 $\pm$ 0.06 &	4.72 &	4.83 &	-- &	4.77 $\pm$	0.06\\
  CD-39 6037   &	6.96 &	6.85 &	-- &	6.91 $\pm$ 0.06 &	6.54 &	6.58 &	-- &	6.56 $\pm$	0.02\\
  CD-43 7161   &	6.34 &	6.35 &	-- &	6.35 $\pm$ 0.01 &	6.09 &	6.17 &	-- &	6.13 $\pm$	0.04\\
  CD-52 976    &	5.96 &	5.91 &	-- &	5.93 $\pm$ 0.03 &	5.55 &	5.55 &	-- &	5.55 $\pm$	0.00\\
  CD-57 1959   &	-- &	5.83 &	5.90 &	5.87 $\pm$ 0.03 &	-- &	5.69 &	5.77 &	5.73 $\pm$	0.04\\
  CD-62 15     &	5.73 &	-- &	-- &	5.73 &	5.41 &	-- &	-- &	5.41\\
  TYC 6002-1724-1     &	5.27 &	5.43 &	-- &	5.35 $\pm$ 0.08 &	5.03 &	5.19 &	-- &	5.11 $\pm$	0.08\\
  TYC 6199-351-1     &	6.48 &	6.63 &	-- &	6.56 $\pm$ 0.08 &	6.32 &	6.54 &	-- &	6.43 $\pm$	0.11\\
  TYC 8161-752-1     &	6.34 &	6.41 &	6.18 &	6.31 $\pm$ 0.09 &	6.14 &	6.25 &	6.05 &	6.15 $\pm$	0.08\\
  TYC 8633-2281-1     &	6.69 &	-- &	-- &	6.69  &	6.64 &	-- &	-- &	6.64\\
  \hline \hline
  \end{tabular}
\end{table}

\newpage
\section{Remarks on individual stars}
\label{remarks}

  \begin{table*}[]
       \caption{Radial velocities for the components of SB2 stars.}
       \centering
       \begin{tabular}{llrr}
       \hline
       STAR & MJD & primary & secondary\\
      & (d) & \kms   & \kms \\
      \hline
      CD --52 2441 & 59177.175649715&  $+29.17 \pm 0.03$     & $+15.63 \pm 0.03$  \\
      TYC 5340--1656--1 &  59169.157938165 &$+7.1\phantom{0}\pm 0.3\phantom{0} $ & $+56.7\phantom{0}\pm 0.3\phantom{0}$\\
      \hline
       \end{tabular}
       \label{tab:sb2}
   \end{table*}

 \subsection{BD--14\,52}
   This star   was investigated in the 437+760 wavelength range. It is an active star.
   According to $Gaia$ DR3 this star is an SB1 binary with a period of 306 days,
  a radial velocity semi-amplitude of 1.15 km/s  and centre of mass velocity
  of 60.2 km/s close to our adopted radial velocity. The orbit has a low eccentricity of 0.16.   

  \subsection{BD--17\,4250}
   This metal-poor star has been analysed in  the 437+760 range, it  is active as indicated by
   \ion{Ca}{ii}-H and -K emission and P-Cygni (and inverse) profiles on H$\alpha$.

      \subsection{BD--15\,4109}
   This star is metal-poor, it was investigated in the 437+760 range. It is  active as
   implied by the \ion{Ca}{ii}-H and -K emission and P-Cygni (and inverse) profiles on H$\alpha$.

  \subsection{CD--23\,11064}

  This star is cool, the \citet{mucciarelli21} calibration provides and effective
  temperature of 3748\,K we ran it at fixed \teff\ = 3750\,K to avoid extrapolation.

 \subsection{CD--24\,613}

 This star is cool and was analysed only in the RVS range with the COOL grid.
 \mygi\ required a shift of +5.6 \kms\ with respect to the $Gaia$ radial velocity.

   \subsection{CD--27\,14182}
  Was  analysed in the 437+760 range. There is a tiny emission in \ion{Ca}{ii}-H and -K lines. 
  No evidence of \ion{Li}{i} doublet.

   \subsection{CD--28\,10387}
   Was analysed in the 437+760 range. Normal H$\alpha$,
   tiny emission in \ion{Ca}{ii}-H and -K lines. 
   No evidence of \ion{Li}{i} doublet.
   
   \subsection{CD--29\,9391}
   Was analysed in the RVS range with the COOL grid.
   Our spectrum required a shift of --1.05 \kms\ with respect to the $Gaia$ radial velocity.

      \subsection{CD--31\,17277}

$Gaia$ DR3 classifies it as Long Period Variable (LPV) but does not
provide a period.

   \subsection{CD--32\,13158}
The star has been  analysed in the 437+760  range. Normal H$\alpha$,
   tiny emission in \ion{Ca}{ii}-H and -K lines. 
   No evidence of \ion{Li}{i} doublet.
The star is of solar metallicity and has a sizeable rotational velocity.
Comparison with the evolutionary tracks of \citet{ekstrom2012} implies
a mass of 6 M\sun\ and an age of 70\,Myr.

\subsection{CD--34\,242}
   This star is cool and was analysed only in the RVS range with the COOL grid.
 \mygi\ required a shift of +2.6 \kms\ with respect to the $Gaia$ radial velocity. 

  \subsection{CD--35\,13334}
   This star is cool and was analysed only in the RVS range with the COOL grid.
 \mygi\ required a shift of +1.8 \kms\ with respect to the $Gaia$ radial velocity. 

    \subsection{CD--35\,13661}
    According to $Gaia$ this star is an SB1 binary with a period of 2 days 
    and a semi-amplitude of 1.35 \kms . The centre of mass radial velocity is +59.02 \kms ,
    very close to our adopted radial velocity. The eccentricity is low, 0.12.
   Analysed in the 437+760  range. H$\alpha$ with P-Cygni and emission on \ion{Ca}{ii}-H and -K lines. No evident feature at the \ion{Li}{i} doublet wavelength.

 \subsection{CD--45\,8357}
   
   This star is classified as an RS CVn binary
   in the $Gaia$ DR3 catalogue. Our spectrum does not
   show  a secondary spectrum (as expected for
   an RS CVn). The radial velocity from our UVES spectrum
   is $-20.5 \pm 0.33$ km/s, to be compared
   to the value provided by $Gaia$ $-32.39\pm 5.24$.
   Although the two values are compatible at 2.2$\sigma$, 
   we consider this as a clear indication that the star's radial velocity is
   varying, as  is the very large error in the $Gaia$ radial velocity, 
   incompatible with a star of this brightness.
   The \ion{Ca}{ii} H\& K line show a very strong emission,
   further supporting the RS CVn classification of this star.
   A chemical analysis of such a star is beyond the scope of
    our paper, however from comparison with synthetic spectra we
    estimate the metallicity to be roughly solar and the
    $v\sin i \sim 20$ \kms

    \subsection{CD--48\,12928}
   
   $Gaia$ DR3 classifies this star as LPV with a period of 289.27 days and an amplitude of 0.16 mag
   The effective temperature from the \citet{mucciarelli21} is 3642\,K we run \mygi\  in the RVS range with the COOL grid and  \teff = 3750\,K to avoid extrapolation. 
   Our UVES spectrum requires a shift of -1.1 \kms\ with respect to the Gaia radial velocity.

   \subsection{CD--50\,823}
   Analysed in the 437+760 range. H$\alpha$ shows emissions, \ion{Ca}{ii}-H and -K lines have emission cores. No \ion{Li}{i}  doublet.

    \subsection{CD--50\,877}
   
   $Gaia$ DR3 classifies this star as LPV, but does not provide a period.
   The effective temperature from the \citet{mucciarelli21} is 3733\,K we run \mygi\  in the RVS range with the COOL grid and  \teff = 3750\,K to avoid extrapolation. 
   Our UVES spectrum requires a shift of -2.0 \kms\ with respect to the $Gaia$ radial velocity.
   
      \subsection{CD--52\,2441}
   
   This star is an SB2 binary, the radial velocities of the two components are provided in Table\,\ref{tab:sb2}.

   \subsection{CD--52\,4849}
     According to the $Gaia$ DR3 catalogue this is an SB1 binary with a period of 3.5 days and a semi-amplitude of 1.4 \kms .
     The centre of mass radial velocity is 150.65 \kms , very close to our adopted radial velocity. The eccenticity of the orbit is 0.26.
   Analysed in the 437+760 range. H$\alpha$ shows emissions, \ion{Ca}{ii}-H and -K lines have emission cores. No \ion{Li}{i}  doublet.

    \subsection{CD--58\,294}
   $Gaia$ DR3 classifies it as LPV with a period of 176.72 d and an amplitude of 0.03 mag.  
   The effective temperature from the \citet{mucciarelli21} is 3707\,K we run \mygi\  in the RVS range with the COOL grid and  \teff = 3750\,K to avoid extrapolation. 
   Our UVES spectrum requires a shift of +0.7 \kms\ with respect to the $Gaia$ radial velocity.

     \subsection{CD--59\,6913}
   $Gaia$ DR3 classifies this star as LPV with a period of 344.88 days and an amplitude of 0.05 mag.
   Analysed in the 437+760 range. H$\alpha$ shows emissions, \ion{Ca}{ii}-H and -K lines have emission cores. No \ion{Li}{i}  doublet.
   
   \subsection{CPD--62\,1126}
   Analysed in the 437+760 range, but only for wavelengths larger than 450\,nm. 
    A weak line is visible in the \ion{Li}{i}  doublet range.

 \subsection{HD\,41020}
   
   $Gaia$ DR3 classifies this star as LPV with a period of 238,79 days and an amplitude of 0.07 mag. 
   No \ion{Li}{i} visible, strong emission in \ion{Ca}{ii}-H and -K, H$\alpha$ also with emissions.

     \subsection{TYC\,5340--1656--1} 

        \begin{figure*}
       \centering
       \includegraphics{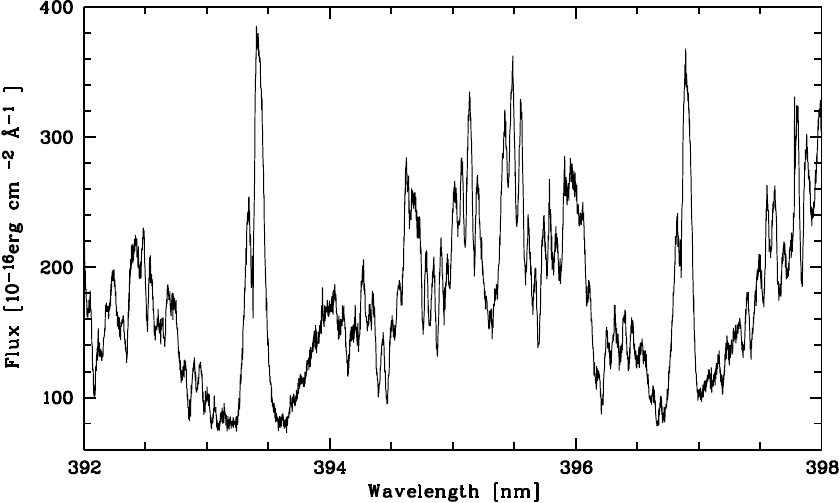}
       \caption{\ion{Ca}{ii} H\& K lines of TYC 5340--1656--1.}
       \label{HK_T5340}
   \end{figure*}
   
   This star is classified as an RS CVn binary (stars with a giant primary of type F--K
   and a dwarf secondary of type G--M), in the $Gaia$ DR3 catalogue.
   Our UVES spectrum shows two almost equally strong line systems, the
   cross correlation peaks provide the radial velocities provided in Table\,\ref{tab:sb2}.
     This fact clearly rules out the interpretation of this system as a classical
   RS CVn system since the two companions are of comparable luminosity.
   They are certainly giants, given the fact that the Balmer lines show no detectable wings.
   However it is interesting that both stars are chromospherically active (a typical
   signature of RS CVn systems) and this is obvious in Fig.\,\ref{HK_T5340} where
   the core emissions of the \ion{Ca}{ii} H and K lines of both companions are clearly visible.
   A chemical analysis of such a star is beyond the scope of our paper, and in any case knowledge of the luminosity ratio of
   the two stars is required. By a quick comparison of the observed spectrum with synthetic spectra
   we expect the metallicity to be roughly solar and the $v\sin i$ of both stars between 20 and 30 km/s. 

    \subsection{TYC\,5422--1192--1}
    
    The star is a G giant member of the Open Cluster NGC 2423. The age of the cluster is 
    estimated to be 310 Myr by \citep{Bossini} and \citet{Dias21} provide a metallicity of +0.12. Our UVES spectrum shows that the star
    is a fast rotator, probably above 100 km/s.

     \subsection {TYC\,5763--1084--1}
   Analysed in the RVS range with the COOL grid, our UVES spectrum requires a shift
   of --1.2 \kms\ with respect to the $Gaia$ radial velocity.

   \subsection{TYC\,6108--150--1}

   According to the $Gaia$ DR3 catalogue this star is an SB1 binary, with a period of 390.6 days and a semi-amplitude of 13.9 \kms .
   The radial velocity of the centre of mass is 103.81 \kms , almost 6\,\kms from our adopted radial
   velocity. The eccentricity is very  low 0.06.
   
   \subsection{TYC\,6195--815--1}
   $Gaia$ DR3 classifies this star as LPV with a period of 365.63 days and an amplitude of 0.09 mag. 
   Analysed in the RVS range with the COOL grid, our UVES spectrum requires a shift
   of +2.2 \kms\ with respect to the $Gaia$ radial velocity.

   \subsection{TYC\,8400--1610--1}
   The effective temperature from the \citet{mucciarelli21} is 3748\,K we run \mygi\  in the RVS range with the COOL grid and  \teff = 3750\,K to avoid extrapolation. 
   Our UVES spectrum requires a shift of --2.3 \kms\ with respect to the $Gaia$ radial velocity.
   
      \subsection{TYC\,8394--14--1}
   
   $Gaia$ DR3 classifies this star as an LPV with a period of 603.23 days
   and an amplitude of 0.05 mag.

   \subsection{TYC\,8584--127--1}
   
        This stars has a measurable rotational velocity, that suggests that
   it is a young evolved massive star, similar to those studied by \citet{lombardo2021}.
   From a fot to an isolated \ion{Fe}{i} line  we derive a
   projected rotational velocity of 11.5 kms$^{-1}$.
   If we compare the position of the star in the CMD with BASTI isochrones \citep{BASTI21} of metallicity --0.40 the implied age
   is 300 Myr and the mass 3.2 M\sun.
   The radial velocity measured from our blue UVES spectrum is 21.67\,kms$^{-1}$, in excellent
   agreement with the $Gaia$ radial velocity ($21.97\pm 0.23$). Our radial velocity has not been zeroed on telluric lines
   so that the associated error is dominated by the centering of the star on the slit and this is of the order
   of 0.5\,km$^{-1}$.

      \begin{figure}
       \centering
       \resizebox{9.5cm}{!}{\includegraphics{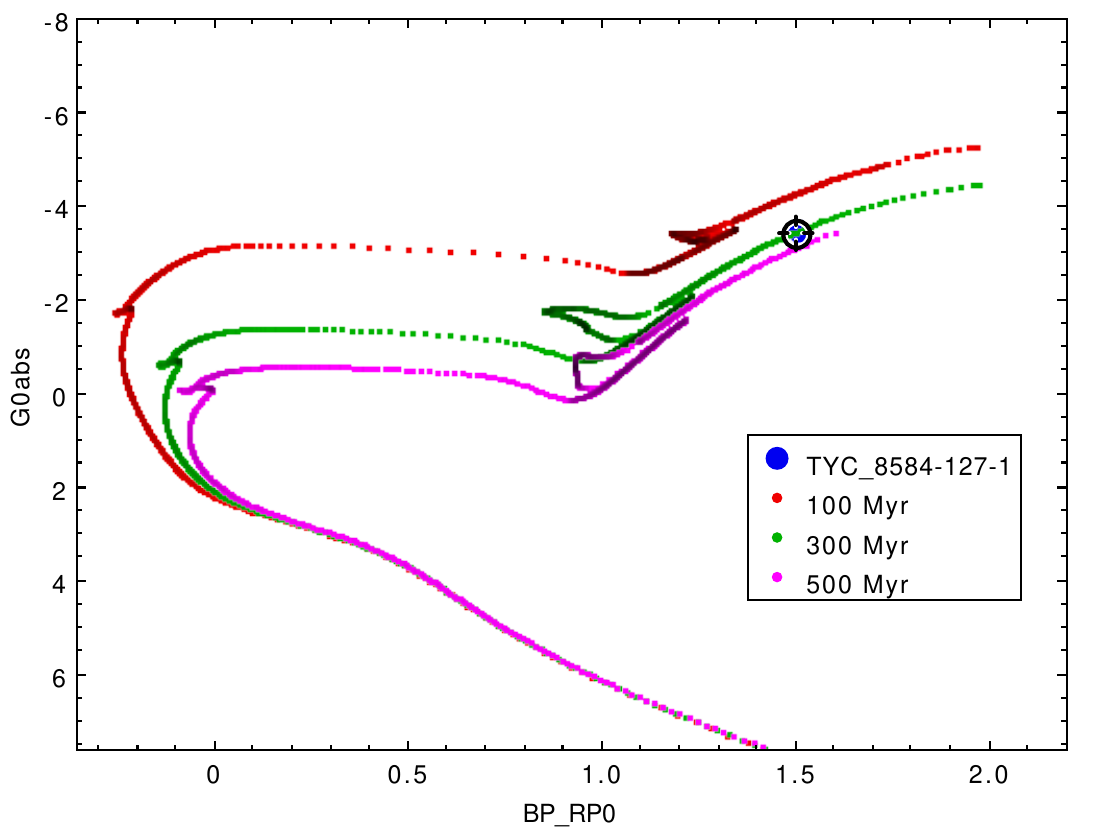}}
       \caption{Position in the $Gaia$ CMD of TYC\,8584--127--1 compared to three BASTI isochrones with metallicity --0.40.}
       \label{fig:TYC_8584-127-1}
   \end{figure}
   
   We ran \mygi\ in the RVS range and found 
   [Fe/H]=$-0.58\pm 0.09$ from 16 \ion{Fe}{i} lines.
It is remarkable that such a young star is moderately metal-poor. Its Galactic orbit is circular on the Galactic plane at roughly the solar radius.
       
\end{appendix}
   
\end{document}